\documentclass[twocolumn]{aastex63}
\usepackage{amsmath}
\usepackage{hyperref}

\newcommand{\paperii}{Paper II}

\newcommand{\mdot}{\dot{M}_w}
\newcommand{\pdot}{\dot{p}_w}
\newcommand{\Lwind}{\mathcal{L}_w}
\newcommand{\Qo}{Q_0}

\newcommand{\RWeaver}{R_{\rm ED}} 
\newcommand{\PWeaver}{P_{\rm ED}}
\newcommand{\prWeaver}{p_{\rm ED}}
\newcommand{\RWeavert}{R_{{\rm ED},\theta}} 
\newcommand{\PWeavert}{P_{{\rm ED},\theta}}
\newcommand{\prWeavert}{p_{{\rm ED},\theta}}

\newcommand{\RMD}{R_{\rm MD}}

\newcommand{\prMD}{p_{\rm MD}}
\newcommand{\RMDa}{R_{{\rm MD},\alpha}}
\newcommand{\PMDa}{P_{{\rm MD},\alpha}}
\newcommand{\prMDa}{p_{{\rm MD},\alpha}}

\newcommand{\Req}{R_{\rm eq}}
\newcommand{\teq}{t_{\rm eq}}
\newcommand{\ReqMD}{R_{\rm eq, MD}}
\newcommand{\teqMD}{t_{\rm eq, MD}}
\newcommand{\ReqED}{R_{\rm eq, ED}}
\newcommand{\teqED}{t_{\rm eq, ED}}

\newcommand{\Rch}{R_{\rm ch}}

\newcommand{\tSt}{t_{\rm St}}
\newcommand{\tStMD}{t_{\rm St, MD}}
\newcommand{\tStED}{t_{\rm St, ED}}

\newcommand{\tss}{t_{\rm ss}}
\newcommand{\tssMD}{t_{\rm ss, MD}}
\newcommand{\tssED}{t_{\rm ss, ED}}

\newcommand{\tot}{t_{\rm catch}}
\newcommand{\totMD}{t_{\rm catch, MD}}
\newcommand{\totED}{t_{\rm catch, ED}}
\newcommand{\tswitch}{t_{\rm switch}}

\newcommand{\RSt}{R_{\rm St}}
\newcommand{\RSp}{R_{\rm Sp}}
\newcommand{\Phii}{P_{i}}
\newcommand{\Thii}{T_{i}}
\newcommand{\tdio}{t_{{\rm d},i,0}}

\newcommand{\trec}{t_{\rm rec}}

\newcommand{\ReqRSt}{\zeta}
\newcommand{\ReqRStMD}{\zeta_{\rm MD}}
\newcommand{\ReqRStED}{\zeta_{\rm ED}}

\newcommand{\nhat}{\mathbf{\hat{n}}}

\newcommand{\tsf}{t_{\rm sf}}

\newcommand{\Ri}{\mathcal{R}_{i}}
\newcommand{\dRi}{\dot{\mathcal{R}}_i}
\newcommand{\Rw}{\mathcal{R}_{w}}
\newcommand{\dRw}{\dot{\mathcal{R}}_w}

\newcommand{\rfree}{\mathcal{R}_{f}}

\newcommand{\nH}{n_{\rm H}}
\newcommand{\nHi}{n_{{\rm H},i}}
\newcommand{\nHbar}{\overline{n}_{\rm H}}
\newcommand{\nHrms}{n_{\rm H, rms}}
\newcommand{\mH}{m_{\rm H}}
\newcommand{\muH}{\mu_{\rm H}}
\newcommand{\sfe}{\varepsilon_{*}}
\newcommand{\rhobar}{\bar{\rho}}
\newcommand{\ecool}{\dot{E}_{\rm cool}}

\newcommand{\tcool}{t_{\rm cool}}

\newcommand{\Msh}{M_{\rm sh}}

\newcommand{\Mcl}{M_{\rm cl}}
\newcommand{\Mst}{M_*}
\newcommand{\Rcl}{R_{\rm cl}}
\newcommand{\Scl}{\Sigma_{\rm cl}}

\newcommand{\ci}{c_i}

\newcommand{\tff}{t_{\rm ff}}

\newcommand{\vw}{{\cal{V}}_w}

\newcommand{\Phot}{P_{\rm hot}}

\newcommand{\pc}{\mathrm{\, pc}}
\newcommand{\pcc}{\mathrm{\, cm}^{-3}}

\newcommand{\Myr}{\mathrm{\, Myr}}

\newcommand{\kms}{\mathrm{\, km\, s^{-1}}}

\newcommand{\Msun}{\, \mathrm{M}_\odot}

\newcommand{\rhobg}{\rho_{\rm bkgnd}}
\newcommand{\rhozo}{\rho_{01}}
\newcommand{\krho}{k_{\rho}}
\newcommand{\kcrit}{k_{\rm crit}}
\newcommand{\etain}{\eta_{\rm in}}
\newcommand{\Grad}{\Gamma_{\rm rad}}
\newcommand{\Qrec}{Q_{\rm rec}}
\newcommand{\nHbartrap}{\overline{n}_{\rm H, trap}}

\newcommand{\pscl}{p_0}

\newcommand{\vout}{v_{\rm out}}
\newcommand{\voutavg}{ \left\langle \vout \right\rangle}
\newcommand{\Abub}{A_w}

\newcommand{\strom}{Str\"{o}mgren }

\newcommand{\WRM}{$\texttt{MWR}$}

\received{XXX}
\revised{XXX}
\accepted{XXX}

\submitjournal{ApJ}

\defcitealias{Lancaster21a}{L21a}

\shorttitle{Co-Evolution of Winds and Photoionized Gas I}
\shortauthors{Lancaster et al.}

\begin{document}

\title{The Co-Evolution of Stellar Wind-blown Bubbles and Photoionized Gas I:\\ Physical Principles and a Semi-Analytic Model}

\correspondingauthor{Lachlan Lancaster}
\email{ltl2125@columbia.edu}

\author[0000-0002-0041-4356]{Lachlan Lancaster}
\thanks{Simons Fellow}
\affiliation{Department of Astronomy, Columbia University,  550 W 120th St, New York, NY 10025, USA}
\affiliation{Center for Computational Astrophysics, Flatiron Institute, 162 5th Avenue, New York, NY 10010, USA}

\author[0000-0001-6228-8634]{Jeong-Gyu Kim}
\affil{Quantum Universe Center, Korea Institute for Advanced Study, Hoegiro 85, Seoul 02455, Republic of Korea}
\affiliation{Division of Science, National Astronomical Observatory of Japan, Mitaka, Tokyo 181-0015, Japan}

\author[0000-0003-2630-9228]{Greg L. Bryan}
\affiliation{Department of Astronomy, Columbia University,  550 W 120th St, New York, NY 10025, USA}
\affiliation{Center for Computational Astrophysics, Flatiron Institute, 162 5th Avenue, New York, NY 10010, USA}

\author[0000-0002-0311-2206]{Shyam H. Menon}
\affiliation{Department of Physics and Astronomy, Rutgers University, 136 Frelinghuysen Road, Piscataway, NJ 08854, USA}
\affiliation{Center for Computational Astrophysics, Flatiron Institute, 162 5th Avenue, New York, NY 10010, USA}

\author[0000-0002-0509-9113]{Eve C. Ostriker}
\affiliation{Department of Astrophysical Sciences, Princeton University, 4 Ivy Lane, Princeton, NJ 08544, USA}
\affiliation{Institute for Advanced Study, 1 Einstein Drive, Princeton, NJ 08540, USA}

\author[0000-0003-2896-3725]{Chang-Goo Kim}
\affiliation{Department of Astrophysical Sciences, Princeton University, 4 Ivy Lane, Princeton, NJ 08544, USA}

\begin{abstract}
We propose a new framework for the simultaneous feedback of stellar winds and photo-ionizing radiation from massive stars, distinguishing the locations where forces are applied, and consequences for internal spatio-temporal evolution of the whole feedback bubble (FB). We quantify the relative dynamical importance of wind-blown bubbles (WBB) versus the photoionized region (PIR) by the ratio of the radius at which the WBB is in pressure equilibrium with the PIR, $\Req$, to the \strom radius, $\RSt$. $\ReqRSt \equiv\Req/\RSt$ quantifies the dynamical dominance of WBBs ($\ReqRSt > 1$) or the PIR ($\ReqRSt < 1$). We calculate $\ReqRSt$ and find that, for momentum-driven winds, $0.1 \lesssim \ReqRSt \lesssim 1$ for the star-forming regions in (i) typical Milky Way-like giant molecular clouds (GMCs), (ii) the most massive of individual OB stars, and (iii) dense, low-metallicity environments, relevant in the early universe. In this regime, both WBBs and the PIR are dynamically important to the expansion of the FB. We develop a semi-analytic Co-Evolution Model (CEM) that takes into account the spatial distribution of forces and the back reactions of both the WBB and PIR. In the $\ReqRSt <1$ regime where the CEM is most relevant, the model differs in the total FB momentum by up to 25\% compared to naive predictions. In the weak-wind limit of $\ReqRSt \ll 1$, applicable to individual OB stars or low-mass clusters, the CEM has factors $\gtrsim 2$ differences in WBB properties. In a companion paper we compare these models to three-dimensional, turbulent hydro-dynamical simulations.
\end{abstract}


\section{Introduction}
\label{sec:intro}

As stars are fundamental building blocks of galaxies, and critical to how we observe (and therefore make physical inferences about) the Universe, it is crucial to have a detailed model of their formation. Unfortunately, there is much we still do not understand about the processes controlling star formation. We do know that these processes take place across many scales: involving the collapse of halo gas into galaxies \citep{Dave12,Pandya20,Pandya23,Voit24a,Voit24b}, the regulation of gas being maintained within that galaxy \citep{Fielding18,CGK2020,OML10,OstrikerShetty2011,PRFM22,CAFG13,KrumholzSFR18}, and the formation and dispersal of dense molecular clouds in which stars are born \citep{KrumholzTan07,KMBBH19,Kruijssen19,chevance20b,JGK18,JGK21,Lancaster21c,STARFORGE21,Menon22,Skinner15}. At each scale, injection of energy and momentum back into the interstellar (ISM) and circum- and inter-galactic media from stars plays a critical role in regulating the collapse of gas to the densities needed to form a star. It is these `feedback' processes, in their various forms, that are perhaps the largest open uncertainty in our understanding of star formation physics \citep{SomervilleDave15}.



On the scales of giant molecular clouds (GMCs) where stars form, feedback is dominated by young massive stars \citep{Murray10,Krumholz14} and energy that originates from the strong, high energy radiation emitted from the atmospheres of these stars can affect the surrounding GMC in multiple ways \citep{Dale15Review,KMBBH19}. Past theoretical studies of feedback effects focus on a single mechanism \citep{Spitzer78,Weaver77} or combine many feedback channels artificially into a single thin-shell evolution equation \citep{KrumholzMatzner09,Murray10,Geen20,Rahner17,Rahner19}. This approximation is made in order to simplify calculations but does not have rigorous theoretical foundations. In reality, however, these mechanisms always act in unison in a way that depends the complex way in which they interact with one another. A complete model of feedback should account for these interactions.

Two mechanisms that are generally understood as potentially dominant \citep{KMBBH19,Chevance23} in the early feedback evolution of `normal' GMCs as we might observe in our own Milky Way are (i) the photo-ionizing or Lyman Continuum (LyC) radiation from these stars which acts to heat the surrounding cloud \citep{Spitzer78,HosokawaInutsuka06,JGK16} and (ii) stellar winds that flow from their atmospheres at thousands of kilometers per second \citep{CAK75,Vink01,VinkSander21}, shock heating and becoming vastly over-pressurized with respect to the surrounding cloud. These mechanisms interact in complex ways, with the photoionized region (PIR) created by the LyC radiation changing the nature of cooling at the interface of the stellar wind-blown bubble (WBB) and compression due to the WBB increasing the recombination rate of the PIR, causing it to contract. In this work we present a framework for considering the co-evolution of feedback from these two mechanisms that provides (i) a simple heuristic for when one will be dominant over another and (ii) a semi-analytic model for evolution under the combined effects of these mechanisms. As the degree to which WBBs retain or lose the mechanical energy they inject (and hence impart momentum to their surroundings), is somewhat uncertain \citep{Lancaster24a}, we present models for different limiting cases.

In \autoref{sec:theory} we review the classical dynamical theories for the evolution of WBBs and the PIR in star-forming clouds before presenting our model for their co-evolution in \autoref{sec:theory_joint}. Key details are illustrated in \autoref{fig:feedback_ratio} and discussed in \autoref{subsec:joint_assumptions}-\autoref{subsec:ed_jfb}. In \autoref{sec:model_discussion} we present a series of sample model calculations and discuss their features, while in \autoref{sec:discussion} we provide an in-depth review of past work on this topic, contrasting with the current work, and outline pathways for future model development. We give a brief summary of our main conclusions in \autoref{sec:conclusion}. Discussion throughout the work is supplemented by appendices, which contain further details.

\begin{deluxetable*}{c|l|l}
    \tablecaption{Explanation of Model Parameters.\label{tab:defs}}
    \tablewidth{\textwidth}
    \tablehead{
    \colhead{Parameters} & \colhead{Meaning} & \colhead{Definitions and References}}
    \startdata
    $\rhobar$ & mean mass density in the GMC &  \\
    $n$ & number density of all particles &  \\
    $\nH$ & number density of hydrogen nuclei &  \\
    $\nHi$ & number density of hydrogen nuclei in photo-ionized gas &  \\
    $\nHbar$ & mean number density of hydrogen nuclei in GMC & \\
    $\mu$ & mean molecular weight &  $\mu \equiv \rhobar/(\overline{n}\mH)$\\
    $\muH$ & mean particle weight per hydrogen nucleus &  $\muH \equiv \rhobar/(\nHbar\mH)$\\ \hline
    $\vw $ & wind velocity         &  \\
    $\mdot$ & wind mass loss rate  &  \\
    $\Lwind$ &  wind mechanical luminosity  & $\Lwind \equiv \frac{1}{2}\mdot\vw^2$ \\
    $\pdot$ & wind momentum input rate & $\pdot \equiv \mdot\vw$ \\
    $\Qo$ & ionizing photon emission rate &  \\
    $\mathcal{R}$ & volume-equivalent radius for volume $V$ & $\mathcal{R} \equiv \left( 3V/4\pi\right)^{1/3}$ \\ \hline
    $\theta$ & fraction of energy lost to cooling for energy-driven-like (ED-like) solutions & $\theta \equiv \ecool/\Lwind$ \\
    $\RWeavert(t)$ & bubble radius for ED-like solutions & \autoref{eq:Rweaver} \\
    $\PWeavert(t)$ & bubble pressure for ED-like solutions & \autoref{eq:Pweaver} \\
    $\prWeavert(t)$ & bubble radial momentum for ED-like solutions & \autoref{eq:prweaver} \\ \hline
    $\alpha_p$ & momentum enhancement factor for  & $\alpha_p \equiv \dot{p}_r/\pdot$, \autoref{eq:alphap_shock}, \\ 
    \nodata & momentum-driven-like (MD-like) solutions & \autoref{eq:alphap_derive} \\ 
    $\RMDa(t)$ & bubble radius for MD-like solutions & \autoref{eq:rEC} \\ 
    $\PMDa(t)$ & bubble pressure for MD-like solutions & \autoref{eq:Phot_EC} \\ 
    $\prMDa(t)$ & bubble radial momentum for MD-like solutions & \autoref{eq:pr_EC} \\ \hline 
    $\voutavg$ & average velocity of diffusive motions into WBB mixing layer & \autoref{eq:vout_def},\\
    \nodata & & \citet{Lancaster24a} \\ 
    $\Abub$ & surface area of WBB & \\ \hline
    $\RSt$ & the \strom radius & \autoref{eq:RSt}, \autoref{eq:Rst_quant},\\ 
    \nodata & & \citet{Stromgren39}\\ 
    $\alpha_B$ & Case-B recombination rate & \citet{Drainebook}\\
    $\ci$ & sound speed in photo-ionized gas & $\approx 10\kms$ \\ 
    $\RSp(t)$ & bubble radius for classical Spitzer photo-ionized bubbles & \autoref{eq:spitzer_sol}  \\
    $\tdio$ & initial dynamical time of photo-ionized bubble & \autoref{eq:tdio_def}, \autoref{eq:tdio_app} \\
    $p_{r,{\rm Sp}}(t)$ & radial momentum for classical photo-ionized bubble & \autoref{eq:pr_spitzer1}\\
    $p_{r,{\rm Sp,adj}}(t)$ & mass-adjusted version of the above & \autoref{eq:pr_spitzer_adj}  \\ \hline
    $\nHbartrap$ & density at which the WBB shell traps all ionizing radiation & \autoref{eq:phot_trap_rhobar}, \autoref{app:trapping} \\ \hline
    $\Req$ & radius at which the WBB is in pressure-equilibrium with the PIR & \\
    $\teq$ & time at which $\Req$ is reached & \\
    $\tot$ & The time for the WBB to catch up to the PIR & \autoref{app:scales} \\
    $R_{\rm eq, M(E)D}$ & $\Req$ for MD-like (or ED-like) solutions & \autoref{eq:RE_MD_def}, \autoref{eq:RE_ED_def} \\
    $t_{\rm eq, M(E)D}$ & $\teq$ for MD-like (or ED-like) solutions & \autoref{eq:tE_MD_def}, \autoref{eq:tE_ED_def} \\
    $\ReqRSt$ & $\Req/\RSt$ defined separately for both MD- and ED-like models & \autoref{eq:eta_def}, \autoref{eq:zetaMD_def} \\
    \nodata &  & \autoref{eq:zetaED_def}, \autoref{app:scales}\\
    $\Rch$ & ``characteristic'' radius at which functional forms of WBB and PIR forces balance & \autoref{eq:Rch_def}
    \enddata
    \tablecomments{Definitions of key parameters used in the main body of the text along with a short explanation and where they are defined or relevant references. Parameters appear in the order they appear in the text and are grouped (horizontal lines) according to the section or subsections they appear.}
\end{deluxetable*}

\section{Review of Bubble Expansion}
\label{sec:theory}

\subsection{Definitions}
\label{subsec:wind_defs}

In this and the following sections we consider the impact of a source of mechanical energy and hydrogen photo-ionizing or Lyman Continuum (LyC) radiation on its surroundings. While in this section we will only consider these mechanisms individually, we will generally refer to the high-pressure bubble that results from any injection of energy as a `feedback bubble.' In general, this consists of a central WBB and the surrounding PIR. We will consider the source of both of these feedback mechanisms to be a cluster of massive stars. Much of this formalism still applies in the context of single massive stars or even winds from Active galactic Nuclei (AGN).

We characterize the background medium into which the source emits energy principally by its mean background mass density, $\rhobar$. The mean number density of hydrogen ions (or nuclei) is then given by $\nHbar \equiv \rhobar/(\muH \mH)$ where $\muH$ is the mean molecular weight per hydrogen nucleus and $\mH$ is the mass of a hydrogen atom. Similarly, the mean density of particles is given by $\overline{n} \equiv \rhobar/(\mu \mH)$, with $\mu$ the mean molecular weight.  The central source emits a wind with constant (in time) velocity $\vw$ and mass loss rate $\mdot$ into this background. Such a wind has a mechanical luminosity $\Lwind = \mdot \vw^2/2$ and a momentum input rate $\pdot = \mdot \vw$. This source can also emit LyC radiation with a LyC photon emission rate, $\Qo$, and average energy per LyC photon of $h\nu_i$.

We will be interested in several summary physical quantities of the FB's evolution, such as its size, the momentum that it carries, and the energy (thermal and kinetic) or pressure of different phases of the gas. We wish to distinguish between the various idealized solutions discussed below and the `true' evolution of these physical quantities (as measured in simulations or discussed in relation to the idealized theory). To this end we note that all variables representing the evolution of a physical quantity according to some idealized theory will be accompanied by appropriate capitalized, Roman-script subscripts. For example, the total momentum carried by the feedback bubble in the `energy-driven' \citet{Weaver77} solution shall be denoted, $\prWeaver$. Unless otherwise noted, this always refers to \textit{radial} momentum measured with respect to the source.

We would generally like to characterize a bubble's size by its radius; however for non-spherically symmetric solutions this is a poorly defined quantity. To circumvent this we define the `effective' or `volume-equivalent' radius, $\mathcal{R}$ associated with some volume $V$ as $\mathcal{R} = \left( 3 V/4 \pi \right)^{1/3}$. 

In the next few sections we will begin by briefly reviewing the classical theories of feedback bubble evolution due to the mechanisms discussed in this work: winds and photoionized gas.

\subsection{Stellar Wind-Blown Bubbles}
\label{subsec:theory_winds}

In the \autoref{subsubsec:classic_wind} and \autoref{subsubsec:ec_wind} we describe energy- and momentum-driven WBB expansion solutions. We use these names to denote the classical dependence of the bubble parameters (radius, pressure, momentum, etc.) on the parameters of the system ($\rhobar$, $t$, and wind feedback parameters). However, recent work has shown \citep{ElBadry19,Lancaster21a,Lancaster24a} that the classical limits of each of these solutions which allow for no cooling losses from the WBB interior \citep{Weaver77} or no retention of energy \citep{Steigman75} are unrealistic. While true WBBs lie somewhere in-between these two scaling regimes, recent work indicates that significant energy-loss is likely to occur \citep{Lancaster21a,Lancaster21b,Lancaster21c,Lancaster24a}. Below, we include parameters in the classical solutions ($\theta$ and $\alpha_p$) in order to allow for energy losses/retention in the energy/momentum-driven models. The relationship between these parameters is explained in \autoref{subsubsec:apt_rel}.

\subsubsection{Energy-Driven Wind-Blown Bubble}
\label{subsubsec:classic_wind}

In the classic work of \citet{Weaver77}, the expansion of a wind into a uniform background medium is considered. In this scenario, the majority of the wind energy is retained within the hot gas of the bubble's interior, allowing it to perform efficient work on its surroundings. This results in the following ``energy-driven'' evolution of the radius, hot gas pressure and momentum
\begin{equation}
    \label{eq:Rweaver}
\begin{split}
    \RWeavert(t) &=
    \left( \frac{125}{154\pi}\frac{(1 - \theta)\Lwind t^3}{\rhobar} \right)^{1/5} \\
    &= (1-\theta)^{1/5} \RWeaver(t)
    \, ,
\end{split}
\end{equation}
\begin{equation}
    \label{eq:Pweaver}
\begin{split}
    \PWeavert(t) &= \frac{5}{22\pi} \left( \frac{125}{154\pi}\right)^{-3/5}
    \left( \frac{(1-\theta)^2\Lwind^2 \rhobar^3}{t^4} \right)^{1/5} \\
    \PWeavert(\RWeavert)&= \frac{5}{22\pi} \left( \frac{125}{154\pi}\right)^{-1/3} 
    \left( \frac{(1-\theta)^2\Lwind^2 \rhobar}{\RWeavert^4} \right)^{1/3} \\
    &= (1-\theta)^{2/5} \PWeaver(t)\, ,
\end{split}
\end{equation}
\begin{equation}
    \label{eq:prweaver}
\begin{split}
    \prWeavert (t) 
    &= \frac{4\pi}{5} \left( \frac{125}{154 \pi}\right)^{4/5}
    \left((1 - \theta)^4\Lwind^4 \rhobar t^7 \right)^{1/5} \\
    \prWeavert (\RWeavert) & = \frac{4\pi}{5} \left( \frac{125}{154\pi}\right)^{1/3} \left((1-\theta)\Lwind \rhobar^2 \RWeavert^7 \right)^{1/3}\\
    &= (1-\theta)^{4/5} \prWeaver(t) \, .
\end{split}
\end{equation}
In the above we have allowed for a constant fraction, $\theta$, of the wind's energy to be lost, resulting in the solutions with $\theta$ subscripts. The last lines of each equation relate these solutions to the unmodified, classical solutions. In the second lines of \autoref{eq:Pweaver} and \autoref{eq:prweaver} we show the dependence of these variables on the radius of the bubble $\RWeavert$.

We mean for the energy losses given by $\theta$ to be representative of losses due to mixing at the interface. \citet{ElBadry19} provided analytic and numerical results that showed that when one approximates turbulent mixing at the interface using a constant diffusivity in an otherwise spherical geometry it results in a constant fraction, $\theta$, of the wind bubble's energy being lost. While one could appeal to energy losses due to gas leakage \citep{HCM09,Rosen14} or other mechanisms to create a similar solution to the above, $\theta$-modified solutions, it is less physically justified that these mechanisms would lead to a constant fraction of energy being lost.

\citet{Weaver77} themselves argue that the classical solution can be modified by thermal conduction at the WBB interface, which leads to evaporative mass-loading of the bubble interior. In certain regimes, this mass-loading can lead to a radiative interior, causing a momentum-driven ($p_r \propto t$) evolution \citep{Weaver77,SilichTT13,maclow88}.  

\subsubsection{Momentum-Driven Wind-Blown Bubble}
\label{subsubsec:ec_wind}

The truly three-dimensional (3D) nature of any real WBB means that the interface between the wind and the shell is subject to an array of 3D instabilities \citep{Vishniac83,Vishniac94,VishniacRyu89,GS96a,GS96b,FoliniWalder06,Ntormousi11,Pittard13,Badjin16}. These instabilities, as well as the inhomogeneous, turbulent background medium with which the wind interacts, lead to mixing of the wind material with the ambient gas, which can be followed by rapid cooling \citep{MVL84,Nakamura06,Rosen14,ElBadry19}.

In \citet{Lancaster21a} we described a theory for how the energy-driven picture of \citet{Weaver77} is altered when allowing for non-spherical expansion. In this picture realistic background structure as well as non-linear growth of instabilities leads to mixing at the interface between the WBB and the surrounding medium, creating intermediate temperature ($T\sim 10^4\, {\rm K}$) gas that can cool efficiently. In particular, we noted that the amount of such cooling can be drastically enhanced by the 3D fractal structure of the interface, created through both large scale mixing and the background density inhomogeneities. The basic proposition of that work was that this interface cooling is severe enough to cause the bubble evolution to follow an effectively momentum-driven solution, rather than an energy-driven solution, from the very earliest stages that cooling is efficient at the interface: after shell formation.

In \citet{Lancaster21a} we posited that the momentum carried by the bubble, $p_r$ followed
\begin{equation}
    \label{eq:pr_EC}
    \prMDa (t)  =  \alpha_p \pdot t \equiv \alpha_p \prMD(t) \, ,
\end{equation}
where $\prMD(t)$ is the idealized `momentum-driven' solution and $\alpha_p$ is the momentum enhancement factor, quantifying the amount by which the WBB carries momentum beyond the momentum input directly by the wind.

Under the additional assumptions of statistical homogeneity and isotropy of the background medium, one can show that the evolution of the WBB's effective radius follows
\begin{equation}
    \label{eq:rEC}
    \RMDa (t) = \left( \frac{3 \alpha_p}{2\pi} \frac{\pdot t^2}{\rhobar}\right)^{1/4} 
    \equiv \left( \alpha_p \right)^{1/4} \RMD (t)
\end{equation}
where $\RMD(t)$ is the idealized radial evolution of the WBB when $\alpha_p = 1$ \citep{Steigman75}.

In both the \citet{Weaver77} and the momentum-driven solutions the thermal pressure in the hot, shocked wind (assumed to be isobaric) is given by the post-shock value, $\Phot = 3 \pdot/(16 \pi \rfree^2)$, where $\rfree$ is the effective radius of the free-wind region \citep[see e.g.][]{Lancaster21a,Weaver77}. As discussed in the Appendix of \citet{Lancaster21a} (and further detailed in \citet{Lancaster24a}), one can show that the relative volume of the free and post-shock wind is related to the momentum enhancement factor, $\alpha_p$, as
\begin{equation}
    \label{eq:alphap_shock}
    \alpha_p = \frac{1}{4} \left[ 3\left(\frac{\Rw}{\rfree}\right)^2 
    + \left(\frac{\Rw}{\rfree}\right)^{-2}\right] \, ;
\end{equation}
here $\Rw$ is the outer radius of the WBB. The above is again generally true for both solutions (excluding geometric effects which are treated more carefully in Appendix A of \citet{Lancaster24a}), which have very different behaviors for $\Rw/\rfree$. We can see from \autoref{eq:alphap_shock} that in the limit that $\alpha_p \to 1$ we have $\rfree \to \Rw$ and thus that, in the idealized, purely momentum-driven solution the entire bubble is made up of the free wind region. 

For the current case of a constant $\alpha_p$ we can use \autoref{eq:alphap_shock} along with the assumption $\alpha_p \gtrsim 1$ to write the pressure of the hot gas as
\begin{equation}
    \label{eq:Phot_EC}
    \PMDa = \frac{3\pdot}{16\pi \Rw^2} \left( \frac{\Rw}{\rfree}\right)^2
    \approx \frac{\alpha_p\pdot}{4\pi \Rw^2} \, .
\end{equation}

\subsubsection{Relationship Between $\theta$ and $\alpha_p$}
\label{subsubsec:apt_rel}

The parameters $\theta$ and $\alpha_p$ are considered constant in the above, idealized solutions. Alternatively, consider $\alpha_p(t) \equiv F_w/\pdot$ to be the factor by which the WBB is exerting a force greater than the ram pressure input by the wind itself at a given time. We can calculate this for the modified energy-driven solution given above using $F_w = 4\pi \RWeavert^2\PWeavert$, using \autoref{eq:Rweaver} and \autoref{eq:Pweaver} we have
\begin{equation}
    \label{eq:apoft}
    \alpha_{p,{\rm ED},\theta} (t) = \frac{10}{11}\left( \frac{125}{154\pi}\right)^{-1/5} 
    \left[ (1-\theta)^4\Lwind^4 \rhobar t^2\right]^{1/5} \, .
\end{equation}
From the above we see that if $\theta$ is to be considered constant then $\alpha_p \propto t^{2/5}$.

Similarly, if we assume the modified momentum-driven solution ($\RMDa$, $\prMDa$, etc.), we can follow the derivation given in the appendix of \citet{Lancaster21a} and the discussion around Equation 25 of that work to calculate the total energy in the bubble, $E_w(t)$. Taking a derivative with respect to time to get the instantaneous fraction of the wind energy that is being retained in the bubble we have
\begin{equation}
    \label{eq:tofap}
    1- \theta_{{\rm MD}, \alpha}(t) = 
    \frac{3}{4} \left(\frac{3}{2\pi} \frac{\alpha_p^5 \pdot}{\rhobar \vw^4 t^2} \right)^{1/4} \, .
\end{equation}
Here then, in order to regard $\alpha_p$ as constant $1- \theta \propto t^{-1/2}$. Note that in order to derive \autoref{eq:tofap} we have ignored turbulent motions in the WBB shell (as befits the idealized treatment here) and assumed $\alpha_p \gtrsim 1.25$ so that the $\mathcal{S}$ quantity introduced in the appendix of \citet{Lancaster21a} is equal to $\alpha_p$ within 5 percent.

In the general case, when energy losses are not necessarily constant in time and it is not valid to assume a momentum-driven-like scaling, one must solve an energy equation (akin to \autoref{eq:Ewdot}, but including other potential energy sinks) along with the other conservation equations of the bubble. In this general case the definitions of $\alpha_p(t)$ ($F_w/\pdot$) and $\theta (t)$ (instantaneous fraction of energy being lost compared to $\Lwind$) used above can still be applied. The above arguments based on the idealized solutions simply illustrates that these quantities are intimately tied to one another. This should be intuitive: more energy retention naturally leads to more momentum input. In the next section we review the results of \citet{Lancaster24a} which explain how $\alpha_p$ is set by turbulent mixing at the WBB interface.

\subsubsection{Conditions for a Momentum-Driven Solution}
\label{subsubsec:ec_conditions}

In \citet{Lancaster21a} it was posited that turbulent mixing at the wind bubble's interface is sufficient to ensure that the bubble enters a momentum-driven limit, $\alpha_p \approx 1$ and $\rfree \approx \Rw$. Section 2.4-2.6 of that work discusses various conditions under which this may actually apply, resulting in conditions on the efficiency of mixing at the wind bubble's interface derived from considering the flux of energy able to be mixed and cooled in the interface region compared to the flux of energy being provided by the wind. As was demonstrated in Figure 18 of \citet{Lancaster21b}, these two fluxes are essentially identical: the mixing layer is able to cool as quickly as it is being provided energy. While \citet{Lancaster21a,Lancaster21b,Lancaster21c} used this as justification for the $\alpha_p \approx 1$ assumption, \citet{Lancaster24a} showed how the matching of these two fluxes provides a new boundary condition on the shocked wind gas, which essentially determines the dynamics of the bubble by dictating to what degree energy can be stored in the bubble interior.

In particular, matching of energy fluxes at the outer wind surface determines $\Rw/\rfree$ and therefore $\alpha_p$ through \autoref{eq:alphap_shock}. \citet{Lancaster24a} use this to derive the relationship between $\alpha_p$ and the conditions of the bubble's interface as 
\begin{equation}
    \label{eq:alphap_derive}
    \alpha_p = \frac{3}{4} \frac{\vw/4}{\voutavg} \frac{4\pi \Rw^2}{\Abub} \, ,
\end{equation}
where $\Abub$ is the surface area of the wind bubble's interface and $\voutavg$ is the velocity at which gas moves out of the bubble and into that interface. The latter is defined as
\begin{equation}
    \label{eq:vout_def}
    \voutavg \equiv \Abub^{-1} \int_{\Abub} \left(\mathbf{v} - \mathbf{W}\right) \cdot \nhat\,  dA \, 
\end{equation}
where $\mathbf{v}$ is the gas velocity, $\mathbf{W}$ is the velocity of the interface, and $\nhat$ is the unit normal vector to the wind bubble's surface, pointing outward.

One clear take away from \autoref{eq:alphap_derive} is that the behavior of a given WBB as `momentum-driven' relies on the relative scaling of $\voutavg$ and $\Rw^2/\Abub$ being constant in time. \autoref{eq:alphap_derive} was validated against simulations in \citet{Lancaster24a}, where it was additionally shown that the apparent resolution independence of the momentum evolution presented in \citet{Lancaster21b} for purely hydrodynamical simulations was due to the relative scaling of $\voutavg$ and $\Rw^2/\Abub$ compensating for each other as a function of numerical resolution. In fact, this resolution-independent, momentum-driven behavior was not seen for simulations which included magnetic fields, mainly due to differences in the fractal structure of the magnetized bubble's interfaces resulting in different behavior of $\Rw^2/\Abub$ both as a function of time and numerical resolution.

Dependence on numerical resolution is fundamentally due to the fact that the scales relevant for resolving dissipative mixing processes across the wind bubble's interface are drastically under-resolved \citep{Lancaster24a}. Due to the large separation of scales, this seems likely to remain the case for global simulations for at least the near future. However, \citet{Lancaster24a} also derive what the evolution of $\alpha_p$ in time \textit{should be} if the correct dissipative scales were resolved (their Section 2.6) and show that (at least when magnetic fields are not dynamically important) $\alpha_p \approx 1$ and constant in time is an appropriate expected evolution.

\subsection{Spitzer Solution for Photoionized Gas Bubbles}
\label{subsec:spitzer}

We review the expansion of bubbles driven by thermal pressure from photoionized gas, which is reviewed in further detail in \citet{JGK16}. We imagine a uniform medium of density $\rhobar$ at the center of which a source of LyC photons begins emitting with a rate $\Qo$. At first, there is an R-type, supersonic expansion of the ionized region \citep{Drainebook}, until the \strom radius \citep{Stromgren39} defined as 
\begin{equation}
    \label{eq:RSt}
\begin{split}
    \RSt &\equiv \left(\frac{3\Qo}{4\pi \alpha_B \nHrms^2} \right)^{1/3} \\
    & = 10.1\, {\rm pc} \, \left(\frac{\Qo}{4\times 10^{50} {\rm s}^{-1}} \right)^{1/2} \left(\frac{\nHbar}{100\pcc} \right)^{-2/3}\, , 
\end{split}
\end{equation}

is reached, with $\nHrms$ the root-mean-square density of hydrogen nuclei in the background medium (which will be approximated by $\nHbar$ below) and $\alpha_B$ the case B recombination rate coefficient \citep{Drainebook}. The region confined by this radius is ionized on the recombination timescale $\trec = \left(\alpha_B \nHbar\right)^{-1}$, on the order of $\sim 10^3 \, {\rm yr}$ for conditions typical of a Milky Way GMC. Since this timescale is short compared with the timescale for the evolution of the feedback bubble overall, it is generally assumed that the PIR's evolution begins at this point at $t=0$.

The subsequent evolution is determined by assuming that all gas within the PIR is maintained in ionization/recombination equilibrium. That is, for a photoionized region with radius $\Ri$ and density of hydrogen nuclei $\nHi$ we have
\begin{equation}
    \label{eq:ionreceq1}
    \frac{4 \pi}{3} \Ri^3 \alpha_B \nHi^2 = \Qo \, .
\end{equation}
We then write down a momentum equation for the evolution of the bubble as
\begin{equation}
    \label{eq:HIImomentum1}
    \frac{d}{dt} \left(\Msh \frac{d\Ri}{dt} \right)
    = 4 \pi \Ri^2 \Phii \, ,
\end{equation}
where $\Msh$ is the mass of the shell of shocked ambient material carried by the PIR and $\Phii = \rho_i \ci^2$ is the thermal pressure of the region, with $\rho_i$ and $\ci$ the density and isothermal sound speed. The sound speed is assumed to be constant and is defined as $\ci^2\equiv k_B T_i/(\mu \mH)$ where $T_i$ is the ionized gas temperature. With a thin shell approximation using $\Msh = 4\pi \Ri^3 \rhobar/3$ and \autoref{eq:ionreceq1} the above becomes
\begin{equation}
    \label{eq:HIImomentum2}
    \frac{d}{dt} \left(\frac{\Ri^3}{3} \frac{d\Ri}{dt} \right)
    = \ci^2\Ri^2 \left(\frac{\RSt}{\Ri} \right)^{3/2} \, .
\end{equation}
Solving this ordinary differential equation (ODE) using the ansatz $\Ri(t) = \RSt\left(1 + at \right)^b$ for some constants $a$ and $b$ gives us the \citet{HosokawaInutsuka06} version of the classic \citet{Spitzer78} solution
\begin{equation}
    \label{eq:spitzer_sol}
    \RSp(t) = \RSt \left(1 + \frac{7}{4} \frac{t}{\tdio} \right)^{4/7} \, ,
\end{equation}
where
\begin{equation}
    \label{eq:tdio_def}
    \tdio \equiv  \frac{\RSp(0)}{\dot{R}_{\rm Sp}(0)}= \frac{\sqrt{3}}{2} \frac{\RSt}{\ci}
\end{equation}
is the initial dynamical expansion time of the bubble.

In using the above solution to infer the momentum imparted to the ambient medium it is natural to take $p_r = \Msh d\RSp/dt$, as we assumed in writing \autoref{eq:HIImomentum1}. Using this assumption we would get 
\begin{equation}
    \label{eq:pr_spitzer1}
    p_{r,{\rm Sp}}(t) = \frac{8\pi}{3\sqrt{3}} \rhobar \ci \RSt^3 
    \left(1 + \frac{7}{4}\frac{t}{\tdio} \right)^{9/7} \, .
\end{equation}
Setting $t=0$ in the above we see that the momentum is finite at the beginning of the evolution of the bubble, which is clearly not physical. This assumption will be particularly bad for large ionized regions, (essentially large $\Qo$), where the time during which the Spitzer solution would apply accurately before the bubble broke out of its nascent cloud would be small. A simple adjustment to this momentum inference is to use the shell mass $\Msh = 4 \pi \Ri^3 (\rhobar - \rho_i)/3$, that is, to subtract out the mass in ionized gas. Though this reduction is not consistent with the derivation of \autoref{eq:HIImomentum2}, we will see that it can be more accurate. This assumption gives us the adjusted inferred momentum
\begin{equation}
    \label{eq:pr_spitzer_adj}
    p_{r,{\rm Sp,adj}}(t) = p_{r,{\rm Sp}}(t)
    \left(1 - \left(\frac{\RSt}{\RSp}\right)^{3/2} \right) \, ,
\end{equation}
which indeed is zero at $t=0$. A similar criteria for the momentum, which does not account for the evolution of the density within the ionized gas, is suggested in \citet{Haid18} and \citet{Pittard22Rad}.

\section{Co-Evolution of Stellar Wind-Blown Bubbles and Photoionized Gas}
\label{sec:theory_joint}

In this section we describe a co-evolution model (CEM) for a WBB and the PIR that always surrounds it in the case of massive O and B type stars. As above, this discussion ignores the dynamical importance of magnetic fields. We ignore the importance of direct radiation pressure in changing the physical structure of the ionized region \citep{Draine11,JGK16,pellegrini11,Rahner17}, and any external pressures or the effects of gravity on slowing the shell \citep{Raga12b,Rahner17,Rahner19,Geen20}. An implementation of the models presented here is provided in a public \href{https://github.com/ltlancas/feedback_SAM}{GitHub repository}.

\subsection{Radiation Trapping}
\label{subsec:trapping}

Before continuing with the key points of our framework and evolution equations, we briefly describe the phenomenon of radiation trapping. Recent work has emphasized the capacity for the shells of WBBs to trap LyC radiation, reducing the impact of photoionized gas pressure as a feedback mechanism \citep{Geen22,Geen23}. In \autoref{app:trapping} we provide a detailed analysis of how this occurs. In particular, we explain why this phenomenon is more important for the expansion of feedback bubbles around individual stars rather than those driven by a cluster of stars into a background density field that is relatively uniform (even if it includes turbulent density fluctuations). Briefly, this is due to an increased ability for radiation to be trapped when considering WBBs expanding into steep background density gradients. We also demonstrate that the timescale on which the LyC radiation of individual stars `breaks out' and forms a cluster-wide \strom sphere (\autoref{eq:tbo}) is relatively short compared to the evolution of the cloud.

We appeal to these calculations to justify our treatment below of the expansion of a feedback bubbles from a cluster of stars in which an initial \strom sphere is established and a WBB expands into it. However, at very high cloud density, the LyC radiation of the cluster-driven feedback bubble may still be trapped in the shell of its WBB. This regime is reached when the mean background density reaches the `trapping density' of (see \autoref{eq:phot_trap_rhobar_app})
\begin{equation}
    \label{eq:phot_trap_rhobar}
\begin{split}
    \nHbartrap &= 3.76 \times 10^{8} \pcc 
    \left(\frac{\Qo}{10^{50} \, {\rm s}^{-1}} \right)^{\frac{11}{3}} \\
    &\times \left(\frac{\Lwind}{10^{37}\, {\rm erg/s}} \right)^{\frac{-14}{3}} 
    \left(\frac{Z}{Z_{\odot}}\right)^{5/3}\, .    
\end{split}
\end{equation}
It is clear that this is only important at extremely high densities. Note that the metallicity dependence included above only accounts for the physics of cooling in the WBBs shell. $\Lwind$ generally decreases at lower metallicity \citep[$\Lwind\propto Z^{0.95}$][]{Vink01} while $\Qo$ modestly increases \citep[$\Qo\propto Z^{-0.04}$][]{SB99}. Combining these metallicity dependencies, this density is even higher in low-metallicity environments ($\nHbartrap\propto Z^{-2.91}$).

\subsection{A Comparison of Scales}
\label{subsec:joint_assumptions}

We first propose a way of thinking about whether winds or photoionized gas are dominant as a feedback mechanism within a given cloud. This discussion is given in greater detail in \autoref{app:scales}. As we discussed above, after the \strom sphere has formed, the photoionized gas is at a finite pressure $\rhobar\ci^2$. Suppose that the pressure in the WBB follows some relation as a function of its radius $\Phot(\Rw)$ that is, for now, agnostic to an energy or momentum-driven solution. For all reasonable choices of $\Phot(\Rw)$ (from e.g. \autoref{eq:Pweaver} or \autoref{eq:Phot_EC}) $\Phot \to \infty$ as $\Rw \to 0$ so that there will be an early time when $\Phot \gg \rhobar \ci^2$. This means that the WBB will have to expand to some finite radius, $\Req$, before it is in pressure equilibrium with the surrounding PIR such that $\Phot(\Req) = \rhobar \ci^2$ .

If $\Req \ll \RSt$ then the WBB will cease to expand into the photoionized gas far before it has significantly disturbed it and therefore winds will be dynamically unimportant. If $\Req \gg \RSt$ then the WBB will reach the \strom sphere and `run it over', as the WBB will still be vastly over-pressurized with respect to the PIR, and winds will be dynamically dominant. If $\Req \lesssim \RSt$ then the WBB will come into equilibrium with the PIR after significantly disturbing it and a model of their true ``co-evolution" will be needed.

If we apply either the momentum-driven or energy-driven solution for the expansion of WBBs we may derive $\Req$ and the timescale at which it occurs, $\teq$. For a momentum-driven solution with some potential non-unity $\alpha_p$ we have
\begin{equation}
    \label{eq:RE_MD_def}
\begin{split}
    \ReqMD &\equiv \sqrt{\frac{\alpha_p \pdot}{4\pi \rhobar \ci^2}} \\
    & = 4.74\, {\rm pc} \, \left(\frac{\alpha_p\pdot}{10^5 \, M_{\odot}\, {\rm km/s/Myr}} \right)^{1/2} \\
    &\times \left(\frac{\nHbar}{100\pcc} \right)^{-1/2} 
    \left( \frac{\ci}{10\, {\rm km/s}}\right)^{-1}\, , 
\end{split}
\end{equation}
which is established at a time
\begin{equation}
    \label{eq:tE_MD_def}
\begin{split}
    \teqMD &\equiv \frac{1}{4\pi \ci^2} \sqrt{\frac{2\pi}{3} \frac{\alpha_p \pdot}{\rhobar}}
    = \frac{\ReqMD}{\sqrt{6}\ci}\\
    &= 1.89 \times 10^5 \, {\rm yr}\, 
    \left(\frac{\alpha_p\pdot}{10^5 \, M_{\odot}\, {\rm km/s/Myr}} \right)^{1/2} \\
    &\times \left(\frac{\nHbar}{100 \pcc} \right)^{-1/2} 
    \left( \frac{\ci}{10\, {\rm km/s}}\right)^{-2}\, .    
\end{split}
\end{equation}
For the energy-driven case with some potentially non-zero $\theta$ we have
\begin{equation}
    \label{eq:RE_ED_def}
\begin{split}
    \ReqED &\equiv     \sqrt{\frac{\sqrt{7} (1-\theta)\Lwind}{22\pi \rhobar \ci^3}}\\
    &= 41.44\, {\rm pc} \left(\frac{(1 - \theta)\Lwind}{10^{38} \, {\rm erg/s}} \right)^{1/2} \\
    &\times \left(\frac{\nHbar}{100\pcc} \right)^{-1/2} 
    \left( \frac{\ci}{10\, {\rm km/s}}\right)^{-3/2}\, ,
\end{split}
\end{equation}
and
\begin{equation}
    \label{eq:tE_ED_def}
\begin{split}
    \teqED &\equiv \frac{7^{3/4}}{5} \sqrt{\frac{(1-\theta)\Lwind}{22\pi \rhobar \ci^5}}
    = \frac{\sqrt{7}}{5} \frac{\ReqED}{\ci}\\
    &= 2.14 \, \times 10^6 \,{\rm yr} \left(\frac{(1-\theta)\Lwind}{10^{38} \, {\rm erg/s}} \right)^{1/2} \\
    &\times \left(\frac{\nHbar}{100\pcc} \right)^{-1/2} 
    \left( \frac{\ci}{10\, {\rm km/s}}\right)^{-5/2}\, .
\end{split}
\end{equation}
We note that in both cases $\teq$ is much longer than the recombination timescale over all parameter regimes of interest, so we may assume that the ionized region maintains ionization equilibrium as the WBB evolves.
We also note that in both cases the time at which the WBB has slowed down to the sound speed of the photoionized gas is comparable to $\teq$, so that the density peak in the shell of shocked photoionized gas begins to be smoothed out by pressure waves at $t \sim \teq$.

The quantity $\Req/\RSt$ is clearly a key determinant in the relative dynamical impact of WBBs and the PIR. As we will use this quantity extensively below we define
\begin{equation}
    \label{eq:eta_def}
    \ReqRSt \equiv \frac{\Req}{\RSt}\, .
\end{equation}
Dimensional versions of this parameter are given in \autoref{eq:zetaMD_def_app} and \autoref{eq:zetaED_def_app} for the MD and ED solutions respectively.

It will be useful for the discussion below to consider what dynamical effect the ionized gas has on the surrounding gas before coming in to force balance with the WBB. In order to examine how quickly the PIR expands into its surroundings we can compare $\teq$ to the dynamical expansion time of the ionized gas in the Spitzer solution, given by \autoref{eq:tdio_def}. Note that, for both cases $\teq \propto \Req/\ci$ so that $\teq/\tdio \propto \Req/\RSt = \ReqRSt$ and in particular $\teq/\tdio \lesssim \ReqRSt$. This means that $\ReqRSt > 1$ really \textit{does} mean that the WBB will `run over' the \strom sphere before it has even had a chance to have a significant dynamical impact on the surroundings.

\begin{figure*}
    \centering
    \includegraphics{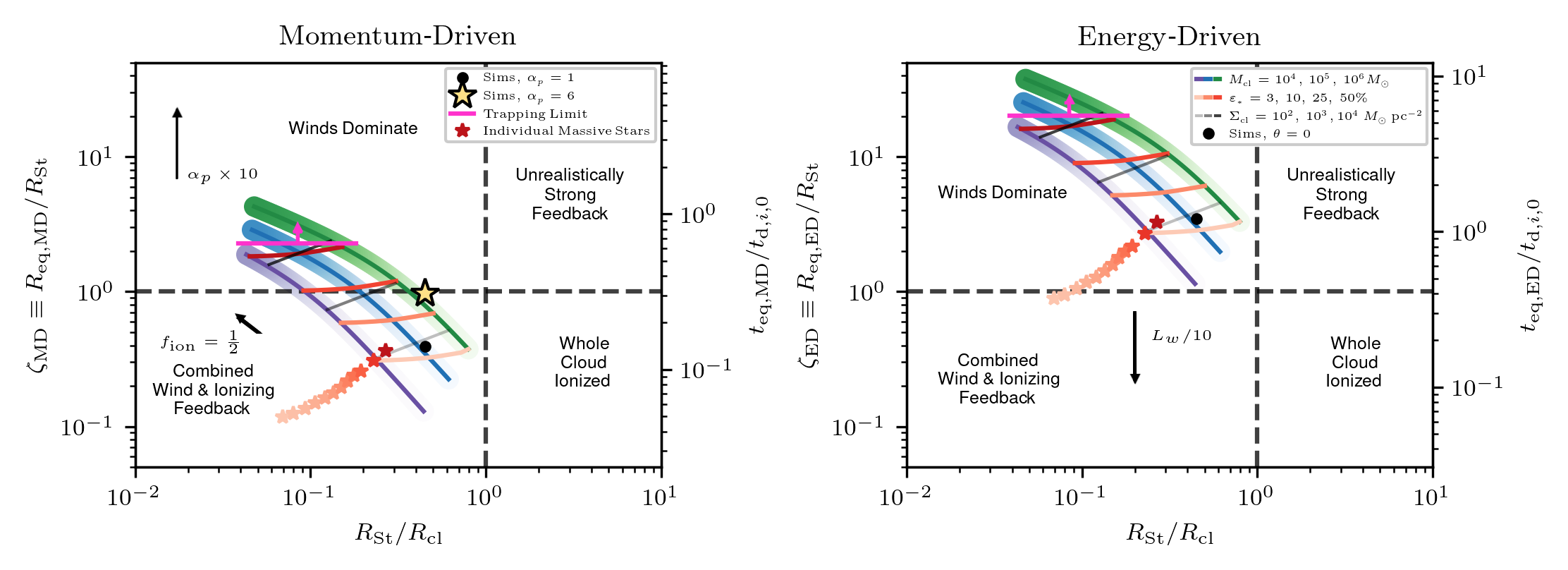}
    \caption{The fraction of the \strom radius occupied by the wind when it comes into force balance with the PIR, $\Req/\RSt$, versus the fraction of the cloud occupied by the \strom Sphere. The case for momentum-driven winds with $\alpha_p = 1$ is shown in the left panel (using \autoref{eq:RE_MD_def}) while the case for energy-driven winds with $\theta = 0$ is shown on the right (using \autoref{eq:RE_ED_def}). Purple, blue, and green lines indicate clouds of varying mass ($M_{\rm cl} = 10^4,\,10^5,\,10^6 \Msun$) and density with a star formation efficiency, $\sfe$, taken from the surface density-dependent relation of \citet{JGK18} (details in text, color gradient along each line, darker is higher density and higher star formation efficiency). Lines of constant $\sfe$ are indicated in shades of red from $3-50\%$, while lines of constant $\Scl$ are indicated in shades of grey. The limit above which the PIR is trapped in the WBB shell (\autoref{eq:phot_trap_rhobar}) is given by the pink line. The position of the simulations presented in this work are indicated with a black circle ($\alpha_p = 1$ on left and $\theta = 0$ at right) or a yellow star ($\alpha_p = 6$, as appropriate for simulations in \paperii). Black arrows in the panels indicate the movement of the lines in this space caused by: an increase in $\alpha_p$ by a factor of 10 (left), a decrease in the photoionizing fraction ($\Qo \to f_{\rm ion}\Qo$) by a factor of 2 (left), or a decrease of wind luminosity by a factor of 10 (right, $\theta = 0.9$). Other than for the individual massive stars shown as red stars ($15$--$60 \Msun$; darker is more massive), we assume IMF-averaged feedback parameters as used in our simulations and detailed in the text.}
    \label{fig:feedback_ratio}
\end{figure*}

In \autoref{fig:feedback_ratio} we display a version of the parameter space of WBB and PIR driven feedback, as described by these parameters. In particular we show $\ReqRSt$ versus the fraction of the cloud radius that the \strom sphere occupies ($\RSt/\Rcl$) for both the momentum-driven (left, $\alpha_p = 1$) and energy-driven (right, $\theta=0$) cases of WBB evolution. Regions to the right side of this plot indicate where the amount of star formation that takes place is enough to completely ionize the cloud, while placement from top to bottom of the plot is an indication of the strength of winds relative to LyC radiation. In \autoref{fig:feedback_ratio} we show values of this parameter space that are occupied for several different assumptions about the type of star forming cloud. In particular we show the evolution of these quantities for clouds of a fixed mass ($\Mcl = 10^4,\, 10^5,\, 10^6\, \Msun$ for purple, blue, and green lines respectively) as they are made increasingly dense $\nHbar = 10-10^6\, \pcc$(darker shading).

\begin{figure*}
    \centering
    \includegraphics{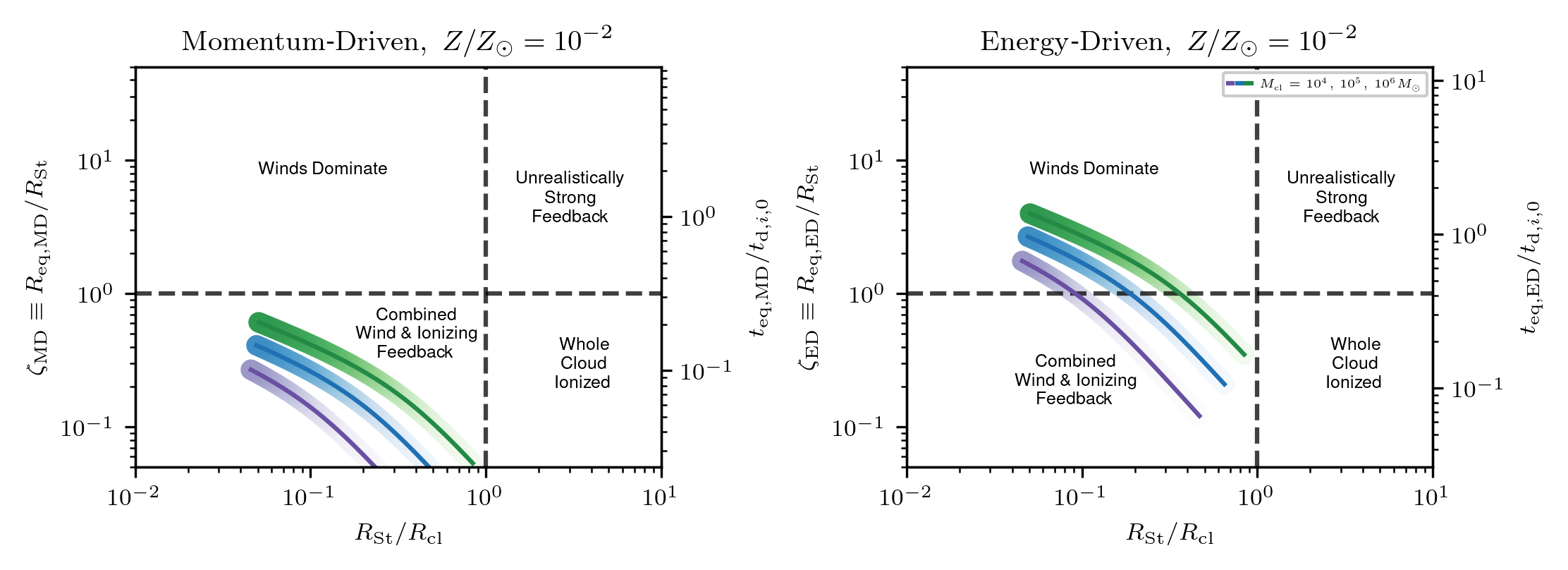}
    \caption{Identical to \autoref{fig:feedback_ratio} but for feedback parameters appropriate for a stellar population with metallicity $Z/Z_{\odot} = 10^{-2}$, details in text. We only include the lines associated with fixed $\Mcl$ as they evolve in density/$\sfe$ determined in the same way as \autoref{fig:feedback_ratio}. These lines are intended to outline the physically realizable region of the parameter space for clusters of massive stars. We see that dense star forming systems with low-metallicity, as may be expected in the early universe, should lie in the region of parameter space where our co-evolution model applies. The trapping limit displayed in \autoref{fig:feedback_ratio} is outside of the region of parameter space displayed here (it lies at $\ReqRSt \gg 50$) so trapping of photo-ionizing radiation is unimportant for these systems.}
    \label{fig:feedback_ratio_lZ}
\end{figure*}

\begin{figure}
    \centering
    \includegraphics[width=\columnwidth]{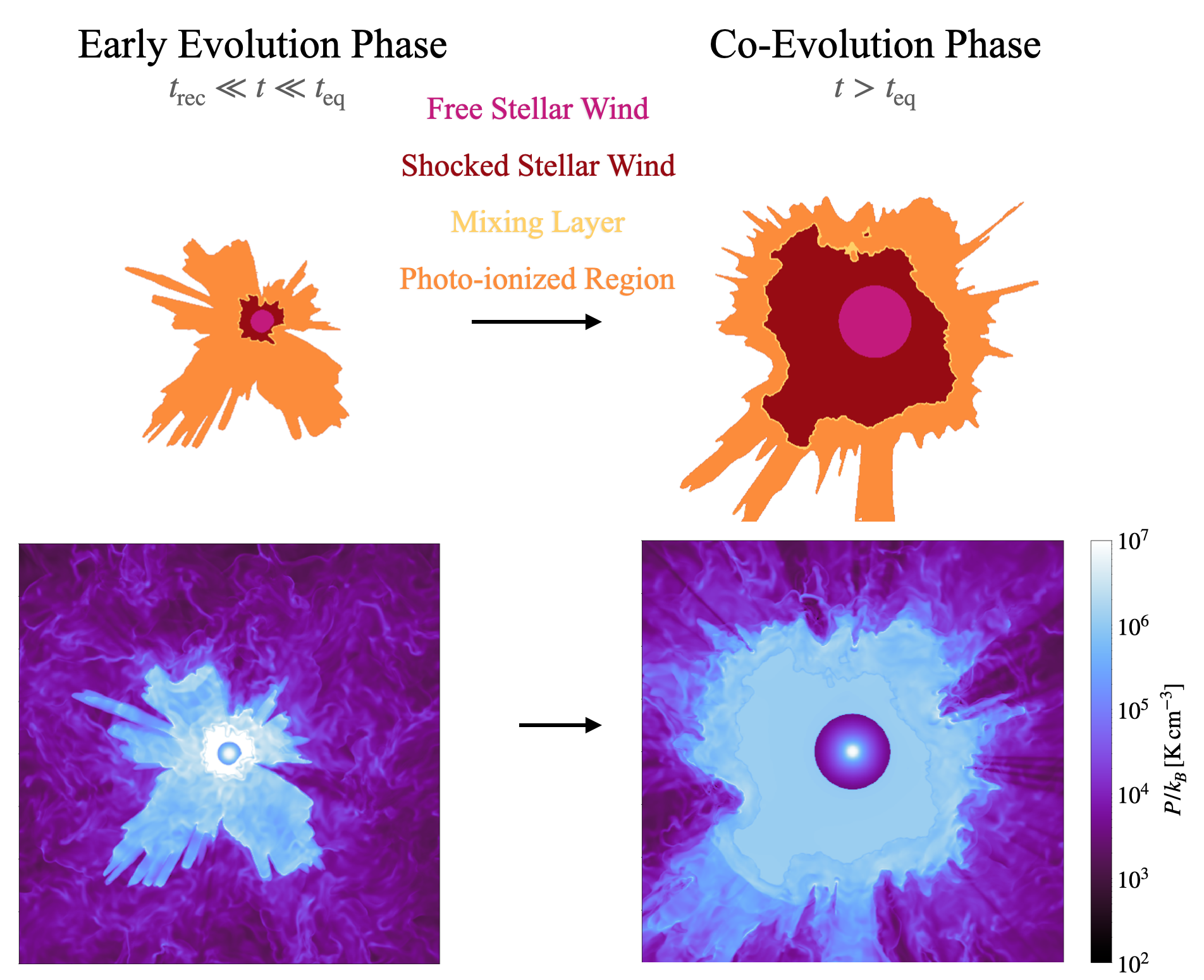}
    \caption{A schematic diagram of the distinct evolutionary stages of a stellar WBB interacting with its surrounding PIR, here specifically for the $\ReqRSt < 1$ regime. The top panels show schematics of the phase distributions (indicated with correspondingly colored text) while the bottom panels show the thermal pressure of the gas. All panels are based on the highest resolution \WRM\ simulation of \paperii. \textit{Left}: Early phase of a \strom Sphere, which is nearly static, with an embedded, over-pressurized and expanding wind bubble. \textit{Right}: At $t>\teq$ the WBB and PIR are in pressure balance and evolve jointly, as described by our models in \autoref{subsec:md_jfb} and \autoref{subsec:ed_jfb}.}
    \label{fig:schematic}
\end{figure}

We define the cloud surface density using its density and mass as $\Scl = \Mcl/(\pi\Rcl^2)$ with $\Rcl = (3\Mcl/4\pi\rhobar)^{1/3}$ and $\rhobar = \muH \mH \nHbar$ with $\muH = 1.4$. We then adopt the surface density dependent star-formation efficiency prediction from Equation 28 of \citet{JGK18}
\begin{equation}
    \label{eq:jgk18_eq28}
    \sfe = \frac{1- \varepsilon_{\rm ej, turb}}{1 + \left(p_*/m_*\right)/v_{\rm ej}}
\end{equation}
where $\varepsilon_{\rm ej,turb}$ is the fraction of the gas that is ejected in the initial turbulent evolution of the cloud and $v_{\rm ej}$ is the characteristic ejection velocity of gas leaving the star forming cloud. We adopt $v_{\rm ej} = 15,\, 23,\, 29\, \kms$ for clouds of mass $10^4,\, 10^5,\, 10^6\, \Msun$, as reported in \citet{JGK18}. Finally, $p_*/m_*$ is the amount of momentum injected per unit stellar mass or the `feedback yield' given by Equation 18 of \citet{JGK18} as
\begin{equation}
    \label{eq:kim18_yield}
    \frac{p_*}{m_*} = \Upsilon_{100} 
    \left( \frac{\Scl}{10^2\,\Msun\, {\rm pc}^{-2}}\right)^{-0.74} \,,
\end{equation}
In order to match the results of \citet{JGK21} at $\Scl \approx 80 \Msun\, {\rm pc}^{-2}$ with $\alpha_{\rm vir}=4$, we adopt $\varepsilon_{\rm ej,turb} = 0.25$ and $\Upsilon_{100} = 337 \, \kms$. An implicit assumption in our use of the SFE prescription of \citet{JGK18} is that winds do not significantly impact the resulting SFE compared to simulations which only include photo-ionizing radiation and radiation pressure. While this is likely a good assumption for $\ReqRSt \lesssim 1$, where winds are less important, it may not be as accurate at $\ReqRSt >1$. However, using the simulation-based, empirical SFE prescription of \citet{Grudic18} (whose simulations include winds) results in a similar qualitative picture.

For calculating $\Req$ and $\RSt$ we additionally assume $\ci = 10 \kms$, $\pdot/\Mst = 9.58\, \kms \Myr^{-1}$, $\Lwind/\Mst = 9.75\times 10^{33}\, {\rm erg}\,{\rm s}^{-1}\,\Msun^{-1}$ and $\Xi = \Qo/\Mst = 4.1\times 10^{46} \sec^{-1}\Msun^{-1}$. These values are used in the simulations presented in \paperii\ and are based on STARBURST99 calculated, IMF-averaged values at solar metallicity during the early evolution of star clusters \citep{SB99}.

In \autoref{fig:feedback_ratio} we display lines of constant cloud surface density, $\Scl$, by taking a fixed $\Scl$ and varying the density, inferring the cloud mass via, $\Rcl = 3\Scl/(4\rhobar)$ and $\Mcl = \pi\Rcl^2\Scl$. We use the same procedure as above to calculate $\sfe$ and the feedback properties. Lines of constant $\Scl$ are not parallel to lines of constant $\sfe$ due to the variation of $v_{\rm ej}$ with $\Mcl$ in the $\sfe$ prescription we adopted.

We show the case of individual massive stars in \autoref{fig:feedback_ratio} using background properties identical to our simulations ($\nHbar = 86.25\pcc$, $\Rcl = 20\, {\rm pc}$) and taking values of $\Mst$, $T_{\rm eff}$, $\Qo$, and $L_{\rm bol}$ from Table 15.1 of \citet{Drainebook} for the main sequence O9.5V-O3V stars. These stars range in mass approximately from $\Mst =15$--$60\,\Msun$. We derive wind parameters $\mdot$ and $\vw$ using the prescriptions of \citet{Vink01} (Eqs. 23-25).

It should be noted that, for the properties of the $10^4 \Msun$ clouds shown in \autoref{fig:feedback_ratio}, the SFE prescription we use implies total stellar masses $\lesssim 500\Msun$ at 
$\nHbar \lesssim 10^3 \pcc$ (the bottom-most part of the purple line). At these total masses the IMF should not be well sampled and we expect stochasticity in the feedback parameters due to this. The low-$\sfe$ part of the purple curve in both panels of \autoref{fig:feedback_ratio} should then be understood to be somewhat blurred out in reality. The range of the expected spread in both dimensions is roughly given by the range of values shown for the individual massive stars.

We also use \autoref{eq:phot_trap_rhobar} to calculate the limit above which all LyC radiation is trapped within the shell of the cluster WBB when it is first formed. By using \autoref{eq:phot_trap_rhobar} in the definitions of $\zeta$ for both the MD and ED cases (\autoref{eq:zetaMD_def_app}, \autoref{eq:zetaED_def}) one can show that $\ReqRSt(\nHbartrap)$ is independent of $\Mst$ for IMF averaged feedback parameters. We follow this procedure to calculate the limit in $\ReqRSt$ above which LyC radiation is trapped, indicated by the horizontal pink line in \autoref{fig:feedback_ratio}.
It is clear that this trapping effect is only important at extremely high densities and SFEs.

Finally, we note that the parameters of the feedback are subject to change, both due to the uncertainty in $\alpha_p$ (or alternatively $\theta$ in an ``energy-driven'' model) and the possible absorption by dust or escape of LyC photons. To display the possible changes in this plot due to these effects we display arrows indicating changes due to an enhancement in $\alpha_p$ by a factor of 10 or a reduction in the fraction of photons that ionize hydrogen by 50\% as arrows in the left panel as well as a reduction in effective wind luminosity due to interface cooling by a factor of 10 in the right panel of \autoref{fig:feedback_ratio}. A factor of 10 in $\alpha_p$ is not unreasonable, as we will see in \paperii\ and a $f_{\rm ion}$ of 50\% is equally likely, as can be seen from Figure 10 of \citet{JGK18}, Figure 2 of \citet{Menon24a}, or from observational studies of massive star-forming clouds in the Milky Way \citep[e.g. Section 4 and Figure 4 of][]{Binder18} and of emission from the diffuse ionized gas in nearby galaxies \citep[e.g. Section 4.3 and Figure 8 of][]{Belfiore22}.

In \autoref{fig:feedback_ratio_lZ} we show the same parameter space spanned by \autoref{fig:feedback_ratio} but for feedback parameters ($\Qo$, $\pdot$, $\Lwind$) appropriate for a stellar population with $Z/Z_{\odot} = 10^{-2}$ as may be appropriate in the early universe. We only show the evolution with density of fixed cloud masses, using the same $\sfe$ prescriptions as above, which may not be perfectly appropriate at low metallicity as they are determined from simulations run at solar metallicity. We determine feedback parameters by scaling a given quantity, $q$, as $q = q_{\odot}(Z/Z_{\odot})^a$. We take the scaling parameters for the wind properties from \citet{Vink01} with $\mdot \propto Z^{0.85}\vw^{-1.23}$, appropriate for hot stars ($T_{\rm eff} > 2.5\times 10^4\,{\rm K}$). We use the wind velocity metallicity scaling of \citet{Leitherer92} $\vw \propto Z^{0.13}$. While these relations are extended slightly beyond the metallicity range they are originally fit to ($Z/Z_{\odot} = 1/33-3$), comprehensive observations of winds at $Z/Z_{\odot} = 10^{-2}$ do not exist to base our calculations off of and these scalings are consistent with more modern calculations \citep{VinkSander21}. We derive $\Qo \propto Z^{-0.04}$ from fitting the scaling to the output of \texttt{STARBURST99} for $Z/Z_{\odot} = 1,\, 1/7$. This is again extrapolated to low $Z$, however the dependence in any case is quite weak and small variations should not affect the broad conclusions of \autoref{fig:feedback_ratio_lZ}. Those conclusions are primarily that, at these metallicities appropriate to the early universe, momentum-driven winds with $\alpha_p = 1$ in dense ($\nH \gtrsim 10^5 \pcc$) star-forming environments also lie in the range of parameters space $0.1 \lesssim \ReqRSt \lesssim 1$ where the co-evolution models developed below should be most useful.

\subsection{Regimes of Wind \& Photoionized Gas Co-Evolution}
\label{subsec:coeval_regimes}

In \autoref{fig:feedback_ratio} and \autoref{fig:feedback_ratio_lZ} the purple, blue, and green lines are meant to be representative of the physically realizable parts of the parameter space for star-forming clouds in the universe. We will break down the part of the space occupied by these curves into three broad regions. The first is where $\ReqRSt>1$, which is relevant in dense, high $\sfe$ clouds as exist in the centers of galaxies \citep{Leroy17,Levy21,Emig20,Sun24,Levy24} or in most environments if winds manage to retain most of their energy (right hand panels of \autoref{fig:feedback_ratio} and \autoref{fig:feedback_ratio_lZ}) though this is highly unlikely based on recent theoretical and observational work \citep{Lopez14,Lancaster21a}. In this regime, the feedback bubble will be dominated by the dynamical evolution of the WBB. Since the high density limit is expected to be consistent with the momentum-driven regime, and since the momentum from radiation is comparable to that from winds \citep{Lancaster21a}, radiation pressure should also be taken into account in the expansion of these bubbles \citep{Skinner15,JGK16,Menon24a}, though we do not account for this below. Although the energy-driven solution with $\theta=0$ is not believed to apply \citep{Lopez11,Lopez14,Rosen14,Lancaster21a}, the right-hand panel of \autoref{fig:feedback_ratio} shows that if WBB solutions \textit{were} energy-driven, winds would completely dominate over photoionization feedback.

The second region of parameter space is roughly characterized by $0.1 < \ReqRSt < 1$ and $0.1 < \RSt/\Rcl< 1$, where both WBBs and the PIR can be dynamically important. At solar metallicity this corresponds to $\sfe \lesssim 10\%$ and densities $\nHbar \lesssim 10^3 \pcc$, for momentum-driven winds. This is the regime of star formation in Galactic GMCs and clouds in nearby star forming galaxies \citep{Sun22,Chevance23}. At metallicities more appropriate for the early universe ($Z = 10^{-2}\, Z_{\odot}$) this region is occupied by extremely dense star formation, which is also observed to occur in the early universe \citep{Pascale23,Adamo24}. The model we develop below will be most useful in this regime, as it more accurately portrays the nuanced interaction of the WBB and PIR which are most relevant here.

The final region of parameter space is $\ReqRSt \ll 1$, occupied by feedback from individual massive stars, or low density star forming regions at low metallicity. We expect the PIR to be dynamically dominant in this regime and we will see in \autoref{sec:model_discussion} that this is indeed the case. In this regime, the model we develop below will be useful in tracking how a weak wind is impacted by its surrounding PIR.

In the next three sections we build a `co-evolution model' (CEM) designed to be applicable in each of these parameter regimes. As indicated by \autoref{fig:schematic}, the model will consist of two distinct phases: (i) an initial Early Evolution Phase, with an embedded, over-pressurized WBB and during which the WBB and PIR are assumed to evolve independently and (ii) a later Co-Evolution Phase where the WBB and PIR are subject to constraints based on their interactions. The transition between these two regimes will depend on the region of parameter space the model lies in $\ReqRSt < 1$ of $\ReqRSt>1$.

We note that the actual dynamical importance of the WBB is roughly proportional to the fraction of the volume of ionized gas that it displaces $\approx \ReqRSt^3$. This can be quite small in the region of parameter space where $\ReqRSt \sim 0.1$, relevant to lower mass clouds. However, allowing for moderate enhancements in $\alpha_p$ ($\ReqRSt^3 \propto \alpha_p^{3/2}$) would imply that the CEM may be needed to accurately describe the evolution of the bulk dynamics of even these HII regions.

The simulations presented in \paperii\ would be found in a part of this parameter space indicated by the black circle in \autoref{fig:feedback_ratio} if $\alpha_p = 1$. In fact, in the actual simulations, $\alpha_p \approx 4-6$, moving the black circle closer to the wind-dominated regime as indicated by the yellow star marker in \autoref{fig:feedback_ratio}. In both cases, the co-evolution model we explain below should provide a more accurate description than previous idealized models.

\subsection{Early Evolution Phase}
\label{subec:early_evol}

After a short period where the feedback bubbles of individual stars join together (see discussion around \autoref{eq:tbo} in \autoref{app:trapping}) the \strom Sphere of the cluster feedback bubble will be established and the wind bubble will expand nearly adiabtically until its shell cools. In the regime where a significant fraction of the LyC radiation can be trapped by the WBB, the WBB is already dynamically dominant and the dynamics of the PIR will quickly follow that of the WBB, as we will see below, so ignoring the dynamics of trapping is allowable here. Since the time for the shell to form and for the \strom Sphere to be ionized are comparable and scale similarly with the mean background density of the system (see \autoref{app:scales}) it is reasonable to take the shell formation time of the cluster-driven wind bubble as the starting point for our early evolution phase.

At this point the WBB and the PIR are likely not in force balance with one another: the WBB will be over-pressurized with respect to the PIR. In fact, to first order, the two feedback bubbles at this point evolve independently of one another, regardless of whether the WBB follows a momentum or energy driven expansion. As each bubble is over-pressurized with respect to its surroundings and (relatively) causally unaware of the other bubble, the solutions of \autoref{subsec:theory_winds} and \autoref{subsec:spitzer} should apply. The only way the WBB affects the PIR at this point is through its dense shell; while it doesn't `trap' all of the LyC radiation, it does recombine more quickly and can make the PIR smaller through this effect. However, as the regime when WBBs and the PIR are both dynamically important (which is of most interest to this model) is far from the trapping limit (where this effect is dominant) the trapping of radiation will generally not be important. For that reason, we leave the incorporation of this effect into the early evolution phase of this model to future work. This is, however, distinct from simpler models for the dynamical impact of feedback bubbles which use a single thin-shell approximation \citep{Murray10,Rahner17,Rahner19,Geen20}.

As the WBB expands it decreases in pressure. In the $\ReqRSt < 1$ regime the WBB will come into equilibrium with the surrounding PIR at roughly $\teq$, at which point the co-evolution phase described in the next sections will apply. In the $\ReqRSt >1$ regime, the WBB may catch up with the PIR ($\Rw >\Ri$) at a time $t<\teq$. In this regime we calculate this `catch up` time at which $R_{w}(\tot) = \RSp(\tot)$, compare it to $\teq$, and choose the smaller of $\tot$ and $\teq$ as the time to switch between the Early and Co-Evolution phases for $\ReqRSt>1$.

\subsection{Momentum-Driven Co-Evolution Phase}
\label{subsec:md_jfb}

We now imagine that the WBB has come into equilibrium with the PIR with no net force being applied across its outer edges. We will assume that the WBB's evolution is given by a momentum-driven solution, with some arbitrary, constant momentum enhancement factor, $\alpha_p$, as described in \autoref{subsubsec:ec_wind}, assuming that the PIR maintains ionization-recombination equilibrium.

We can obtain a new momentum equation for this evolution by working from \autoref{eq:HIImomentum1} and using force balance to replace $\Phii = \alpha_p \pdot/4\pi \Rw^2$ giving us 
\begin{equation}
    \label{eq:HIImomentum_joint1}
\begin{split}
    \frac{d}{dt} \left(\frac{4\pi}{3} \rhobar \Ri^3 \frac{d\Ri}{dt}\right)
    &= 4\pi \Ri^2\Phii\\
    &=\alpha_p\pdot
    \left( \frac{\Ri}{\Rw} \right)^2\, ,
\end{split}
\end{equation}
where we have again assumed $\Msh = 4\pi \Ri^3\rhobar/3$. In order to solve the above differential equation we must then have a relationship between $\Ri$ and $\Rw$. This relationship will come from (i) the force balance we've derived above and (ii) ionization-recombination equilibrium\footnote{\autoref{eq:HIImomentum_joint1} would result in the co-evolution model of \citet{Geen20} if we ignored the deceleration term that results from applying the time derivative on the left-hand side.}.

The force balance condition derived above, along with $\Phii = \rho_i \ci^2$, gives us $\rho_i$ in terms of $\Rw$ as 
\begin{equation}
    \label{eq:force_cond}
    \rho_i = \frac{\alpha_p\pdot}{4\pi \ci^2 \Rw^2} \, .
\end{equation}

In the context of the co-evolution, where all gas in the WBB is hot enough to be collisionally ionized, the ionization-recombination equilibrium condition is modified from \autoref{eq:ionreceq1} to
\begin{equation}
    \label{eq:ionreceq2}
    \frac{4\pi}{3} \left(\Ri^3 - \Rw^3 \right) \alpha_B \nHi^2
    = \Qo \, .
\end{equation}
which we can re-write as
\begin{equation}
    \label{eq:ireq_cond}
    \rho_i= \rhobar\left( \frac{\RSt^3}{\Ri^3 - \Rw^3} \right)^{1/2} \, .
\end{equation}
Note that the WBB enhances $\rho_i$ relative to the classical solution, due to the presence of $\Rw$ in the denominator.

Equating \autoref{eq:force_cond} and \autoref{eq:ireq_cond} we have a relationship for $\Ri$ in terms of $\Rw$ as
\begin{equation}
    \label{eq:RiRw_rel}
    \Ri = \Rw \left( 1 + \frac{\Rw }{\Rch} \right)^{1/3} \, ,
\end{equation}
where 
\begin{equation}
    \label{eq:Rch_def}
    \Rch \equiv  
    \frac{\alpha_B}{12\pi (\muH\mH \ci^2)^2} \frac{\pdot^2\alpha_p^2}{\Qo}
\end{equation}
is the characteristic radius at which the force exerted by a momentum-driven WBB is equal to the thermal pressure force exerted by the PIR if they were being exerted in the same location \citep{KrumholzMatzner09,JGK16,Lancaster21a}. A dimensional version of this equation is given in \autoref{eq:Rch_app}. Note that, as both $\ReqMD$ and $\Rch$ quantify the relative importance of the momentum-driven WBB and photoionized gas there should be a relationship between them and indeed it is particularly straight forward:
\begin{equation}
    \label{eq:Rch_Req_rel}
    \Rch = \frac{\ReqMD^4}{\RSt^3}\, .
\end{equation}
Together, \autoref{eq:HIImomentum_joint1} and \autoref{eq:RiRw_rel} constitute an ordinary differential equation (ODE) in one variable, either $\Ri$ or $\Rw$.

\begin{figure}
    \centering
    \includegraphics{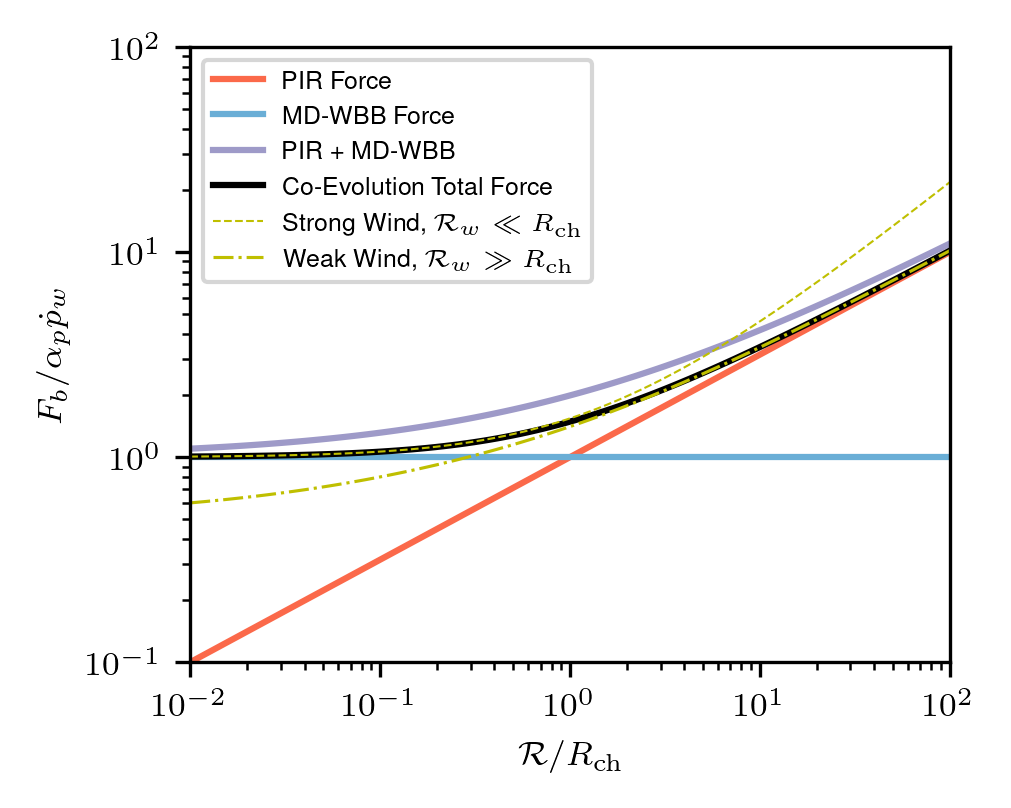}
    \caption{The force exerted by the FB in the co-evolution phase of the momentum-driven CEM. We show the forces exerted by the PIR (red), the momentum-driven WBB (blue), the sum of these (purple), the CEM (black), and the CEM in limits of strong and weak winds (yellow dashed and dashed-dotted respectively). These are shown as a fraction of the MD-WBB force and as a function of $\mathcal{R}/\Rch$, where $\mathcal{R}$ is the outer radius of the FB: $\Ri$ in the PIR and CEM Forces (and limits) and $\Rw$ in the MD-WBB. We see that expressing the force exerted by the bubble as simply the sum of the PIR and MD-WBB forces (purple) over-estimates the true force exerted by the bubble by 35\% when $\mathcal{R}/\Rch\approx 1$.}
    \label{fig:force_comp}
\end{figure}

In \autoref{fig:force_comp} we compare the relative contribution of the WBB and the PIR in the co-evolution phase showing the force given by the right hand side of \autoref{eq:HIImomentum_joint1} (black line) compared to the individual contributions of the wind ($\alpha_p \pdot$) and the Spitzer bubble force $F_{b,{\rm Sp}} = 4\pi\rhobar \ci^2\RSt^2(\Ri/\RSt)^{1/2}$ as fractions of the wind force. We show these as functions of $\mathcal{R}/\Rch$ where $\mathcal{R}$ is the outer radius of the FB: $\Ri$ in the case of the PIR-only force and the CEM, $\Rw$ in the case of the momentum-driven WBB (MD-WBB) only force.

We may take limits of the force exerted by the PIR as given by \autoref{eq:HIImomentum_joint1} in the weak or strong wind regime. The weak wind limit is characterized by $(\Rw/\Ri)^3\ll 1$ (the WBB doesn't disturb the evolution too much). Using \autoref{eq:RiRw_rel} this is equivalent to $\Rw \gg \Rch$ in which case we have
\begin{equation}
    \label{eq:approxF1}
    F_b \approx F_{b,{\rm Sp}} + \frac{\alpha_p \pdot}{2} \frac{\Rw}{\Ri} \, .
\end{equation}
Similarly, in the opposite limit corresponding to a strong wind ($\Rw \ll \Rch$), we have
\begin{equation}
    \label{eq:approxF2}
    F_b \approx \alpha_p \pdot 
    \left( 1 + \frac{2}{3}\frac{\Rw}{\Rch}\right)
\end{equation}
so that the momentum input is just slightly above the momentum-driven solution. While this limit formally implies $\Rw < \ReqMD$ for $\ReqRStMD < 1$, and therefore would not really apply in the Co-Evolution Phase, this regime should apply for a large part of the evolution in the $\ReqRStMD \gg 1$, strong-wind regime and shows the amount that the PIR adds to the dynamics in this limit.

In \autoref{fig:force_comp} we graphically compare these different approximations to the total force along with the force exerted in the Spitzer and momentum-driven WBB solutions. We see that the simple approximation $F_b = \alpha_p \pdot + F_{b,{\rm Sp}}$ tends to over-estimate the force exerted by $\sim 35\%$ when $\Ri/\Rch \approx1$. This is because the approximation does not account for how the compression caused by the wind bubble decreases the size of the PIR by increasing its density and therefore its recombination rate (see discussion after \autoref{eq:force_low_eta}). The difference between this simple approximation and the semi-analytic model is small enough that the approximation is still useful for quick estimates. This additionally lends some justification to models which make similar approximations in which the force, $F_b(r)$, of different feedback mechanisms are treated as acting at the same location ($r$) and not materially affecting the functional form of one another \citep{KrumholzMatzner09,Murray10}.

We will proceed by solving the above ODE (\autoref{eq:HIImomentum_joint1}) numerically. In order to evolve the system we will need initial conditions on $\Rw$ or $\Ri$, and one of their time derivatives. We assume the Co-Evolution phase begins at $\tswitch$. For $\ReqRStMD < 1$ we will choose $\tswitch = \teqMD$ and therefore take the initial condition on $\Rw$ as $\ReqMD$. For $\ReqRStMD > 1$ we will choose $\tswitch$ to be the minimum of $\teqMD$ and the time at which the MD-WBB cathces up with the PIR, $\RMDa(\totMD) = \RSp(\totMD)$, $\tswitch = {\rm min}(\teqMD, \, \totMD)$. We will choose as it's radius the WBB's radius at this time, $\RMDa(\tswitch)$.

For both regimes the initial condition on the velocity of the ionization front 
\begin{equation}
    \dRi (t=\tswitch) = \frac{\alpha_p \pdot \tswitch + p_{r,{\rm Sp,adj}}(\tswitch)}{4\pi\rhobar \Ri(t=\tswitch)^3 / 3} \, ,
\end{equation}
where the adjusted Spitzer momentum is taken from \autoref{eq:pr_spitzer_adj} and $\Ri(t=\tswitch)$ is determined from \autoref{eq:RiRw_rel} with $\Rw$ as above. That is, we take the velocity of the bubble's front to be determined by the momentum that should have been injected up to that point divided by the mass carried by the feedback bubble. In \autoref{app:mdje} we describe how we solve these equations in dimensionless form. The dimensionless form of these equations constitute a one-parameter family of solutions which we parameterize with
\begin{equation}
    \label{eq:zetaMD_def}
\begin{split}
    \ReqRStMD &\equiv \frac{\ReqMD}{\RSt} \\
    &= 0.47  \left(\frac{\Qo}{4\times 10^{50} \, {\rm s}^{-1}} \right)^{-1/3}\\
    & \times \left( \frac{\alpha_p \pdot}{10^5 \, M_{\odot}\, {\rm km/s/Myr}}\right)^{1/2} \\
    &\times \left( \frac{\nHbar}{100 \pcc} \right)^{1/6}
    \left( \frac{\ci}{10 \, \kms}\right)^{-1}\, .  
\end{split}
\end{equation}
Note that \citet{KrumholzMatzner09} similarly obtain a one-parameter family of solutions for bubble evolution driven by a combination of radiation pressure and ionized gas pressure.  Their parameter is $\Rch/\RSt$, which from \autoref{eq:Rch_Req_rel} is equal to $\ReqRStMD^4$.

In this formalism we instantaneously transition between the early evolution phase and the co-evolution phase at $t= \tswitch$. We therefore instantaneously require the satisfaction of \autoref{eq:RiRw_rel}. This will require one of the main state variables ($\Ri$, $\Rw$, $p_r$) to evolve discontinuously at this point. With the above choices we have made $\Ri$ the discontinuous variable.

\subsection{Energy-Driven Co-Evolution Phase}
\label{subsec:ed_jfb}

We next derive a similar model to that of \autoref{subsec:md_jfb} that assumes that the WBB follows an energy-driven solution as in \autoref{subsubsec:classic_wind} but allowing for some constant fraction of energy lost, $\theta$. As discussed at the beginning of \autoref{sec:theory}, we expect $1-\theta \ll 1$ in any realistic solution. We will still apply ionization-recombination equilibrium in the form of \autoref{eq:ionreceq2} but now the condition of force-balance will involve the balancing of the thermal pressure of the ionized gas against the thermal pressure of the WBB which will now in general depend on the history of the amount of energy stored in the WBB over time. We therefore need to additionally solve an energy equation for the WBB interior which is given by \citep{Weaver77,ElBadry19}
\begin{equation}
    \label{eq:Ewdot}
    \frac{d E_w}{dt} = (1-\theta)\Lwind - \Phot \frac{d V_w}{dt} \, ,
\end{equation}
with
\begin{equation}
    \label{eq:Ew_def}
    E_w = \frac{3}{2} \Phot V_w
\end{equation}
and $V_w = 4\pi \Rw^3/3$. In the above we have assumed that the bubble's interior energy is dominated by the thermal energy of the shocked wind material that is at thermal pressure $\Phot$ with adiabatic index $\gamma = 5/3$ and that the only way for the bubble to change its energy content is through (i) input from the mechanical energy source (gain) and (ii) mechanical work on its surroundings (loss) , and (iii) cooling at the WBB interface as parameterized by $\theta$ (loss).

Our new momentum equation is then derived similarly to \autoref{eq:HIImomentum_joint1} but by using $P_i = \Phot$. 
This modified momentum equation, along with \autoref{eq:ionreceq2}, \autoref{eq:Ewdot}, \autoref{eq:Ew_def} constitute four equations in four unknowns: $\Rw$, $\Ri$, $E_w$, and $\Phot$. We again need initial conditions on each of these as well as an initial condition on $\dRi$ as the momentum equation is second order.

Similarly to \autoref{subsec:md_jfb}, we take this Co-Evolution Phase to begin at $\tswitch$. For for $\ReqRStED < 1$ we take $\tswitch = \teqED$ with $\Rw = \ReqED$, $\Phot = \PWeaver(\teqED)$, and $\Ri$ given by \autoref{eq:ionreceq2} with $\rho_i = \Phot/\ci^2$ as required by pressure equilibrium. For $\ReqRStED >1$ we take $\tswitch = {\rm min}\left(\totED,\, \teqED \right)$. $\totED$ is the equivalent definition of the catch up time assuming that the WBB initially follows $\RWeavert$ rather than $\RMDa$.

We take the initial velocity of the ionization front to be given by the total momentum that should have been injected up until $t=\tswitch$ divided by the total mass in the shell. This gives 
\begin{equation}
    \dRi (t=\tswitch) = \frac{\prWeavert(\tswitch) + p_{r,{\rm Sp,adj}}(\tswitch)}{4\pi\rhobar \Ri(t=\tswitch)^3 / 3} \, .
\end{equation}
In \autoref{app:edje} we discuss the system of equations provided here and how we solve them in dimensionless form. This dimensionless system again constitutes a one-parameter family of solutions which we parameterize with
\begin{equation}
    \label{eq:zetaED_def}
\begin{split}
    \ReqRStED &\equiv \frac{\ReqED}{\RSt} \\
    &= 4.08  \left(\frac{\Qo}{4\times 10^{50} \, {\rm s}^{-1}} \right)^{-1/3}\\
    & \times \left( \frac{\alpha_p \pdot}{10^5 \, M_{\odot}\, {\rm km/s/Myr}}\right)^{1/2} \\
    &\times \left( \frac{\nHbar}{100 \pcc} \right)^{1/6}
    \left( \frac{\ci}{10 \, \kms}\right)^{-1}\, .  
\end{split}
\end{equation}

\begin{figure*}
    \centering
    \includegraphics[width=\textwidth]{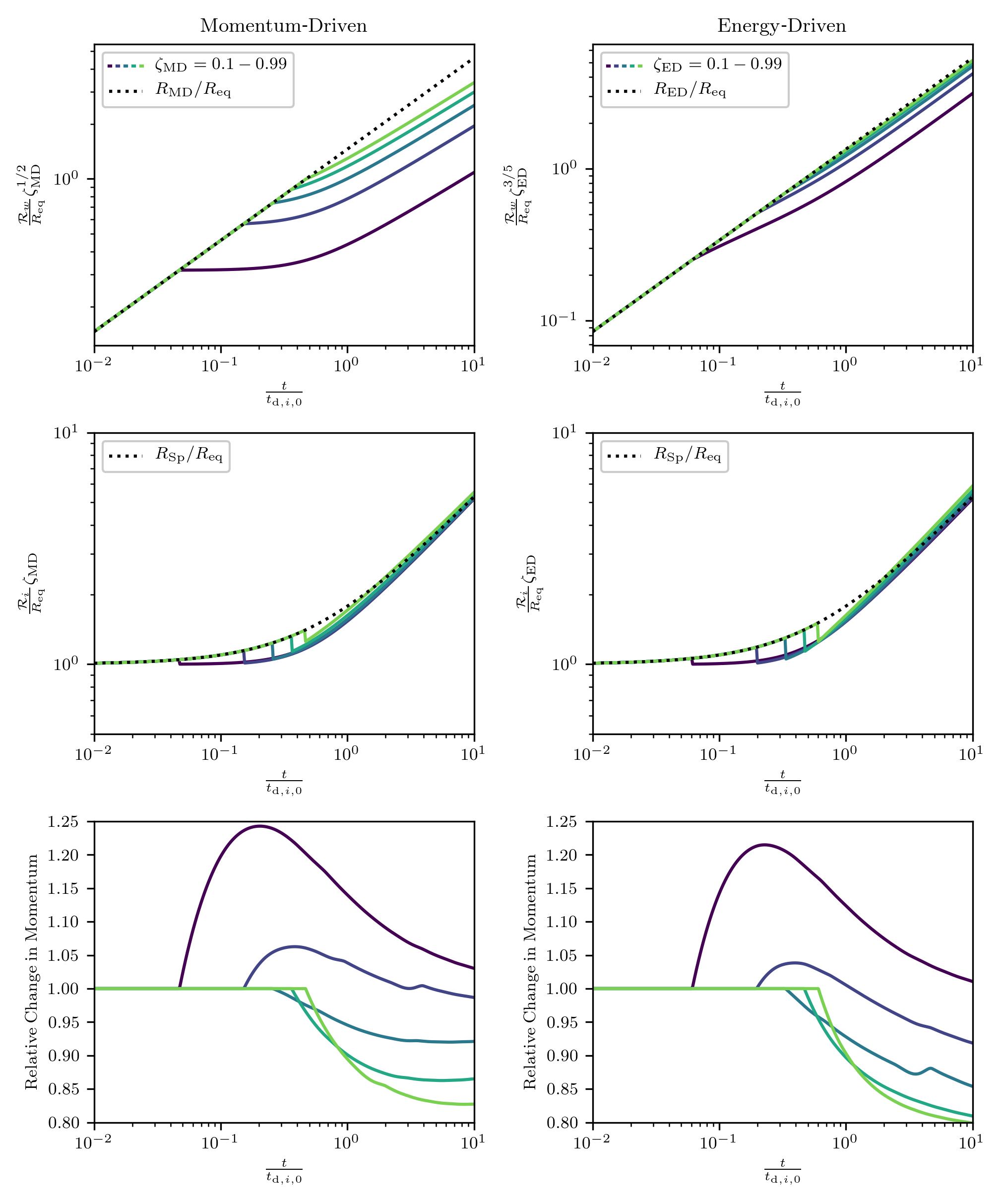}
    \caption{Solutions to the dimensionless form of the co-evolution model (CEM) for several different values of the controlling free-parameter $\ReqRSt$. The momentum-driven CEM is shown in the left hand panels and the energy-driven CEM is shown in the right hand panels. In all panels CEM solutions appear in solid colored lines ranging from $\ReqRSt = 0.10,\,0.32,\, 0.55,\,0.77,\, 0.99$. Reference solutions are shown as dotted black lines. \textit{Top panels}: The dimensionless wind bubble radius, $\xi_w \equiv \Rw/\Req$ computed using \autoref{eq:Rwind_dimensionless} (left) and \autoref{eq:dimensionless_wind_ed} (right), divided by $\ReqRStMD^{-1/2}$, $\ReqRStED^{-3/5}$ for the energy and momentum driven cases respectively. \textit{Middle panel}: Dimensionless ionized gas radius, $\xi_i\equiv \Ri/\Req$, divided by $\ReqRSt^{-1}$. The Spitzer solution, \autoref{eq:dimensionless_spitz}, is shown as a black dotted line. \textit{Bottom panel}: The relative change in total momentum carried by the bubble, as computed using \autoref{eq:printerp1}, divided by the sum of the idealized WBB (\autoref{eq:pwind_dimensionless} on left and \autoref{eq:dimensionless_prw_ed} on right) and PIR momentum (\autoref{eq:pr_spitz_adj_dimensionless}).}
    \label{fig:numerical_solution}
\end{figure*}

\section{Model Exploration}
\label{sec:model_discussion}

While the models we described in \autoref{subsec:coeval_regimes} - \autoref{subsec:ed_jfb} are applicable in the $\ReqRSt > 1$ regime, we will only explore models where $\ReqRSt < 1$, as this is the regime in which the CEMs make the largest difference to FB dynamics. In \autoref{fig:numerical_solution} we provide an exploration of the two different co-evolution models (CEMs) derived in \autoref{subsec:md_jfb} (left panels) and \autoref{subsec:ed_jfb} (right panels) in dimensionless form. In particular we show the evolution of the wind bubble and ionization front radius (top and middle panels) relative to $\Req$ for the respective models and the relative change in total momentum carried by the bubbles relative to the sum of the individual idealized solutions (bottom panels).

In the top panels we see that, for weaker wind models (smaller $\ReqRSt$), the WBB comes in to pressure equilibrium with the PIR earlier, marked by the change in the derivative of $\Rw$. At this point the WBB begins to feel the pressure exerted on it by the PIR and slows down, causing significant changes away from the unimpeded behavior (given by the black dotted line) for low $\ReqRSt$ values. The suppression of $\Rw$ is up to a factor of 3 in the MD-CEM. This deviation is less pronounced for the ED-CEM since, in this model, the containment of the wind bubble caused by the pressure force of the PIR leads to a build-up of thermal energy within the WBB which pushes back against the PIR, leading to renewed expansion of the WBB.

The evolution of $\Ri$ shown in the middle panels of \autoref{fig:numerical_solution} is less affected by the presence of the WBB, as indicated by the smaller relative deviations from the classical solution, given by the black dotted line. In fact, the deviation is at most 20\% for the models displayed here. As all of these solutions lie in the region of parameter space given by $\ReqRSt < 1$, this is somewhat to be expected. This tells us that the classical theory of HII region expansion is likely not significantly changed by the inclusion of WBBs, provided we are in the $\ReqRSt \ll 1 $ regime.

It is evident from the bottom panels that at small values of $\ReqRSt$ the momentum from the CEM solution is actually larger than the momentum given by the separate idealized solutions. This can be straightforwardly explained if we examine the pressure force term at $\teq$ in the limit of small $\ReqRSt$, this gives
\begin{equation}
    \label{eq:force_low_eta}
    \frac{4\pi \rho_i\Ri^2}{4\pi \rhobar \RSt^2} \approx 1 + \frac{1}{2}\ReqRSt^3 \, .
\end{equation}
So that it is natural to expect a momentum enhancement. Physically there are two mechanisms at play in determining the value of $4\pi \rho_i \Ri^2$. The first effect is the compression by the wind bubble, which both increases the force by increasing $\rho_i$ and decreases the force by increasing the recombination rate and drawing in the ionization-front. On balance, this effect tends to decrease the force since $\Ri \propto \rho_i^{-2/3}$ so that $4\pi \rho_i \Ri^2 \propto \rho_i^{-1/3}$. The second effect is that the wind bubble provides a volume of gas that is already ionized, liberating LyC photons to ionize gas further out, this tends to increase $\Ri$ and the force of the bubble. \autoref{eq:force_low_eta} tells us that the second effect wins out at low-values of $\ReqRSt$, though it is clear from \autoref{fig:numerical_solution} that the opposite is true at near-unity values of $\ReqRSt$. Overall, the momentum evolution of the joint feedback bubble is still within 25\% of the naive value given by the sum of the idealized solutions in all models presented.

\begin{figure*}
    \centering
    \includegraphics{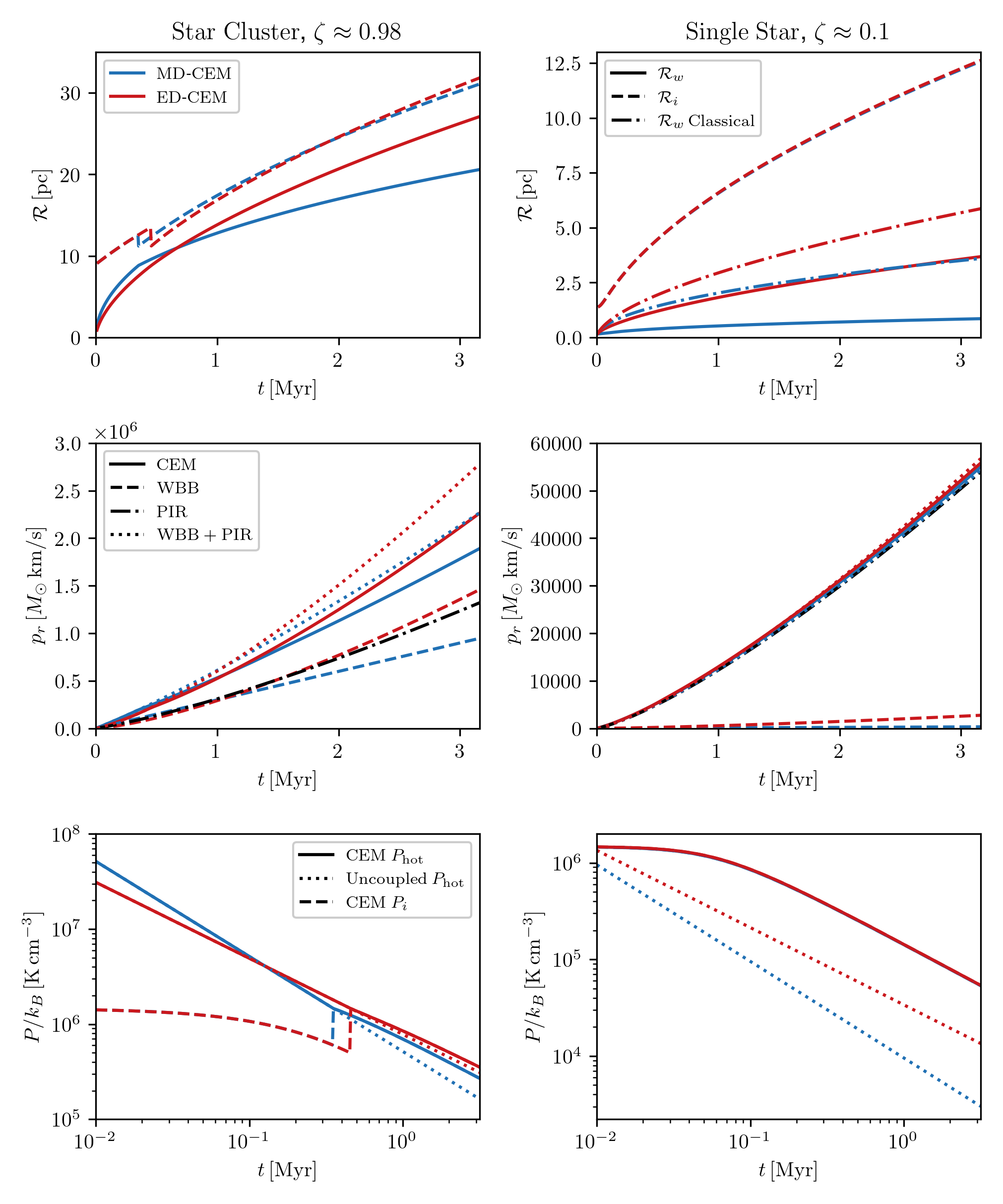}
    \caption{A comparison of physical variables for the two co-evolution models discussed in the text: the momentum-driven CEM shown in blue (\autoref{subsec:md_jfb}) and the energy driven CEM shown in red (\autoref{subsec:ed_jfb}). Panels at the left show the evolution of a FB from a star cluster ($\ReqRSt \approx0.98$) while panels at the right show models with parameters relevant to the FB around an individual massive star ($\ReqRSt \approx0.1$). \textit{Top panels}: Wind bubble (solid) and ionization front (dashed) radius over time. \textit{Middle panels}: Total momentum carried by the CEM models (solid) compared to the WBB (dashed) and PIR (dash-dotted) classical solutions and  their sum (dotted). \textit{Bottom Panels}: Pressure in the hot (solid, dotted) and ionized (dashed) gas over time.}
    \label{fig:physical_solution}
\end{figure*}

\begin{deluxetable}{ccc}
    \tablecaption{Feedback Parameters.\label{tab:fb_params}}
    \tablewidth{0pt}
    \tablehead{
    \colhead{Parameter} & \colhead{Star Cluster } & \colhead{Single Star }\\
     & ($5000\,\Msun$) & ($15\,\Msun$) }
    \startdata
    $\Lwind\, [{\rm erg}\, {\rm s^{-1}}]$ &  $4.875\times10^{37}$  & $7.642\times10^{34}$ \\
    $\mdot\, [\Msun\, {\rm Myr}^{-1}]$ & $14.825$  & $0.043$\\
    $\vw\, [{\rm km}\, {\rm s}^{-1}]$ & $3230$ & $2368$ \\
    $\pdot\, [\Msun\kms\, {\rm Myr}^{-1}]$ & $4.79\times 10^{4}$ & $1.02\times 10^2$ \\
    $\Qo\, [{\rm s}^{-1}]$ & $2.05\times 10^{50}$ & $7.59\times 10^{47}$ \\
    $\nHbar\, [ {\rm cm}^{-3}]$ & $86.25$ & $86.25$ \\
    \enddata
    \tablecomments{Parameters of feedback used in \autoref{fig:physical_solution}.}
\end{deluxetable}

While the dimensionless version of the solutions are useful to understand scalings, it is more physically insightful to inspect a dimensional evaluation of these solutions. To that end, in \autoref{fig:physical_solution} we present a comparison of the momentum-driven and energy-driven versions of the CEMs for parameters of $\rhobar$, $\Lwind$, $\pdot$, and $\Qo$ appropriate for (i) the simulations presented in \paperii\, of feedback from a massive star cluster and (ii) the FB around an individual massive star with $\Mst \approx 15\,\Msun$ as represented by the palest red star in \autoref{fig:feedback_ratio}. We outline the feedback parameters used for each of these cases in \autoref{tab:fb_params}.

We assume that $92\%$ of $\Lwind$ is lost to cooling so that $\theta = 0.92$ for the model presented in \autoref{fig:physical_solution}, consistent with levels of energy loss in recent simulations \citep{Lancaster21b}. In order to contrast against a momentum-driven model with a similar value of $\ReqRSt$ we take $\alpha_p = 6.25$ (which is comparable to the values realized in the simulations given in \paperii) so that both models have $\ReqRSt \approx 0.98$. In particular, both models have $\Req \approx 8.8 \pc $ and $\RSt \approx 9.0\pc$. The single star models are evaluated in order to illustrate the behavior of the CEMs in the low $\ReqRSt$ limit, to this end with take $\alpha_p = 1$ for the MD-CEM and $\theta = 0.98$ for the ED-CEM. These choices result in $\ReqRSt \approx 0.1$ for both models, in particular $\RSt = 1.4\pc$ and $\Req = 0.165\pc$ in both models.

In \autoref{fig:physical_solution} we show the model evolutions for the star-cluster CEMs in the left panels and the single star CEMs in the right panels. The models illustrate the radial evolution of the ionized gas and WBB (top panel), the momentum evolution (middle panel), and the pressure in both the hot WBB and the PIR (bottom panel) for both the MD- and ED-CEM.

In the models with parameters applicable to the star cluster feedback simulated in \paperii, we see that, while we have tuned these models to be nearly comparable, the energy-driven CEM still quickly overtakes the momentum-driven one at $t>\teq$ in both $\Rw$ and total momentum evolution. At these high values of $\ReqRSt$ it is also clear from the middle panel that both CEMs decrease the total dynamical impact of both feedback mechanisms, as one can tell from comparison to the reference solutions (the sum of \autoref{eq:prweaver} or \autoref{eq:pr_EC} and \autoref{eq:pr_spitzer_adj}) shown as dotted lines. It is also clear, comparing the dashed and dot-dashed lines, that the WBB and PIR contribute nearly equally in this scenario. We use the momentum driven version of this co-evolution model as a point of comparison against our three-dimensional hydro-dynamical simulations in \paperii.

We use the model parameters of a single star, shown in the right columns of \autoref{fig:physical_solution}, in order to illustrate the behavior of the CEMs in the low $\ReqRSt$ limit. For this reason, this treatment assumes the background is of uniform density and ignores the issue of trapping of ionizing radiation in the isothermal sphere-like density profiles surrounding individual stars (see \autoref{app:trapping} for more details). In the low $\ReqRSt$ models displayed in \autoref{fig:physical_solution} we see that the expansion of the PIR, given by the dashed lines in the top right panel, is relatively unaffected by the presence of the WBB, as we would expect for this weak wind limit. This is also seen in the momentum evolution where the CEM evolution (solid, colored lines) are relatively indistinguishable from one another or the classical PIR evolution (dash-dotted black line). The WBB evolution, however, is drastically impacted by the back-reaction of the PIR, as indicated by the difference in evolution of $\Rw$ between the CEMs (solid, colored lines) and their corresponding classical energy- or momentum-driven solutions (dash-dotted, colored lines). In these models the pressure of the hot gas in the WBB as a function of time (solid lines in bottom right panel) is essentially determined by the pressure of the PIR, which is indistinguishable from the hot gas pressure evolution in these plots. While these diagnostics emphasize that the PIR is the main source of dynamical feedback in these regimes, it also shows that the CEMs can be used to more accurately determine the size and dynamics of WBBs in these scenarios, which should also be applicable to low-mass, low-density clouds.

\section{Discussion}
\label{sec:discussion}

Here we review the past works that have addressed this problem and provide an outlook for the future development of these semi-analytic models.

\subsection{Past Treatments}
\label{subsec:theory_review}

\citet{CapriottiKozminski01} gave a review of the dynamics of WBBs that focused on the collisionless aspect of WBB interiors and in particular how this aspect of the dynamics, along with the potential for mixing, could lead to momentum-driven like behavior for the WBB. They also derive the evolution of photoionized gas in a fashion similar to \citet{Spitzer78} and \citet{HosokawaInutsuka06}. However, they did not consider the perturbation that the wind has on the dynamics of the photoionized gas. Using this simplified description of the feedback bubble dynamics they compare the kinetic energy injected by the WBB up until its expansion becomes subsonic in the PIR ($\dot{\mathcal{R}}_w = \ci$) to the kinetic energy of the neutral shell surrounding the PIR just after it has begun to expand ($\Ri/\RSt \approx 1.38$) and conclude that the latter is $\sim 10^4 \times$ the former for typical parameters (photoionized gas dominates the dynamics). This comparison ignores the question of the time scales over which each of these conditions is met and how they compare to one another and the dynamical time scale of the star forming region. In particular, as we discuss in \autoref{sec:theory_joint}, under dense (but not unreasonable $\Scl \gtrsim 10^3 \, \Msun\, {\rm pc}^{-2}$) star-formation environments, the WBB quickly over-runs the photoionized gas before it has a chance to significantly expand, even if the WBB is in a momentum-driven $\alpha_p \sim 1$ regime. The treatment of \citet{CapriottiKozminski01} also ignores the dynamical impact of the WBB after it has come into force balance with the PIR, which we have demonstrated can be quite significant.

\citet{Raga12c} derive an ODE for the evolution of a thin-shell of photoionized gas around an energy-driven WBB using the balance of ram-pressure and thermal pressure at the bubble's edge (similar to \citet{Spitzer78} and \citet{Geen20}, see below). The same authors had previously emphasized \citep{Raga12b} that such an approach ignores the inertia of the shell and can lead to significant errors. In order to relax the assumption of a thin-shell they derive an additional ODE for a `thick-shell' that is roughly equivalent in approach to our derivation given in \autoref{subsec:md_jfb} but again applies the balancing of ram pressure with thermal pressure at the edge of the ionized gas. Due to the use of this condition their solutions eventually relax to stable points where the bubbles are in pressure equilibrium with their surroundings. Both models presented in \citet{Raga12c} assume that (i) the photoionized gas is immediately in pressure equilibrium with the WBB ($\teq = 0$) and (ii) that the WBB is energy-driven which both over-emphasize the dynamical importance of the WBB by (i) restricting the size of the photoionized gas region unrealistically and (ii) not accounting for cooling in the WBB.

The \texttt{WARPFIELD} model of \citet{Rahner17,Rahner19} provides a one-dimensional framework for solving for the dynamical evolution of the feedback bubble from a massive star cluster. The model assumes that the bubble is made up of a single thin-shell where all forces (from stellar winds, gravity, direct and indirect radiation pressure) are applied. \texttt{WARPFIELD} accounts for the time-dependence of cooling in the WBB interior caused by a conductively evaporative flow, as in \citet{Weaver77}, but does not include a model for turbulently enhanced cooling \citep{ElBadry19,Lancaster21a} which could make the WBBs act in a more momentum-driven manner much earlier in the evolution even when conduction is weak. The structure of the shell driven by the bubble is solved on-the-fly under the assumption of hydro-static equilibrium \citep{Abel05,Pellegrini07,Draine11,JGK16} which impacts the amount of ionizing radiation caught in the shell and the coupling of radiation pressure on dust grains. While the thermal pressure of the photoionized gas is used to solve for the hydro-static equilibrium shell, the force due to this thermal pressure gradient is not included in the momentum evolution equation \citep[see][Eqs 5-9]{Rahner17}. Upon comparison to \autoref{fig:physical_solution}, we can see that this is would not result in an accurate description of the dynamics for typical, massive GMCs like those meant to be represented by the left-hand-side panels. In these clouds, feedback from LyC radiation contributes nearly equally or more to that of the WBB.

\citet{Geen20} follows a similar approach to that of \texttt{WARPFIELD}. However, like \citet{Raga12c}, instead of a momentum equation, \citet{Geen20} uses a condition on the balance of the ram-pressure of the surrounding medium and the pressure of the feedback bubble at a single radius (from various feedback mechanisms) to evolve the feedback bubble's dynamics. Included in this evolution equation is a generic velocity term, ``$v_0$," that is meant to mimic the effects of accretion and gravity. As pointed out by \citet{Raga12b}, the balancing of thermal and ram-pressure approach ignores the inertia of the shell which can have a large impact on the bubble dynamics (e.g. the difference between the \citet{Spitzer78} and \citet{HosokawaInutsuka06} solutions). \citet{Geen20} additionally assume that the WBB is able to efficiently cool, entering a momentum-driven regime with $\alpha_p = 1$, as soon as the swept-up shell of the WBB is able to cool. Like \citet{Raga12c}, the model of \citet{Geen20} assumes that the two phases are immediately in force-balance with one another, equivalent to assuming that the co-evolution phase of our model applies instantaneously. Again, this assumption drastically under-estimates the impact of the photoionized gas at early times when considering expansion into an \textit{unstratified} background density as it assumes that the photoionized gas is effectively trapped by the WBB's shell. However, as discussed at length in \autoref{app:trapping}, this is likely a correct assumption when considering expansion of the bubble into a steeply stratified background density profile, like that of an isothermal sphere which is considered in the majority of the work of \citet{Geen20}. This latter case likely applies at early times in a cluster's evolution around individual massive stars before a cluster feedback bubble (from many stars) is able to be driven. \citet{Geen22} provide a further investigation of the effect of WBBs and photoionized gas expanding particularly into an isothermal sphere density background and provide a calculation similar to that of \autoref{app:trapping} for when photoionized gas should or shouldn't be trapped by the WBB's shell.

\subsection{Prospects for Future Work}
\label{subsec:problems}

There are four main avenues for improvement to the semi-analytic models discussed here. The first regards the question of cooling of WBBs. As we can see from \autoref{fig:feedback_ratio}, whether or not one draws the conclusion that WBBs are dynamically important hinges strongly on the question of the WBBs being momentum or energy-driven. This is principally a question of diffusion processes across the WBBs surface (both thermal and turbulent heat dissipation) which remain incompletely understood \citep{Lancaster24a}. The ideal version of these semi-analytic models would include a parameterized model for this heat dissipation that is solved on-the-fly and included in the WBB energy equation (akin to \autoref{eq:Ewdot}). This is done for the case of only conductive heat dissipation in a spherical scenario by the \texttt{WARPFIELD} models, but as \citet{Lancaster21a,Lancaster21c} has shown, cooling in turbulently-mixed intermediate-temperature gas can certainly dominate energy losses.

Second, while we focus here on the interaction of photo-ionizing radiation and WBBs, a full description of the dynamics and structure of HII regions should include further feedback physics (e.g. direct and indirect radiation pressure). While this has been accounted for in other feedback simulations of star-forming clouds \citep[e.g.][]{Skinner15,STARFORGE21,JGK18,Menon22,Menon23}, a full accounting for the effects of these various mechanisms on a wide range of environments is yet to be carried out. Semi-analytical models \citep[e.g.][]{Rahner17,Rahner19,Toddlers23} have nominally included these effects, but usually make simplifying assumptions such as the thin-shell approximation which prevent a faithful representation of the gas density and ionization structure, as we discuss in \autoref{subsec:theory_review}.

Third, in order to best match observations and simulations, these semi-analytic models must take into account inhomogeneities in the background medium into which they expand. As we show in \paperii, while our models are able to broadly match the characteristics of the simulations, a degree of disagreement is baked-in due to the assumption of spherical symmetry in the CEMs. Future semi-analytical models should ideally attempt to account for the inhomogeneity that is inherent to the molecular clouds in which stars form if they aim to produce accurate dynamical or observational predictions.

The CEM models we have developed here imply that the effect of WBBs on their surrounding PIR, and vice-versa, should have strong observational consequences. These should be most drastic for the WBBs, which can be effectively contained by the PIR in the regions of parameter space probed here (\autoref{fig:numerical_solution}). The compression of the PIR by the WBB should also lead to density inhomogeneities in the PIR which will have consequences for models of nebular emission \citep[e.g.][]{MendezDelgado24}. We leave detailed predictions for observational signatures from our semi-analytic models to future work. The abundance of state-of-the-art data available from current and upcoming experiments \citep{SOFIA_FEEDBACK,LVM_Science,Kreckel24_Orion,SIGNALS} should make this a very fruitful avenue for learning about massive star feedback physics.

\section{Conclusion}
\label{sec:conclusion}

We end here with a short summary of our main conclusions:
\begin{itemize}
    \item After reviewing classical feedback models in \autoref{sec:theory}, we motivate in \autoref{sec:theory_joint} a new approach to the simultaneous impact of both photo-ionizing gas pressure and stellar wind-blown bubbles that is centered on the comparison of the radius at which the WBB comes into pressure equilibrium with the photoionized gas, $\Req$, and the classical \strom radius, $\RSt$. When $\Req \ll \RSt$ photoionized gas pressure will be dynamically dominant while for $\Req\gg \RSt$ winds will dominate.

    \item We consider values of $\Req/\RSt$ for star forming clouds across all different star-forming regimes: from that of dense star formation in the early universe to Milky-Way like GMCs. Recent work has suggested that wind feedback will be close to the momentum-driven limit, due to strong interface cooling.  In this case, both typical Milky-Way star-forming GMCs (\autoref{fig:feedback_ratio}) and dense star forming, low-metallicity environments (\autoref{fig:feedback_ratio_lZ}) fall in the regime $\Req \lesssim \RSt$ in which both WBB and photoionized gas are dynamically important. Only in the densest star forming environments (or if there is minimal cooling of shocked wind gas) will winds dominate feedback.
    
    \item Motivated by the fact that both winds and photoionization are important, in \autoref{subsec:coeval_regimes} and subsequent sections we develop Co-Evolution Models that take into account the dynamical back-reaction of WBBs on photoionized gas and vice versa. We develop a model for both momentum-driven and energy-driven winds. These models consist of an `early phase' during which WBBs and the PIR evolve independently and a `co-evolution phase' during which they are affected by one another. The early phase takes up a larger fraction of the evolution for larger $\ReqRSt$.

    \item In \autoref{sec:model_discussion} we explore these models and their consequences for the dynamical evolution (momentum, bubble radii, pressure) of the ionized gas in star-forming regions. We conclude that, for typical parameters, the back reaction of WBBs and the PIR on one another acts to decrease the dynamical impact of both (see \autoref{fig:physical_solution}). Overall, however, we find that in the $\Req<\RSt$ regime where these models are most relevant, the evolution of the radius of the photo-ionized gas only mildly differs from the classical Spitzer solution. Moreover, the total momentum injected by the bubble is within 25\% of the sum from wind and photoionized gas considered separately.

    \item In \autoref{sec:discussion} we provide a thorough comparison of this work to past models developed in the literature and outline prospects for future improvements. Principal among these would be to remove the assumption of a constant density PIR and take into account the effects of direct radiation pressure as has been previously considered in other spherical models \citep[e.g.][]{JGK16}.

    \item In \autoref{subsec:trapping} and \autoref{app:trapping} we consider the trapping of LyC radiation in the shells of WBBs and find that this is only important in the expansion of feedback bubbles around individual massive stars (where the surrounding density profile is steep) and in extremely dense star forming environments. We also motivate that the breakout of photoionized gas from the shells of individual stars is relatively quick compared to the evolutionary timescale of GMCs, so that the \strom sphere corresponding to the feedback from several stars should form relatively quickly.

    \item The models developed here are tested against three-dimensional, inhomogeneous radiation magneto-hydrodynamical simulations in \paperii. There we find remarkably good agreement between the models and general properties of the simulations. The main disagreements between the models and simulations are demonstrated to originate from the models' (i) lack of inhomogeneous gas structure and (ii) lack of time-variable momentum input rates ($\alpha_p(t)$).
\end{itemize}

\acknowledgments

The authors would like to thank Ulrich Steinwandel, Laura Sommovigo, Brent Tan, Drummond Fielding, Romain Teyssier, Mike Grudi\'{c}, Erin Kado-Fong, and Mordecai-Mark Mac Low for useful discussions. The authors would like to thank the anonymous referee for a thorough and thoughtful review. L.L. gratefully acknowledges the support of the Simons Foundation under grant 965367. This work was supported in part by grant 510940 from the Simons Foundation to E.~C.\ Ostriker. C.-G.K. acknowledges support from a NASA ATP award No. 80NSSC22K0717. J.-G.K acknowledges support from the EACOA Fellowship awarded by the East Asia Core Observatories Association, and from KIAS Individual Grant QP098701 at Korea Institute for Advanced Study. GLB acknowledges support from the NSF (AST-2108470, AST-2307419), NASA TCAN award 80NSSC21K1053, and the Simons Foundation through the Learning the Universe Collaboration.
\software{
{\tt scipy} \citep{scipy},
{\tt numpy} \citep{harrisNumpy2020}, 
{\tt matplotlib} \citep{matplotlib_hunter07},
{\tt adstex} (\url{https://github.com/yymao/adstex})
}

\appendix

\section{Trapping of Photoionized Gas}
\label{app:trapping}

Many recent theoretical works on the dynamics of feedback bubbles driven by both winds and photoionized gas have emphasized the ``trapping" of the photo-ionizing or Lyman Continuum (LyC) radiation by the shell of material swept-up by the wind bubble \citep{Rahner17,Rahner19,Geen22,Geen23,Toddlers23}.  While such a situation is often observed in simulated feedback bubbles of single stars in backgrounds with stratified densities \citep{Geen23} and is convenient for theoretical modeling (as it allows the feedback bubble dynamics to be treated in a single, thin-shell formalism) it is often not seen in simulations with a cluster of massive stars \citep[e.g. \paperii\ and][]{Dale14} or uniform density backgrounds \citep{Ngoumou15,Haid18}. We provide an explanation for this in this appendix by analyzing the density of the shell of gas swept-up by the WBB and therefore its capacity to trap LyC radiation.

We follow the derivation of \citet{KooMcKee92a,KooMcKee92b} for the expansion of a `fast,' constant mechanical luminosity, $\Lwind$ ($\etain = 1$ in their formalism), WBB in a power-law background density profile of the form
\begin{equation}
    \label{eq:density_power_law}
    \rhobg(r) = \rhozo r^{-\krho} \, ,
\end{equation}
so that the mass enclosed within a radius $r$ is
\begin{equation}
    \label{eq:mass_poewr_law}
    M_{\rm bkgnd}(<r) = \frac{4\pi \rhozo r^{3-\krho}}{3-\krho} \, .
\end{equation}
In the above, the dimensionality of the constant, $\rhozo$ is dependent upon the value of $\krho$ that is assumed. For a uniform background, $\krho = 0$, $\rhozo$ is the mean background density. We will assume $0\leq \krho < 3$, so that the source is not in a cavity and that $M_{\rm bkgnd}(r) \to 0$ as $r\to 0$.

\citet{KooMcKee92b} give the expansion of a WBB in such a background density profile as
\begin{equation}
    \label{eq:Revol_power_law}
    \Rw(t) = \left[ \frac{(3 - \krho)\Grad\xi\Lwind t^3}{3\rhozo}\right]^{\frac{1}{5-\krho}}\, ,
\end{equation}
where $\Grad$ is the fraction of energy injected by the bubble that remains after accounting for losses to radiative cooling and $\xi$ is an order unity parameter (akin to $\alpha$ in Section II of \citet{Weaver77}). Note that choosing $\krho = 0$ and $\Grad\xi = 125/154\pi$ reduces the above to the classical result, \autoref{eq:Rweaver}. Note that the scaling law of the bubble's radial expansion is
\begin{equation}
    \label{eq:eta_scl_def}
    \eta \equiv \frac{d\ln \Rw}{d\ln t} = \frac{3}{5-\krho} \, .
\end{equation}
One can see from the above (as is also noted in \citet{Geen20}) that for the case of an isothermal sphere with $\krho=2$ (as might be expected in the cores around proto-stars \citet{LeeHennebelle18}) we have $\eta = 1$, i.e. a constant velocity radial expansion. For the momentum-conserving case a similar derivation would lead us to $\eta = 2/(4-\krho)$, which also leads to constant velocity expansion when $\krho =2$.

We next want to investigate the cooling time of the shocked, ambient medium that is swept-up by the expanding bubble. Following \citet{KooMcKee92a,KooMcKee92b} for the case $\gamma=5/3$ the cooling time of the gas that has just been shocked at time $t$ is
\begin{equation}
    \label{eq:tcool_shock}
    \tcool(t) = \frac{\muH^2\mH}{\mu}\frac{k_B T}{\rho_{\rm sh}\Lambda(T)} \approx \frac{\muH^2\mH}{4\mu}\frac{k_B T^{1+\alpha}}{\rhobg(\Rw(t))\Lambda_1} = \frac{C_1}{\rhozo} \dRw^{2(1 + \alpha)}\Rw^{\krho} \ \, ,
\end{equation}
where $\mu = \rho /\mH n$ is the mean mass per particle and $\muH = \rho/\mH \nH$ is the mean mass per hydrogen nucleus, both in units of $\mH$. In the above we have taken a simplified power-law cooling function $\Lambda(T) = \Lambda_1 T^{-\alpha}$. To simplify the presentation below, our default choice of these values are those used in \citet{KooMcKee92a,KooMcKee92b}: $\alpha=1/2$ and $\Lambda_1 = 1.6 \times 10^{-19} \, {\rm erg}\, {\rm cm}^3\, {\rm s}^{-1}\, {\rm K}^{1/2}$, though a more modern and accurate estimate is that given by \citet{Drainebook} with $\alpha= 0.7$ and $\Lambda_1 = 1.74 \times 10^{-18} \, {\rm erg}\, {\rm cm}^3\, {\rm s}^{-1}\, {\rm K}^{0.7}$. These assumptions both roughly approximate the CIE cooling function for cosmic abundances in the temperature range $10^5\,{\rm K} \leq T \leq 10^7 \, {\rm K}$ as is relevant for such shocks. In order for one to easily apply either cooling function we will keep most of the derivation below general. \citet{maclow88}, who use a cooling function that is virtually identical to that of \citet{Drainebook}, additionally take the cooling function to be linearly proportional to the metallicity relative to solar. This can be accounted for in our model by taking $\Lambda_1 \to \left(Z/Z_{\odot}\right)\Lambda_1$. The parameter $C_1$ is then defined as
\begin{equation}
    C_1 \equiv \frac{\muH^2 \mH k_B}{4\mu \Lambda_1} 
    \left[\frac{3}{16} \frac{\mu\mH}{k_B} \right]^{1 + \alpha} \, .
\end{equation}
For the choice of parameters from \citet{KooMcKee92a,KooMcKee92b} this yields $C_1 = 5.97 \times 10^{-35}\, {\rm g}\, {\rm cm}^{-6} \, {\rm s}^4$. The above also applied the strong jump conditions $\rho_{\rm sh} = 4 \rhobg(\Rw(t))$ and $T = 3\mu\mH \dRw^2(t)/16 k_B$. Using \autoref{eq:Revol_power_law}, we can then calculate the ratio of cooling time to the age of the system as
\begin{equation}
    \label{eq:tscale_ratio}
    \frac{\tcool}{t} = \frac{C_1\eta^{2(1 + \alpha)}}{\rhozo} 
    \left[ \frac{(3 - \krho)\Grad\xi\Lwind}{3\rhozo} \right]^{\frac{2(1+\alpha)+\krho}{5-\krho}}
    t^{-\frac{(9 + 4\alpha) - 2(3+\alpha)\krho}{5-\krho}} \, .
\end{equation}
Our following treatment then splits on whether or not this ratio decreases ($\krho < \kcrit$) or increases ($\krho > \kcrit$) in time, with $\kcrit = (9+4\alpha)/(6+2\alpha) = 59/37,\, 11/7$ for $\alpha = 0.7,\, 0.5$ respectively ($\kcrit \approx 1.6$ for both $\alpha$ values). In the case that the above ratio decreases in time, the shocked ISM surrounding the bubble will begin as adiabatic and will likely be hot enough to be collisionally ionized, so that its ability to absorb LyC photons would be very low. Eventually, it will become radiative and collapse to form a thin, dense shell which should have a very high capacity to absorb LyC radiation. In the latter case, the shell will begin as dense and radiative, with a high capacity to absorb LyC photons, and eventually transition to a phase where its forward shock becomes adiabatic, decreasing its ability to trap LyC photons.

We can obtain an estimate for when the WBB shell transitions from adiabatic to radiative ($\krho < \kcrit$) or vice versa ($\krho>\kcrit$) by setting \autoref{eq:tscale_ratio} equal to unity and solving for, $t$ with the assumption $\Grad\xi \propto {\rm const.}$. This gives us the `transition time' as
\begin{equation}
    \label{eq:shell_formation_power_law}
    t_{\rm trans} = \left[ \frac{C_1 \eta^{2(1+\alpha)}}{\rhozo} \right]^{\frac{5-\krho}{(9 + 4\alpha) - 2(3+\alpha)\krho}}
    \left[ \frac{(3-\krho)\Grad\xi \Lwind}{3\rhozo}\right]^{\frac{2(1+\alpha)+\krho}{(9 + 4\alpha) - 2(3+\alpha)\krho}} \, .
\end{equation}
In the $\krho < \kcrit$ case this is the shell formation time, $\tsf$.

We next want to calculate the ability of the WBB shell to trap LyC radiation by calculating its recombination rate, assuming it is in ionization-recombination equilibrium. Assuming that the shell of gas around the WBB is at a constant density (discussed further below) and nearly fully ionized its recombination rate is 
\begin{equation}
    \label{eq:Qrec_power_law}
    \Qrec (t) = V_{\rm sh} n_{\rm sh}^2 \alpha_B 
    = \frac{M_{\rm sh}}{\rho_{\rm sh}}\left( \frac{\rho_{\rm sh}}{\muH \mH}\right)^2\alpha_B
    = \frac{4\pi \rhozo \Rw^{3-\krho}}{(3-\krho)(\muH\mH)^2} \rho_{\rm sh} \alpha_B
    = \frac{4\pi \alpha_B \rhozo^2\Rw^{3-2\krho}\dRw^2}{(3-\krho)(\muH\mH \ci )^2} \, 
\end{equation}
where in the last equality we have taken the constant density of the shell to be $\rho_{\rm sh} = \dRw^2\rhobg(\Rw)/\ci^2$ as is true for an isothermal shock, as should be relevant when the shell is radiative and able to absorb LyC photons. Using \autoref{eq:Qrec_power_law} and \autoref{eq:Revol_power_law} one can show that $\dot{Q}_{\rm rec} < 0$ when $\krho > 5/4$ (also noted in \citet{Geen22}) and, in the momentum-driven case $\dot{Q}_{\rm rec}<0$ when $\krho > 1$. That is, the shell's ability to trap LyC radiation decreases in time for steep density gradients.

It is instructive to investigate $\Qrec(t_{\rm trans})$ as for $5/4<\krho < \kcrit$ this will be when the shell has its maximum ability to trap LyC radiation, while for $\krho > \kcrit$ (where it was higher at earlier times) and $\krho < 5/4$ (where it will be higher at later times) it will be at its minimum value. We can calculate this using \autoref{eq:Qrec_power_law} and \autoref{eq:shell_formation_power_law} as
\begin{equation}
    \label{eq:qrec_tsf_general}
    \Qrec(t_{\rm trans}) = \frac{4\pi \alpha_B \rhozo^2\eta^2}{(3 - \krho)(\muH\mH\ci)^2}
    \left[\frac{C_1 \eta^{2(1+\alpha)}}{\rhozo}\right]^{\frac{5-4\krho}{(9 + 4\alpha) - 2(3+\alpha)\krho}}
    \left[\frac{(3 - \krho)\Grad\xi\Lwind}{3\rhozo}\right]^{\frac{11 + 6\alpha + 8\krho + 4\krho\alpha}{(9 + 4\alpha) - 2(3+\alpha)\krho}} \, .
\end{equation}

By comparing \autoref{eq:qrec_tsf_general} to $\Qo$ we can then obtain a condition on the parameters of the problem ($\rhozo$, $\krho$,  and $\Lwind$) where the shell traps the LyC radiation. For the simplest case of $\krho = 0$, $\rhozo = \rhobar$, and the \citet{KooMcKee92a,KooMcKee92b} choice of the cooling function ($\alpha = 1/2$) this gives us a simple limit on $\rhobar$ above which the LyC radiation is trapped as soon as the shell forms. With the additional assumptions of $\Grad = 1$ and $\xi =(0.88)^{11/14} \approx 0.90$ (as in Section II of \citet{Weaver77}) the number density of hydrogen atoms at which the LyC radiation will be trapped at the initial shell formation is
\begin{equation}
    \label{eq:phot_trap_rhobar_app}
\begin{split}
    \nHbartrap &= \frac{1}{\muH\mH} \left[ \frac{25\left( \muH \mH \ci\right)^2\Qo}{12\pi \alpha_B} \right]^{\frac{11}{3}}
    \left[\xi \Lwind \right]^{\frac{-14}{3}}
    \left[\frac{125}{27 C_1} \right]^{\frac{5}{3}} \\
    &= 1.48 \times 10^{7} \pcc 
    \left(\frac{\Qo}{10^{50} \, {\rm s}^{-1}} \right)^{11/3} 
    \left(\frac{\Lwind}{2 \times 10^{37}\, {\rm erg/s}} \right)^{-14/3} 
    \left(\frac{Z}{Z_{\odot}} \right)^{5/3}\, ,
\end{split}
\end{equation}
where we have additionally assumed $\ci = 10\kms$ and $\alpha_B = 3.11\times 10^{-13} \pcc \, {\rm s}^{-1}$ (explained in \autoref{app:scales}) and included the metallicity scaling that comes from the physics of cooling in the shell. In the above we have intentionally chosen values for $\Lwind$ and $\Qo$ that are appropriate for a star cluster roughly of a mass $\Mcl = 2\times 10^3 \Msun$ as we wish to estimate when the LyC radiation from a cluster-driven FB will be trapped. In this uniform density environment it is clear that very high densities are needed in order to trap the LyC radiation at the beginning of shell formation. However if $\Qo,\, \Lwind \propto \Mcl$, then $\nHbartrap \propto \Mcl^{-1}$ and this density could be achievable in high SFE environments \citep{Leroy17,Levy21,Levy24,Emig20,Sun24,Pascale23,Adamo24}. We additionally note that, as $C_1 \propto \Lambda_1^{-1} \propto (Z/Z_{\odot})^{-1}$ we have $\nHbartrap \propto (Z/Z_{\odot})^{5/3}$. This somewhat counter-intuitive scaling is due to the fact that the shell forms earlier at higher metallicity ($\tsf \propto (Z/Z_{\odot})^{-5/11}$) at which point the bubble has accumulated less mass in its surrounding shocked ISM, and therefore this mass is less able to trap LyC radiation when it cools and condenses.

We can then consider the opposite regime of an isothermal sphere, $\krho = 2$, which, as we argued above, should be relevant to the cores around forming proto-stars. We will maintain the simpler cooling function ($\alpha = 1/2$). In this case, $\Qrec(t_{\rm trans})$ represents the minimum capacity for the WBB shell to absorb LyC photons in time. It is important to note that the pure power-law profile of \autoref{eq:density_power_law} necessitates through \autoref{eq:Qrec_power_law} that, no matter the value of $\Qo$, there will always be a time early enough when all of the LyC radiation is trapped, i.e. $\Qrec(t) > \Qo$. This may not be the case if we actually consider the time at which any kind of shell is formed\footnote{This is roughly when the inertia of the swept-up gas is equal to the inertia of the free wind.}, or a density profile with a central, constant density region at some small radius.

Before we consider the $\Qrec(t_{\rm trans})$ let us consider the transition timescale itself. Specifying \autoref{eq:shell_formation_power_law} to $\krho = 2$ we have
\begin{equation}
    t_{\rm trans} (\krho = 2) = \frac{\rhozo}{C_1} \left[\frac{3\rhozo}{\Grad \xi \Lwind} \right]^{5/3}
    =  244\, {\rm Myr} 
    \left( \frac{\rhozo/\muH\mH}{100\, \pcc {\rm pc}^2}\right)^{8/3} 
    \left( \frac{\Lwind}{6\times 10^{35}\, {\rm erg/s}}\right)^{-5/3} \, .
\end{equation}
In the above, we have used $\Grad \xi = 1.81$ as calculated from Equation 3.10 of \citet{KooMcKee92b} for $\gamma_{\rm sw} = 5/3$, $\gamma_{\rm sa} = 1$ and $\krho = 2$. We have also chosen a value of $\Lwind$ appropriate for a single star of mass $\Mst = 30 \,\Msun$ (see Appendix A of \citet{Geen21}) and assumed a density scale for the mass profile that takes the number density of hydrogen atoms at a distance of $1\, {\rm pc}$ from the stars as $100\, {\rm cm}^{-3}$. From the above it is clear that the point at which the bubble transitions to being adiabatic (and would therefore better allow LyC radiation to escape) is long after all other timescales relevant to the problem (see \autoref{app:scales}).

The quantity $\Qrec(t_{\rm trans})$ is therefore physically uninteresting in this scenario\footnote{Note that, for $\krho = 2$, $\Qrec(t_{\rm trans}) \propto \rhozo^{-1/3}$ and that a density limit associated with this transition would then correspond to the density \textit{below which} the shell still traps all radiation at the transition time.}. Instead, let us calculate the radius at which the recombination rate is equal to $\Qo$. This will be the point at which the shell is no longer able to contain the radiation and it ``breaks out" in a ``champagne"-like flow \citep{TT79}. Note that, as discussed in \citet{Franco90}, since $\krho = 2 > 3/2$, the radiation should formally escape to infinity after this point. Using \autoref{eq:Qrec_power_law} we can calculate this breakout radius as
\begin{equation}
    \label{eq:Rbo_isothermal}
    R_{\rm bo} = \frac{4\pi \alpha_B \rhozo^2}{\Qo\left( \muH\mH\ci\right)^2} \left[\frac{\Grad\xi\Lwind}{3\rhozo} \right]^{2/3} = 4.89\, {\rm pc}
    \left(\frac{\Qo}{7\times 10^{48}\, {\rm s}^{-1}} \right)^{-1}
    \left( \frac{\Lwind}{6\times 10^{35}\, {\rm erg/s}}\right)^{2/3}
    \left( \frac{\rhozo/\muH\mH}{100\, \pcc {\rm pc}^2}\right)^{4/3} \, ,
\end{equation}
and the time at which is occurs as
\begin{equation}
    \label{eq:tbo}
    t_{\rm bo} = R_{\rm bo} 
    \left[ \frac{\Grad \xi\Lwind}{3\rhozo}\right]^{-1/3}
     = 9.7\times 10^4\, {\rm yr}
     \left(\frac{\Qo}{7\times 10^{48}\, {\rm s}^{-1}} \right)^{-1}
    \left( \frac{\Lwind}{6\times 10^{35}\, {\rm erg/s}}\right)^{1/3}
    \left( \frac{\rhozo/\muH\mH}{100\, \pcc {\rm pc}^2}\right)^{5/3} \, ,
\end{equation}
In the above we have made the same assumptions on $\ci$ and $\alpha_B$ as \autoref{eq:phot_trap_rhobar_app}, but we have chosen values for $\Qo$ and $\Lwind$ that are appropriate to an individual massive star of $\Mst = 30\, \Msun$ (see Appendix A of \citet{Geen21}). We should note that this breakout will likely occur more rapidly if the background field is not a perfect isothermal sphere but contains some turbulent structure, as is the case in \citet{Geen21}.

As we have ignored the details of the density structure of the shell, the estimate of \autoref{eq:Rbo_isothermal} is likely only correct within a factor of a few. To illustrate the range of systematic uncertainty we compare this estimate to a few comparable models. \citet{Geen22} provide a similar calculation to ours in their Equation 25. Instead of using the density at the forward-shock of the bubble they use the density that matches the pressure at the wind bubble interior $\rho_{\rm sh} = P_w/\ci^2$, this leads to identical scaling dependencies on all relevant factor but differs by a constant factor of $\sim 1.5$. A factor of 2 difference is inferred by taking the interior density as estimated using the pressure drop estimate across the shell from Equation 3.9 of \citet{KooMcKee92b}. With these uncertainties in mind, \autoref{eq:Rbo_isothermal} compares reasonably well to the results of \citet{Geen22} which calculates the breakout radius by manually integrating the structure of the WBB shell as the bubble evolves in an isothermal sphere background.

Taken together, \autoref{eq:phot_trap_rhobar_app} and \autoref{eq:Rbo_isothermal} explain why photoionized gas is initially trapped around young massive stars, but is not generally trapped when considering feedback from a cluster of stars in a relatively uniform density environment. This suggests an evolution of feedback bubbles from clusters of massive stars where each star's LyC radiation is initially trapped by the shell of its WBB before it breaks out. Each of these breakouts would eventually join together to ionize a large portion of the cloud as the individual WBBs also merge. This breakout and merging process would happen relatively quickly as indicated by \autoref{eq:tbo}.

\section{Relevant Scales}
\label{app:scales}

In this Appendix we give a detailed analysis of the time and spatial scales relevant to the interaction of a WBB and the PIR surrounding it. The reference feedback parameters for the scaling used below are meant to roughly trace those relevant for a $5\times10^3\,\Msun$ mass cluster in a $10^5\,\Msun$ mass cloud with a radius of roughly $20\pc$, at solar metallicity, similar to the simulations performed in \paperii. The feedback parameters ($\Qo$, $\Lwind$, $\pdot$) are representative of average values over the first $2\Myr$ of the star cluster's evolution.

The first timescale to introduce, generally the shortest, is the recombination timescale
\begin{equation}
    \label{eq:trec_def}
    \trec \equiv \left(\nHbar \alpha_B \right)^{-1} = 1.02 \times 10^3 \, {\rm yr} \left( \frac{\nHbar}{100\, {\rm cm}^{-3}}\right)^{-1}
    \left( \frac{\alpha_B}{3.11 \times 10^{-13}\, {\rm s}^{-1}\, {\rm cm}^3}\right)^{-1} \, .
\end{equation}
This is the timescale on which the \strom Sphere is established (the R-type evolution of the ionization front). We will use the above reference value for the case B recombination rate in all our calculations below, it is taken from the relation of \citet{Glover10} for $T= 8\times 10^3\, {\rm K}$ which is the relation used in the code used in \paperii\, and the average temperature of the ionized gas as measured in those simulations. It is over this timescale that the \strom Radius,
\begin{equation}
    \label{eq:Rst_quant}
    \RSt \equiv \left(\frac{3\Qo}{4\pi \nHbar^2 \alpha_B} \right)^{1/3} = 10.1 \pc 
    \left( \frac{\Qo}{4\times 10^{50}\, {\rm s}^{-1}}\right)^{1/3}
    \left(\frac{\nHbar}{100\pcc} \right)^{-2/3}\, ,
\end{equation}
is ionized. The characteristic timescale for the PIR to expand under its own thermal pressure, in the absence of other dynamical effects,  is 
\begin{equation}
    \label{eq:tdio_app}
    \tdio = \frac{\RSp(t=0)}{\dot{R}_{\rm Sp}(t=0)} 
    = \frac{\sqrt{3}}{2} \frac{\RSt}{\ci} 
    = 8.59 \times 10^5 \, {\rm yr} 
    \left( \frac{\Qo}{4\times10^{50} \, {\rm s}^{-1}} \right)^{1/3}
    \left(\frac{\nHbar}{100  \pcc} \right)^{-2/3}
    \left( \frac{\ci}{10 \kms} \right)^{-1} \, .
\end{equation}

The next relevant timescale is the shell-formation timescale of the WBB which we can get from \autoref{eq:shell_formation_power_law} by taking $\krho = 0$, $\eta = 3/5$, $\Grad = 1$, and $\xi = 0.9$ (justified in the previous appendix) and the simplified cooling function of \citet{KooMcKee92a,KooMcKee92b} ($\alpha = 0.5$). This gives us
\begin{equation}
    \label{eq:tsf}
    \tsf = \left[ \frac{27C_1}{125\rhobar}\right]^{5/11}
    \left[\frac{\xi\Lwind}{\rhobar} \right]^{3/11} = 
    5.30\times 10^2 \, {\rm yr} 
    \left(\frac{\Lwind}{10^{38} \, {\rm erg/s}} \right)^{2/11}
    \left( \frac{\nHbar}{100\, \pcc}\right)^{-8/11} 
    \left(\frac{Z}{Z_{\odot}}\right)^{-5/11}\, .
\end{equation}
Note we have included the scaling with metallicity here which comes from $C_1\propto \Lambda_1^{-1} \propto (Z/Z_{\odot})^{-1}$. This timescale is comparable both in absolute value and in scaling with density to $\trec$. A natural starting point for the co-evolution of the WBB and surrounding photoionized gas is one where the \strom Sphere has been ionized and the WBB's shell has formed. As we justify in \autoref{app:trapping}, for the parameters of interest to our co-evolution model, the shell does not significantly disturb the ionized gas at this point.

Since the initial thermal pressure in the PIR is finite, $\Phii(t=0) = \rhobar \ci^2$, and the pressure in the shocked-wind in either \autoref{eq:Pweaver} or \autoref{eq:Phot_EC} formally becomes infinite as $t \to 0$, there will be a time period during which the WBB expands, unimpeded, into the surrounding PIR. We can get a rough estimate of when this evolution ends by calculating when the force exerted by the shocked wind matches the force that would be exerted on the WBB by the PIR at the background density, $4\pi \Rw^2 \rhobar \ci^2$. In the case of a momentum driven bubble with the pressure given by \autoref{eq:Phot_EC}, this gives us force equality at a radius of
\begin{equation}
    \label{eq:ReqMD_app}
    \ReqMD \equiv \sqrt{\frac{\alpha_p \pdot}{4\pi \rhobar \ci^2}}
    = 4.74\, {\rm pc} \, \left(\frac{\alpha_p\pdot}{10^5 \, M_{\odot}\, {\rm km/s/Myr}} \right)^{1/2} 
    \left(\frac{\nHbar}{100\pcc} \right)^{-1/2} 
    \left( \frac{\ci}{10\, {\rm km/s}}\right)^{-1}\, ,
\end{equation}
and a time
\begin{equation}
    \label{eq:teqMD_app}
    \teqMD \equiv \frac{1}{6 \ci^2} \sqrt{\frac{3}{2\pi} \frac{\alpha_p \pdot}{\rhobar}}
    = \frac{\ReqMD}{\sqrt{6}\ci}
    = 1.89 \times 10^5 \, {\rm yr}\, 
    \left(\frac{\alpha_p\pdot}{10^5 \, M_{\odot}\, {\rm km/s/Myr}} \right)^{1/2} 
    \left(\frac{\nHbar}{100 \pcc} \right)^{-1/2} 
    \left( \frac{\ci}{10\, {\rm km/s}}\right)^{-2} \, ,
\end{equation}
One can show that the time it takes for the WBB expansion to become sub-sonic in the photo-ionized gast $\dot{R}_{\rm MD}(\tssMD) = \ci$ is $\tssMD = 3\teqMD/2$.

We can also write down the time it would take for the momentum-driven wind bubble to reach $\RSt$ if it were not affected by the PIR at all. One can show that setting \autoref{eq:rEC} equal to $\RSt$ gives us this `Wind \strom time'
\begin{equation}
    \label{eq:tSt}
    \tStMD \equiv \teqMD \left(\frac{\RSt}{\ReqMD} \right)^2
    = 8.67 \times 10^5 \, {\rm yr}\,
    \left(\frac{\alpha_p\pdot}{10^5 \, M_{\odot}\, {\rm km/s/Myr}} \right)^{-1/2} 
    \left(\frac{\nHbar}{100 \pcc} \right)^{-5/6} 
    \left( \frac{\Qo}{4\times10^{50}\, {\rm s}^{-1}}\right)^{2/3} \, .
\end{equation}
Both $\tssMD$ and $\tStMD$ will be relevant to consider in the strong wind regime ($\Req \gg \RSt$) in order to determine over what timescale the HII region can come to be dominated by the WBB.

We can of course derive similar quantities to the above for the case of an energy-driven WBB with the pressure given by \autoref{eq:Pweaver}. These are
\begin{equation}
    \label{eq:ReqED_app}
    \ReqED = \sqrt{\frac{\sqrt{7}\Lwind}{22\pi \rhobar \ci^3}} = 
    41.44\, {\rm pc} \left(\frac{\Lwind}{10^{38} \, {\rm erg/s}} \right)^{1/2} 
    \left(\frac{\nHbar}{100\pcc} \right)^{-1/2} 
    \left( \frac{\ci}{10\, {\rm km/s}}\right)^{-3/2}\, ,
\end{equation}
\begin{equation}
    \label{eq:teqED_app}
    \teqED = \frac{7^{3/4}}{5} \sqrt{\frac{\Lwind}{22\pi \rhobar \ci^5}}
    = \frac{\sqrt{7}}{5} \frac{\ReqED}{\ci}
    = 2.14 \, \times 10^6 \,{\rm yr} \left(\frac{\Lwind}{10^{38} \, {\rm erg/s}} \right)^{1/2} 
    \left(\frac{\nHbar}{100\pcc} \right)^{-1/2} 
    \left( \frac{\ci}{10\, {\rm km/s}}\right)^{-5/2}\, ,
\end{equation}
$\tssED = (3/\sqrt{7})^{5/2} \teqED$, and
\begin{equation}
    \tStED = \teqED \left(\frac{\RSt}{\ReqED} \right)^{5/3}
    = 2.05\times 10^5\,{\rm yr}  \left(\frac{\Lwind}{10^{38} \, {\rm erg/s}} \right)^{-1/3} 
    \left(\frac{\nHbar}{100\pcc} \right)^{-7/9} 
    \left( \frac{\Qo}{4\times 10^{50}\, {\rm s}^{-1}}\right)^{5/9} \, .
\end{equation}
It is clear from the above scalings that, as one would expect, the energy-driven bubble remains over-pressurized with respect to the photoionized gas for much longer (potentially the entire cloud evolution) and has the potential to over-take the \strom radius quite quickly.

In order to see if this evolution makes up a significant fraction of the overall feedback bubble's evolution we can compare $\teq$ and $\tSt$ to the free-fall time of a cloud, $\tff$, since clouds are generally thought to be fully disrupted on the order of a few free-fall times \citep{chevance20b,chevance20,Lancaster21c,JGK21,Menon24b}. The free-fall time is given as
\begin{equation}
    \label{eq:tffdef}
    \tff \equiv \sqrt{\frac{3\pi}{32 G \rhobar}} 
    = 4.35 \times 10^6 \, {\rm yr} \, 
    \left( \frac{\nHbar}{100  \pcc} \right)^{-1/2} \, .
\end{equation}
So we see that $\teq/\tff$ is independent of the mean density of the cloud, $\rhobar$, and only depends on the properties of the wind and the temperature of the PIR. This ratio takes on values of $\teq/\tff \approx 0.03,\, 0.49$ for the momentum and energy-driven solutions respectively at fiducial values used above. Furthermore, we see that, for these same fiducial values, $\teq/\tSt \sim 0.12,\, 22.5$ for momentum and energy driven solutions respectively.

From the above arguments we can conclude that there is a significant fraction of the bubble's evolution that occurs while both the WBB and the PIR are in pressure equilibrium. The timescale for this evolution to be reached is also long compared with the recombination timescale for the gas ($\trec/\teq \sim 2\times 10^{-2},\, 4\times 10^{-4}$, in MD and ED solutions respectively). It is then reasonable to assume that the PIR maintains ionization-recombination equilibrium during this evolution.

We can also compare $\teq$ to $\tdio$ (\autoref{eq:tdio_app}) as $\teq/ \tdio \ll 1$ would imply that the dynamical evolution of the PIR is unimportant before it comes in to pressure equilibrium with the WBB. From our calculations above we have these ratios as
\begin{equation}
    \frac{\teqMD}{\tdio} = 0.22 
    \left(\frac{\alpha_p\pdot}{10^5 \, M_{\odot}\, {\rm km/s/Myr}} \right)^{1/2} 
    \left( \frac{\Qo}{4\times 10^{50} \, {\rm s}^{-1}} \right)^{-1/3}
    \left(\frac{\nHbar}{100 \pcc} \right)^{1/6} 
    \left( \frac{\ci}{10\, {\rm km/s}}\right)^{-1} \, ,
\end{equation}
\begin{equation}
    \frac{\teqED}{\tdio} = 2.49 
    \left(\frac{\Lwind}{10^{38} \, {\rm erg/s}} \right)^{1/2} 
    \left( \frac{\Qo}{4\times10^{50} \, {\rm s}^{-1}} \right)^{-1/3}
    \left(\frac{\nHbar}{100 \pcc} \right)^{1/6} 
    \left( \frac{\ci}{10\, {\rm km/s}}\right)^{-1/2} \, .
\end{equation}
So we see that $\teqMD/ \tdio$ is small but not negligible, and $\teqED/\tdio$ is generally large. Considering the PIR as static until $\teq$ is then generally not valid.

The comparison of $\teq$ to $\tdio$ is also important for the validity of an equilibrium being quickly established between the WBB and the PIR. As the WBB rapidly expands to reach equilibrium with the PIR, it sweeps up mass that is photoionized and heated to the PIR's temperature, $\Thii$. This shell of PIR will be in pressure equilibrium with the WBB on its interior edge and overly-dense (and therefore overly pressurized) with respect to the rest of the PIR on its outer edge. The mass in this shell will be redistributed throughout the region roughly over the sound-crossing time, $\tdio$, but as we saw above, this is a few times larger than the equilibrium time, $\teq$. So a true equilibrium, with the PIR at a nearly constant density, the shell having re-distributed its mass, will only properly occur after a few sound crossing times (depending on the ratio of $\Req$ to $\RSt$). All of this to say, our model assumes that the equilibrium is set up relatively quickly, and this may not be the case, as we see in \paperii.

This leads us to the consider how the PIR and WBB should evolve both before and after they are in equilibrium with one another. In order to avoid coupling their evolutions beforehand we will assume that they follow their idealized solutions up until some time $\tswitch$. After this point the bubbles will evolve together in that the WBB will be impeded by the dynamical pressure of the PIR and the PIR will be made smaller by the compression (and enhanced recombination rate) due to the WBB. For $\Req<\RSt$ it makes sense for this transition to occur at $\teq$, when it is first reasonable to expect the bubbles to be in equilibrium with one another.

For $\Req > \RSt$, this choice is more complicated as the WBB may catch up with the photo-ionized gas $R_w(\tot) = \RSp(\tot)$ for some catch-up time, $\tot < \teq$. We can calculate this assuming $\Rw(t)$ is given by either $\RMDa(t)$ of $\RWeavert(t)$, as appropriate. We might also expect to treat the WBB as being in equilibrium with the PIR once its expansion has become subsonic in the PIR at $\tss$. To that end, for $\Req>\RSt$, we choose $\tswitch = {\rm min}\left( \tot,\, \tss\right)$ in both models below.

Finally, we write the characteristic radius, which is defined in \autoref{eq:Rch_def} as the radius at which the forces due to separately evolving wind and photoionized gas driven bubbles are equal, quantitatively as
\begin{equation}
    \label{eq:Rch_app}
    \Rch \equiv \frac{\alpha_B}{12\pi(\muH\mH \ci^2)^2}\frac{\alpha_p^2\pdot^2}{\Qo} = \frac{\Req^4}{\RSt^3} = 4.84\times 10^{-1}\pc 
    \left(\frac{\alpha_p\pdot}{10^5 \, M_{\odot}\, {\rm km/s/Myr}} \right)^{2} 
    \left( \frac{\Qo}{4\times10^{50} \, {\rm s}^{-1}} \right)^{-1}
    \left( \frac{\ci}{10\, {\rm km/s}}\right)^{-4} \, .
\end{equation}

\section{Dimensionless Equations for Momentum-Driven Co-Evolution Model}
\label{app:mdje}

In order to quickly explore the parameter space it is helpful to describe the evolution equations of this co-evolution model (CEM) in dimensionless form. This includes the Spitzer and momentum-driven wind bubbles and their momentum behavior as well as the  co-evolution phase equation,  \autoref{eq:HIImomentum_joint1}. To that end we can define
\begin{equation}
    \label{eq:dimensionless_def}
    \xi \equiv \frac{\mathcal{R}}{\ReqMD} \, \, , \, \, 
    \chi \equiv \frac{t}{\tdio} \, \, , \, \, 
    \pscl \equiv \frac{4\pi}{3} \rhobar \frac{\ReqMD^4}{\tdio}
\end{equation}
where $\tdio$ is the initial dynamical expansion time of the \strom Sphere, defined in \autoref{app:scales}, and $\pscl$ defines a reference radial momentum. In this way we have $\xi_i \equiv \Ri/\Req$ and $\xi_w \equiv \Rw/\Req$. We will additionally define the one free-parameter of the model which quantifies the relative strength of LyC radiation and winds. We choose this to be
\begin{equation}
    \label{eq:zetaMD_def_app}
    \ReqRStMD \equiv \frac{\ReqMD}{\RSt} 
    = 0.47 \, \left(\frac{\Qo}{4\times 10^{50} \, {\rm s}^{-1}} \right)^{-1/3}
    \left( \frac{\alpha_p \pdot}{10^5 \, M_{\odot}\, {\rm km/s/Myr}}\right)^{1/2} \left( \frac{\nHbar}{100 \pcc} \right)^{1/6}
    \left( \frac{\ci}{10 \, \kms}\right)^{-1}
    \, .
\end{equation}
In the simulations discussed in \paperii we have $\ReqRStMD = 0.70, \, 0.99$ for $\alpha_p = 3, \,6$ respectively.

This parameter is also related to the transition time between the early evolution and co-evolution for $\ReqRStMD < 1$, in dimensionless form:
\begin{equation}
    \chi_{\rm switch} = \chi_{\rm eq} \equiv \frac{\teq}{\tdio} = \frac{\sqrt{2}}{3}\frac{\Req}{\RSt} = \frac{\sqrt{2}}{3} \ReqRStMD \, .
\end{equation}

With these definitions the radial and momentum evolution of the wind bubble can be written in dimensionless form as 
\begin{equation}
    \label{eq:Rwind_dimensionless}
    \xi_{{\rm MD}} = \sqrt[4]{\frac{9}{2}} \ReqRStMD^{-1/2} \chi^{1/2}\, ,
\end{equation}
\begin{equation}
    \label{eq:pwind_dimensionless}
    \frac{\prMD}{\pscl} = \frac{9}{4}\ReqRStMD^{-2} \chi\, .
\end{equation}
We can also write the Spitzer solution, \autoref{eq:spitzer_sol}, in dimensionless form as
\begin{equation}
    \label{eq:dimensionless_spitz}
    \xi_{\rm Sp} = \ReqRStMD^{-1} \left(1 + \frac{7}{4} \chi\right)^{4/7} \, .
\end{equation}

For $\ReqRStMD > 1$ we take the switch time as the minimum of $\tot$ and $\tss$, as described in \autoref{app:scales}. In dimensionless form we may write $\chi_{\rm ss} = \tss/\tdio = 3\chi_{\rm eq}/2 = \ReqRStMD/\sqrt{2}$. The above relations allow us to write the dimensionless catch-up time $\chi_{\rm ot} = \tot/\tdio$ as
\begin{equation}
    \sqrt[4]{\frac{9}{2}} \ReqRStMD^{1/2} \chi_{\rm ot}^{1/2}= \left(1 + \frac{7}{4} \chi_{\rm ot}\right)^{4/7} \, .
\end{equation}
There is no analytic solution to this equation, so we solve it numerically. There is a solution in the limit that $\chi_{\rm ot} \gg 1$, but that is the opposite limit in which we expect the equation to be useful (when WBBs are strong and therefore $\tot < \tdio$).

The dimensionless version of the Spitzer momentum can also be written in either the form of \autoref{eq:pr_spitzer1} which becomes
\begin{equation}
    \label{eq:pr_spitz_dimensionless}
    \frac{p_{r,{\rm Sp}}}{\pscl} = \ReqRStMD^{-4} \left(1 +\frac{7}{4}\chi \right)^{9/7}
\end{equation}
or the form of \autoref{eq:pr_spitzer_adj} (which correctly approaches zero momentum at $t=0$) which becomes
\begin{equation}
    \label{eq:pr_spitz_adj_dimensionless}
    \frac{p_{r,{\rm Sp,adj}}}{\pscl} = \ReqRStMD^{-4} \left(1 +\frac{7}{4}\chi \right)^{9/7} 
    \left[1 - \left(1 + \frac{7}{4} \chi \right)^{-6/7}\right] \, .
\end{equation}
The dependence on $\ReqRStMD$ in all of the Spitzer-related equations so far is simply an artifact of our choice of reference variables ($\Rch$ and $\tdio$).

We can then also write the co-evolution equation, \autoref{eq:HIImomentum_joint1}, as 
\begin{equation}
    \label{eq:dimensionless_ode}
    \frac{d}{d\chi} \left(\xi_i^3 \frac{d\xi_i}{d\chi} \right)
    = \frac{9}{4} \ReqRStMD^{-2} \left(1 + \ReqRStMD^{-3}\xi_w \right)^{2/3} \, .
\end{equation}

For $\ReqRStMD < 1$ the condition on the wind radius at the beginning of the co-evolution phase is then $\xi_w = 1$, which gives us $\xi_{i} = \xi_w\left(1 + \ReqRStMD^{-3}\xi_w\right)^{1/3}$. As we noted above, this will cause a discontinuity in $\Ri$ as the $\xi$ condition given here is not required to match the early phase evolution of the ionized gas radius, which follows the Spitzer solution.

For $\ReqRStMD < 1$, the initial condition on $d\Ri/dt$ can be written using \autoref{eq:pwind_dimensionless} and \autoref{eq:pr_spitz_adj_dimensionless} as
\begin{equation}
    \label{eq:psi0_ic}
    \frac{d\xi_i}{d\chi} = \xi_i^{-3} \left[
    \frac{3}{2\sqrt{2}}\ReqRStMD^{-1}
    + \ReqRStMD^{-4} \left(1 + \frac{7\sqrt{2}}{12}\ReqRStMD \right)^{9/7}
    \left( 1 -\left(1 + \frac{7\sqrt{2}}{12}\ReqRStMD \right)^{-6/7}\right)\right] \, .
\end{equation}

In the $\ReqRStMD>1$ case all values are set similarly to above $\xi_w = \xi_w(\chi_{\rm switch})$, $\xi_i$ is set using this along with \autoref{eq:RiRw_rel} and $d\xi_i/d\chi$ is set based on conservation of momentum.

Once we have a solution to the above system there remains the question, to some extent of its exact interpretation. In particular, how we should determine the total momentum carried by the bubble. In keeping with our choice of initial condition above, the most natural choice for the momentum from our solution is
\begin{equation}
    \label{eq:printerp1}
    p_{r,1} = \Msh \frac{d\Ri}{dt} 
    = \pscl \xi_i^3 \dot{\xi}_i
\end{equation}
where we have assumed a solution for $\xi(\chi)$ is given.

\section{Dimensionless Equations for Energy-Driven Co-Evolution Model}
\label{app:edje}

As in \autoref{app:mdje} we here derive a dimensionless form for the dynamical system described in \autoref{subsec:ed_jfb}. We begin with a full accounting of the evolution equations. These are the WBB energy equation
\begin{equation}
    \frac{dE_w}{dt} = (1-\theta)\Lwind - \Phot \frac{dV_w}{dt}
\end{equation}
with $\theta$ a constant representing losses due to interface mixing (as noted in the main text $1-\theta \ll 1$ for realistic systems) and relationships between $E_w$, $\Phot$ and $V_w$ given by
\begin{equation}
    E_w = \frac{3}{2} \Phot V_w \,\, , \,\,
    V_w = \frac{4\pi}{3} \Rw^3 \, ,
\end{equation}
as in Section III of \citet{Weaver77}. The momentum equation is then written as
\begin{equation}
    \frac{d}{dt} \left(\frac{4\pi}{3}\rhobar \Ri^3 \frac{d\Ri}{dt} \right) = 
    4\pi \Ri^2 \Phot \, .
\end{equation}
and the ionization-recombination equilibrium condition can be written as 
\begin{equation}
    \frac{\rho_i}{\rhobar} = \frac{\Phot}{\rhobar\ci^2} = \left( \frac{\RSt^3}{\Ri^3 - \Rw^3} \right)^{1/2}\, ,
\end{equation}
where the first equality comes from pressure equilibrium across the WBB surface $\Phot = \rho_i\ci^2$.

We then make the following definitions of dimensionless parameters:
\begin{equation}
    \xi \equiv \frac{R}{\ReqED} \,\, , \,\, 
    \chi \equiv\frac{t}{\tdio} \,\, , \,\, 
    \pscl \equiv \frac{4\pi}{3} \frac{\ReqED^4}{\tdio} \,\, , \,\, 
    \mathcal{M} = \frac{v}{\ci} \,\, , \,\, 
    \tilde{E} = \frac{E_w}{(1-\theta)\Lwind \tdio} \,\, , \,\, 
    \tilde{P} = \frac{\Phot}{\rhobar \ci^2}
\end{equation}
and the one free-parameter of the problem is
\begin{equation}
    \label{eq:zetaED_def_app}
    \ReqRStED \equiv \frac{\ReqED}{\RSt} = 4.08 \,
    \left(\frac{\Lwind(1-\theta)}{10^{38}\, {\rm erg}\, {\rm s}^{-1}} \right)^{1/2}
    \left(\frac{\nHbar}{100\pcc} \right)^{1/6}
    \left(\frac{\Qo}{4\times 10^{50}\, {\rm s}^{-1}} \right)^{-1/3}
    \left(\frac{\ci}{10\, \kms} \right)^{-3/2} \, .
\end{equation}

We first would like to write the classical solutions for the radius and momentum evolution of each bubble in terms of these new dimensionless parameters. The first quantity to consider is the dimensionless equilibration time
\begin{equation}
    \chi_{\rm eq} = \frac{\teqED}{\tdio} = \frac{\sqrt{7}}{5} \frac{\ReqED}{\ci} \frac{2}{\sqrt{3}} \frac{\ci}{\RSt} = \frac{2}{5} \sqrt{\frac{7}{3}} \ReqRStED \, ,
\end{equation}
which is also the time at which the co-evolution phase will begin when $\ReqRStED < 1$.

The ionized gas evolution follows exactly the same form as that given in \autoref{eq:dimensionless_spitz}, \autoref{eq:pr_spitz_dimensionless}, and \autoref{eq:pr_spitz_adj_dimensionless}. The energy-driven wind bubble radial evolution takes the form
\begin{equation}
    \label{eq:dimensionless_wind_ed}
    \xi_w = \left(\frac{5}{2} \sqrt{\frac{3}{7}} \right)^{3/5} \ReqRStED^{-3/5} \chi^{3/5} \, ,
\end{equation}
and the WBB momentum takes the dimensionless from
\begin{equation}
    \label{eq:dimensionless_prw_ed}
    \frac{\prWeaver}{\pscl} = \frac{3}{5} \left(\frac{5}{2}\sqrt{\frac{3}{7}} \right)^{12/5} \ReqRStED^{-12/5} \chi^{7/5} \, .
\end{equation}

With the above (and \autoref{eq:dimensionless_spitz}) we can write the catch-up time in dimensionless form as
\begin{equation}
    \left(\frac{5}{2} \sqrt{\frac{3}{7}} \right)^{3/5} \ReqRStED^{2/5} \chi_{\rm ot}^{3/5} = \left(1 + \frac{7}{4} \chi_{\rm ot}\right)^{4/7} \, .
\end{equation}
This, again, has no analytic solution, so we solve it numerically. The time at which the expansion becomes subsonic in the PIR, the other potential switching time for $\ReqRStED>1$, can be written in dimensionless form as $\chi_{\rm ss} = (3/\sqrt{7})^{5/2}\chi_{\rm eq} =18 \ReqRStED/5\cdot7^{3/4}$.

We can then re-write the evolution equations in dimensionless form as: pressure and energy equations
\begin{equation}
    \tilde{E} = \frac{2}{11}\sqrt{\frac{7}{3}} \ReqRStED \xi_w^3 \tilde{P}\, ,
\end{equation}
\begin{equation}
    \frac{d\ln \tilde{E}}{d\chi} = \tilde{E}^{-1} - 2 \frac{d\ln \xi_w}{d\chi}
\end{equation}
the momentum equation
\begin{equation}
    \frac{d\mathcal{M}_i}{d\chi} = \frac{3\sqrt{3}}{2\ReqRStED \xi_i} \left(\tilde{P} - \mathcal{M}_i^2 \right)
\end{equation}
with the mach number of the ionization front expansion defined as
\begin{equation}
    \mathcal{M}_i = \frac{1}{\ci}\frac{d \Ri}{dt}  = \frac{2\ReqRStED}{\sqrt{3}} \frac{d\xi_i}{d\chi}\, ,
\end{equation}
which serves as a dynamical evolution equation for $\xi_i$. We finally need a dynamical equation for the evolution of $\xi_w$ or $\Rw$ as the derivative is used in the energy evolution equation above. This is obtained by restating the ionization-recombination equilibrium equation as
\begin{equation}
    \xi_w^3 = \xi_i^3 - \tilde{P}^{-2} \ReqRStED^{-3}\, .
\end{equation}
We can then take a derivative of this with respect to $\chi$ to obtain
\begin{equation}
    \frac{d\ln \xi_w}{d\chi} = \frac{\left(\frac{\xi_i}{\xi_w} \right)^3\frac{d\ln \xi_i}{d\chi} + \frac{A}{\tilde{E}}}{1 + 5A}
\end{equation}
where 
\begin{equation}
    A \equiv \frac{2}{3\xi_w^3 \tilde{P}^2\ReqRStED^3} \, .
\end{equation}

For $\ReqRStED < 1$ this phase of evolution begins at $t=\teqED$. In this case the initial conditions for the state variables should then be $\xi_w = 1$, $\tilde{P} = \PWeaver(\teqED)/\rhobar\ci^2 = 1$ by definition, $\tilde{E}$ is given by the above relations with $\xi_w = \tilde{P}= 1$, $\xi_i$ is given by ionization-recombination equilibrium and the velocity of the ionization front or $\mathcal{M}_i$ is given by momentum conservation as in \autoref{subsec:ed_jfb}. In dimensionless form this results in the initial condition on the derivative of $\xi_i$ as
\begin{equation}
     \frac{d\xi_i}{d\chi} =  \xi_i^{-3}\left[\frac{3}{2} \sqrt{\frac{3}{7}} \ReqRStED^{-1}
    + \ReqRStED^{-4} \left(1 +\frac{7}{10}\sqrt{\frac{7}{3}} \ReqRStED \right)^{9/7} 
    \left[1 - \left(1 + \frac{7}{10}\sqrt{\frac{7}{3}} \ReqRStED \right)^{-6/7}\right] \right] \, .
\end{equation}

As in \autoref{app:mdje}, for $\ReqRStED> 1$, the above parameters are set based on the idealized values at $\chi_{\rm switch}$. For example, the additional parameter of $\tilde{P}$ is set as $\PWeaver(t_{\rm switch})/\rhobar \ci^2$. Just as in the momentum-driven case, with these choices we have chosen the ionization front radius, $\Ri$, to be discontinuous across the transition from the early phase to the co-evolution phase.

%

\bibliography{bibliography}{}

\begin{thebibliography}{}
\expandafter\ifx\csname natexlab\endcsname\relax\def\natexlab#1{#1}\fi
\providecommand{\url}[1]{\href{#1}{#1}}
\providecommand{\dodoi}[1]{doi:~\href{http://doi.org/#1}{\nolinkurl{#1}}}
\providecommand{\doeprint}[1]{\href{http://ascl.net/#1}{\nolinkurl{http://ascl.net/#1}}}
\providecommand{\doarXiv}[1]{\href{https://arxiv.org/abs/#1}{\nolinkurl{https://arxiv.org/abs/#1}}}

\bibitem[{{Abel} {et~al.}(2005){Abel}, {Ferland}, {Shaw}, \& {van
  Hoof}}]{Abel05}
{Abel}, N.~P., {Ferland}, G.~J., {Shaw}, G., \& {van Hoof}, P.~A.~M. 2005,
  \apjs, 161, 65, \dodoi{10.1086/432913}

\bibitem[{{Adamo} {et~al.}(2024){Adamo}, {Bradley}, {Vanzella}, {Claeyssens},
  {Welch}, {Diego}, {Mahler}, {Oguri}, {Sharon}, {Abdurro'uf}, {Hsiao}, {Xu},
  {Messa}, {Lassen}, {Zackrisson}, {Brammer}, {Coe}, {Kokorev}, {Ricotti},
  {Zitrin}, {Fujimoto}, {Inoue}, {Resseguier}, {Rigby}, {Jim{\'e}nez-Teja},
  {Windhorst}, {Hashimoto}, \& {Tamura}}]{Adamo24}
{Adamo}, A., {Bradley}, L.~D., {Vanzella}, E., {et~al.} 2024, arXiv e-prints,
  arXiv:2401.03224, \dodoi{10.48550/arXiv.2401.03224}

\bibitem[{{Badjin} {et~al.}(2016){Badjin}, {Glazyrin}, {Manukovskiy}, \&
  {Blinnikov}}]{Badjin16}
{Badjin}, D.~A., {Glazyrin}, S.~I., {Manukovskiy}, K.~V., \& {Blinnikov}, S.~I.
  2016, \mnras, 459, 2188, \dodoi{10.1093/mnras/stw790}

\bibitem[{{Belfiore} {et~al.}(2022){Belfiore}, {Santoro}, {Groves},
  {Schinnerer}, {Kreckel}, {Glover}, {Klessen}, {Emsellem}, {Blanc}, {Congiu},
  {Barnes}, {Boquien}, {Chevance}, {Dale}, {Kruijssen}, {Leroy}, {Pan},
  {Pessa}, {Schruba}, \& {Williams}}]{Belfiore22}
{Belfiore}, F., {Santoro}, F., {Groves}, B., {et~al.} 2022, \aap, 659, A26,
  \dodoi{10.1051/0004-6361/202141859}

\bibitem[{{Binder} \& {Povich}(2018)}]{Binder18}
{Binder}, B.~A., \& {Povich}, M.~S. 2018, \apj, 864, 136,
  \dodoi{10.3847/1538-4357/aad7b2}

\bibitem[{{Capriotti} \& {Kozminski}(2001)}]{CapriottiKozminski01}
{Capriotti}, E.~R., \& {Kozminski}, J.~F. 2001, \pasp, 113, 677,
  \dodoi{10.1086/320809}

\bibitem[{{Castor} {et~al.}(1975){Castor}, {Abbott}, \& {Klein}}]{CAK75}
{Castor}, J.~I., {Abbott}, D.~C., \& {Klein}, R.~I. 1975, \apj, 195, 157,
  \dodoi{10.1086/153315}

\bibitem[{{Chevance} {et~al.}(2023){Chevance}, {Krumholz}, {McLeod},
  {Ostriker}, {Rosolowsky}, \& {Sternberg}}]{Chevance23}
{Chevance}, M., {Krumholz}, M.~R., {McLeod}, A.~F., {et~al.} 2023, in
  Astronomical Society of the Pacific Conference Series, Vol. 534, Protostars
  and Planets VII, ed. S.~{Inutsuka}, Y.~{Aikawa}, T.~{Muto}, K.~{Tomida}, \&
  M.~{Tamura}, 1, \dodoi{10.48550/arXiv.2203.09570}

\bibitem[{{Chevance} {et~al.}(2020){Chevance}, {Kruijssen}, {Hygate},
  {Schruba}, {Longmore}, {Groves}, {Henshaw}, {Herrera}, {Hughes}, {Jeffreson},
  {Lang}, {Leroy}, {Meidt}, {Pety}, {Razza}, {Rosolowsky}, {Schinnerer},
  {Bigiel}, {Blanc}, {Emsellem}, {Faesi}, {Glover}, {Haydon}, {Ho}, {Kreckel},
  {Lee}, {Liu}, {Querejeta}, {Saito}, {Sun}, {Usero}, \& {Utomo}}]{chevance20b}
{Chevance}, M., {Kruijssen}, J.~M.~D., {Hygate}, A. P.~S., {et~al.} 2020,
  \mnras, 493, 2872, \dodoi{10.1093/mnras/stz3525}

\bibitem[{{Chevance} {et~al.}(2022){Chevance}, {Kruijssen}, {Krumholz},
  {Groves}, {Keller}, {Hughes}, {Glover}, {Henshaw}, {Herrera}, {Kim}, {Leroy},
  {Pety}, {Razza}, {Rosolowsky}, {Schinnerer}, {Schruba}, {Barnes}, {Bigiel},
  {Blanc}, {Dale}, {Emsellem}, {Faesi}, {Grasha}, {Klessen}, {Kreckel}, {Liu},
  {Longmore}, {Meidt}, {Querejeta}, {Saito}, {Sun}, \& {Usero}}]{chevance20}
{Chevance}, M., {Kruijssen}, J.~M.~D., {Krumholz}, M.~R., {et~al.} 2022,
  \mnras, 509, 272, \dodoi{10.1093/mnras/stab2938}

\bibitem[{{Dale}(2015)}]{Dale15Review}
{Dale}, J.~E. 2015, \nar, 68, 1, \dodoi{10.1016/j.newar.2015.06.001}

\bibitem[{{Dale} {et~al.}(2014){Dale}, {Ngoumou}, {Ercolano}, \&
  {Bonnell}}]{Dale14}
{Dale}, J.~E., {Ngoumou}, J., {Ercolano}, B., \& {Bonnell}, I.~A. 2014, \mnras,
  442, 694, \dodoi{10.1093/mnras/stu816}

\bibitem[{{Dav{\'e}} {et~al.}(2012){Dav{\'e}}, {Finlator}, \&
  {Oppenheimer}}]{Dave12}
{Dav{\'e}}, R., {Finlator}, K., \& {Oppenheimer}, B.~D. 2012, \mnras, 421, 98,
  \dodoi{10.1111/j.1365-2966.2011.20148.x}

\bibitem[{{Draine}(2011{\natexlab{a}})}]{Drainebook}
{Draine}, B.~T. 2011{\natexlab{a}}, {Physics of the Interstellar and
  Intergalactic Medium}

\bibitem[{{Draine}(2011{\natexlab{b}})}]{Draine11}
---. 2011{\natexlab{b}}, \apj, 732, 100, \dodoi{10.1088/0004-637X/732/2/100}

\bibitem[{{Drory} {et~al.}(2024){Drory}, {Blanc}, {Kreckel}, {S{\'a}nchez},
  {Mej{\'\i}a-Narv{\'a}ez}, {Johnston}, {Jones}, {Pellegrini}, {Konidaris},
  {Herbst}, {S{\'a}nchez-Gallego}, {Kollmeier}, {de Almeida},
  {Barrera-Ballesteros}, {Bizyaev}, {Brownstein}, {i Saguer}, {Cherinka},
  {Cioni}, {Congiu}, {Cosens}, {Dias}, {Donor}, {Egorov}, {Egorova}, {Froning},
  {Garc{\'\i}a}, {Glover}, {Greve}, {H{\"a}berle}, {Hoy}, {Ibarra}, {Li},
  {Klessen}, {Krishnarao}, {Kumari}, {Long}, {M{\'e}ndez-Delgado}, {Popa},
  {Ramirez}, {Rix}, {S{\'a}nchez}, {Sankrit}, {Sattler}, {Sayres}, {Singh},
  {Stringfellow}, {Wachter}, {Watkins}, {Wong}, \& {Wofford}}]{LVM_Science}
{Drory}, N., {Blanc}, G.~A., {Kreckel}, K., {et~al.} 2024, \aj, 168, 198,
  \dodoi{10.3847/1538-3881/ad6de9}

\bibitem[{{El-Badry} {et~al.}(2019){El-Badry}, {Ostriker}, {Kim}, {Quataert},
  \& {Weisz}}]{ElBadry19}
{El-Badry}, K., {Ostriker}, E.~C., {Kim}, C.-G., {Quataert}, E., \& {Weisz},
  D.~R. 2019, \mnras, 490, 1961, \dodoi{10.1093/mnras/stz2773}

\bibitem[{{Emig} {et~al.}(2020){Emig}, {Bolatto}, {Leroy}, {Mills},
  {Jim{\'e}nez Donaire}, {Tielens}, {Ginsburg}, {Gorski}, {Krieger}, {Levy},
  {Meier}, {Ott}, {Rosolowsky}, {Thompson}, \& {Veilleux}}]{Emig20}
{Emig}, K.~L., {Bolatto}, A.~D., {Leroy}, A.~K., {et~al.} 2020, \apj, 903, 50,
  \dodoi{10.3847/1538-4357/abb67d}

\bibitem[{{Faucher-Gigu{\`e}re} {et~al.}(2013){Faucher-Gigu{\`e}re},
  {Quataert}, \& {Hopkins}}]{CAFG13}
{Faucher-Gigu{\`e}re}, C.-A., {Quataert}, E., \& {Hopkins}, P.~F. 2013, \mnras,
  433, 1970, \dodoi{10.1093/mnras/stt866}

\bibitem[{{Fielding} {et~al.}(2018){Fielding}, {Quataert}, \&
  {Martizzi}}]{Fielding18}
{Fielding}, D., {Quataert}, E., \& {Martizzi}, D. 2018, \mnras, 481, 3325,
  \dodoi{10.1093/mnras/sty2466}

\bibitem[{{Folini} \& {Walder}(2006)}]{FoliniWalder06}
{Folini}, D., \& {Walder}, R. 2006, \aap, 459, 1,
  \dodoi{10.1051/0004-6361:20053898}

\bibitem[{{Franco} {et~al.}(1990){Franco}, {Tenorio-Tagle}, \&
  {Bodenheimer}}]{Franco90}
{Franco}, J., {Tenorio-Tagle}, G., \& {Bodenheimer}, P. 1990, \apj, 349, 126,
  \dodoi{10.1086/168300}

\bibitem[{{Garcia-Segura} {et~al.}(1996{\natexlab{a}}){Garcia-Segura},
  {Langer}, \& {Mac Low}}]{GS96b}
{Garcia-Segura}, G., {Langer}, N., \& {Mac Low}, M.~M. 1996{\natexlab{a}},
  \aap, 316, 133

\bibitem[{{Garcia-Segura} {et~al.}(1996{\natexlab{b}}){Garcia-Segura}, {Mac
  Low}, \& {Langer}}]{GS96a}
{Garcia-Segura}, G., {Mac Low}, M.~M., \& {Langer}, N. 1996{\natexlab{b}},
  \aap, 305, 229

\bibitem[{{Geen} {et~al.}(2023){Geen}, {Bieri}, {de Koter}, {Kimm}, \&
  {Rosdahl}}]{Geen23}
{Geen}, S., {Bieri}, R., {de Koter}, A., {Kimm}, T., \& {Rosdahl}, J. 2023,
  \mnras, 526, 1832, \dodoi{10.1093/mnras/stad2667}

\bibitem[{{Geen} {et~al.}(2021){Geen}, {Bieri}, {Rosdahl}, \& {de
  Koter}}]{Geen21}
{Geen}, S., {Bieri}, R., {Rosdahl}, J., \& {de Koter}, A. 2021, \mnras, 501,
  1352, \dodoi{10.1093/mnras/staa3705}

\bibitem[{{Geen} \& {de Koter}(2022)}]{Geen22}
{Geen}, S., \& {de Koter}, A. 2022, \mnras, 509, 4498,
  \dodoi{10.1093/mnras/stab3245}

\bibitem[{{Geen} {et~al.}(2020){Geen}, {Pellegrini}, {Bieri}, \&
  {Klessen}}]{Geen20}
{Geen}, S., {Pellegrini}, E., {Bieri}, R., \& {Klessen}, R. 2020, \mnras, 492,
  915, \dodoi{10.1093/mnras/stz3491}

\bibitem[{{Glover} {et~al.}(2010){Glover}, {Federrath}, {Mac Low}, \&
  {Klessen}}]{Glover10}
{Glover}, S.~C.~O., {Federrath}, C., {Mac Low}, M.~M., \& {Klessen}, R.~S.
  2010, \mnras, 404, 2, \dodoi{10.1111/j.1365-2966.2009.15718.x}

\bibitem[{{Grudi{\'c}} {et~al.}(2021){Grudi{\'c}}, {Guszejnov}, {Hopkins},
  {Offner}, \& {Faucher-Gigu{\`e}re}}]{STARFORGE21}
{Grudi{\'c}}, M.~Y., {Guszejnov}, D., {Hopkins}, P.~F., {Offner}, S. S.~R., \&
  {Faucher-Gigu{\`e}re}, C.-A. 2021, \mnras, 506, 2199,
  \dodoi{10.1093/mnras/stab1347}

\bibitem[{{Grudi{\'c}} {et~al.}(2018){Grudi{\'c}}, {Hopkins},
  {Faucher-Gigu{\`e}re}, {Quataert}, {Murray}, \& {Kere{\v{s}}}}]{Grudic18}
{Grudi{\'c}}, M.~Y., {Hopkins}, P.~F., {Faucher-Gigu{\`e}re}, C.-A., {et~al.}
  2018, \mnras, 475, 3511, \dodoi{10.1093/mnras/sty035}

\bibitem[{{Haid} {et~al.}(2018){Haid}, {Walch}, {Seifried}, {W{\"u}nsch},
  {Dinnbier}, \& {Naab}}]{Haid18}
{Haid}, S., {Walch}, S., {Seifried}, D., {et~al.} 2018, \mnras, 478, 4799,
  \dodoi{10.1093/mnras/sty1315}

\bibitem[{{Harper-Clark} \& {Murray}(2009)}]{HCM09}
{Harper-Clark}, E., \& {Murray}, N. 2009, \apj, 693, 1696,
  \dodoi{10.1088/0004-637X/693/2/1696}

\bibitem[{{Harris} {et~al.}(2020){Harris}, {Jarrod Millman}, {van der Walt},
  {Gommers}, {Virtanen}, {Cournapeau}, {Wieser}, {Taylor}, {Berg}, {Smith},
  {Kern}, {Picus}, {Hoyer}, {van Kerkwijk}, {Brett}, {Haldane}, {Fern{\'a}ndez
  del R{\'\i}o}, {Wiebe}, {Peterson}, {G{\'e}rard-Marchant}, {Sheppard},
  {Reddy}, {Weckesser}, {Abbasi}, {Gohlke}, \& {Oliphant}}]{harrisNumpy2020}
{Harris}, C.~R., {Jarrod Millman}, K., {van der Walt}, S.~J., {et~al.} 2020,
  arXiv e-prints, arXiv:2006.10256.
\newblock \doarXiv{2006.10256}

\bibitem[{{Hosokawa} \& {Inutsuka}(2006)}]{HosokawaInutsuka06}
{Hosokawa}, T., \& {Inutsuka}, S.-i. 2006, \apj, 646, 240,
  \dodoi{10.1086/504789}

\bibitem[{{Hunter}(2007)}]{matplotlib_hunter07}
{Hunter}, J.~D. 2007, Computing in Science and Engineering, 9, 90,
  \dodoi{10.1109/MCSE.2007.55}

\bibitem[{{Kapoor} {et~al.}(2023){Kapoor}, {Baes}, {van der Wel}, {Gebek},
  {Camps}, {Nersesian}, {Meidt}, {Smith}, {Vicens}, {D'Eugenio}, {Martorano},
  {Barrientos}, \& {Sartorio}}]{Toddlers23}
{Kapoor}, A.~U., {Baes}, M., {van der Wel}, A., {et~al.} 2023, \mnras, 526,
  3871, \dodoi{10.1093/mnras/stad2977}

\bibitem[{{Kim} {et~al.}(2020){Kim}, {Ostriker}, {Fielding}, {Smith}, {Bryan},
  {Somerville}, {Forbes}, {Genel}, \& {Hernquist}}]{CGK2020}
{Kim}, C.-G., {Ostriker}, E.~C., {Fielding}, D.~B., {et~al.} 2020, \apjl, 903,
  L34, \dodoi{10.3847/2041-8213/abc252}

\bibitem[{{Kim} {et~al.}(2016){Kim}, {Kim}, \& {Ostriker}}]{JGK16}
{Kim}, J.-G., {Kim}, W.-T., \& {Ostriker}, E.~C. 2016, \apj, 819, 137,
  \dodoi{10.3847/0004-637X/819/2/137}

\bibitem[{{Kim} {et~al.}(2018){Kim}, {Kim}, \& {Ostriker}}]{JGK18}
---. 2018, \apj, 859, 68, \dodoi{10.3847/1538-4357/aabe27}

\bibitem[{{Kim} {et~al.}(2021){Kim}, {Ostriker}, \& {Filippova}}]{JGK21}
{Kim}, J.-G., {Ostriker}, E.~C., \& {Filippova}, N. 2021, \apj, 911, 128,
  \dodoi{10.3847/1538-4357/abe934}

\bibitem[{{Koo} \& {McKee}(1992{\natexlab{a}})}]{KooMcKee92a}
{Koo}, B.-C., \& {McKee}, C.~F. 1992{\natexlab{a}}, \apj, 388, 93,
  \dodoi{10.1086/171132}

\bibitem[{{Koo} \& {McKee}(1992{\natexlab{b}})}]{KooMcKee92b}
---. 1992{\natexlab{b}}, \apj, 388, 103, \dodoi{10.1086/171133}

\bibitem[{{Kreckel} {et~al.}(2024){Kreckel}, {Egorov}, {Egorova}, {Blanc},
  {Drory}, {Kounkel}, {M{\'e}ndez-Delgado}, {Rom{\'a}n-Z{\'u}{\~n}iga},
  {S{\'a}nchez}, {Stringfellow}, {Stutz}, {Zari}, {Barrera-Ballesteros},
  {Bizyaev}, {Brownstein}, {Congiu}, {Fern{\'a}ndez-Trincado}, {Garc{\'\i}a},
  {Hillenbrand}, {Ibarra-Medel}, {Jin}, {Johnston}, {Jones}, {Kim},
  {Kollmeier}, {Kong}, {Krishnarao}, {Kumari}, {Li}, {Long},
  {Mata-S{\'a}nchez}, {Mej{\'\i}a-Narv{\'a}ez}, {Popa}, {Rix}, {Sattler},
  {Serna}, {Singh}, {S{\'a}nchez-Gallego}, {Wofford}, \&
  {Wong}}]{Kreckel24_Orion}
{Kreckel}, K., {Egorov}, O.~V., {Egorova}, E., {et~al.} 2024, \aap, 689, A352,
  \dodoi{10.1051/0004-6361/202449943}

\bibitem[{{Kruijssen} {et~al.}(2019){Kruijssen}, {Schruba}, {Chevance},
  {Longmore}, {Hygate}, {Haydon}, {McLeod}, {Dalcanton}, {Tacconi}, \& {van
  Dishoeck}}]{Kruijssen19}
{Kruijssen}, J.~M.~D., {Schruba}, A., {Chevance}, M., {et~al.} 2019, \nat, 569,
  519, \dodoi{10.1038/s41586-019-1194-3}

\bibitem[{{Krumholz} {et~al.}(2018){Krumholz}, {Burkhart}, {Forbes}, \&
  {Crocker}}]{KrumholzSFR18}
{Krumholz}, M.~R., {Burkhart}, B., {Forbes}, J.~C., \& {Crocker}, R.~M. 2018,
  \mnras, 477, 2716, \dodoi{10.1093/mnras/sty852}

\bibitem[{{Krumholz} \& {Matzner}(2009)}]{KrumholzMatzner09}
{Krumholz}, M.~R., \& {Matzner}, C.~D. 2009, \apj, 703, 1352,
  \dodoi{10.1088/0004-637X/703/2/1352}

\bibitem[{{Krumholz} {et~al.}(2019){Krumholz}, {McKee}, \& {Bland
  -Hawthorn}}]{KMBBH19}
{Krumholz}, M.~R., {McKee}, C.~F., \& {Bland -Hawthorn}, J. 2019, \araa, 57,
  227, \dodoi{10.1146/annurev-astro-091918-104430}

\bibitem[{{Krumholz} \& {Tan}(2007)}]{KrumholzTan07}
{Krumholz}, M.~R., \& {Tan}, J.~C. 2007, \apj, 654, 304, \dodoi{10.1086/509101}

\bibitem[{{Krumholz} {et~al.}(2014){Krumholz}, {Bate}, {Arce}, {Dale},
  {Gutermuth}, {Klein}, {Li}, {Nakamura}, \& {Zhang}}]{Krumholz14}
{Krumholz}, M.~R., {Bate}, M.~R., {Arce}, H.~G., {et~al.} 2014, in Protostars
  and Planets VI, ed. H.~{Beuther}, R.~S. {Klessen}, C.~P. {Dullemond}, \&
  T.~{Henning}, 243--266, \dodoi{10.2458/azu_uapress_9780816531240-ch011}

\bibitem[{{Lancaster} {et~al.}(2024){Lancaster}, {Ostriker}, {Kim}, {Kim}, \&
  {Bryan}}]{Lancaster24a}
{Lancaster}, L., {Ostriker}, E.~C., {Kim}, C.-G., {Kim}, J.-G., \& {Bryan},
  G.~L. 2024, arXiv e-prints, arXiv:2405.02396,
  \dodoi{10.48550/arXiv.2405.02396}

\bibitem[{{Lancaster} {et~al.}(2021{\natexlab{a}}){Lancaster}, {Ostriker},
  {Kim}, \& {Kim}}]{Lancaster21c}
{Lancaster}, L., {Ostriker}, E.~C., {Kim}, J.-G., \& {Kim}, C.-G.
  2021{\natexlab{a}}, \apjl, 922, L3, \dodoi{10.3847/2041-8213/ac3333}

\bibitem[{{Lancaster} {et~al.}(2021{\natexlab{b}}){Lancaster}, {Ostriker},
  {Kim}, \& {Kim}}]{Lancaster21a}
---. 2021{\natexlab{b}}, \apj, 914, 89, \dodoi{10.3847/1538-4357/abf8ab}

\bibitem[{{Lancaster} {et~al.}(2021{\natexlab{c}}){Lancaster}, {Ostriker},
  {Kim}, \& {Kim}}]{Lancaster21b}
---. 2021{\natexlab{c}}, \apj, 914, 90, \dodoi{10.3847/1538-4357/abf8ac}

\bibitem[{{Lee} \& {Hennebelle}(2018)}]{LeeHennebelle18}
{Lee}, Y.-N., \& {Hennebelle}, P. 2018, \aap, 611, A88,
  \dodoi{10.1051/0004-6361/201731522}

\bibitem[{{Leitherer} {et~al.}(1992){Leitherer}, {Robert}, \&
  {Drissen}}]{Leitherer92}
{Leitherer}, C., {Robert}, C., \& {Drissen}, L. 1992, \apj, 401, 596,
  \dodoi{10.1086/172089}

\bibitem[{{Leitherer} {et~al.}(1999){Leitherer}, {Schaerer}, {Goldader},
  {Delgado}, {Robert}, {Kune}, {de Mello}, {Devost}, \& {Heckman}}]{SB99}
{Leitherer}, C., {Schaerer}, D., {Goldader}, J.~D., {et~al.} 1999, \apjs, 123,
  3, \dodoi{10.1086/313233}

\bibitem[{{Leroy} {et~al.}(2017){Leroy}, {Schinnerer}, {Hughes}, {Kruijssen},
  {Meidt}, {Schruba}, {Sun}, {Bigiel}, {Aniano}, {Blanc}, {Bolatto},
  {Chevance}, {Colombo}, {Gallagher}, {Garcia-Burillo}, {Kramer}, {Querejeta},
  {Pety}, {Thompson}, \& {Usero}}]{Leroy17}
{Leroy}, A.~K., {Schinnerer}, E., {Hughes}, A., {et~al.} 2017, \apj, 846, 71,
  \dodoi{10.3847/1538-4357/aa7fef}

\bibitem[{{Levy} {et~al.}(2021){Levy}, {Bolatto}, {Leroy}, {Emig}, {Gorski},
  {Krieger}, {Lenki{\'c}}, {Meier}, {Mills}, {Ott}, {Rosolowsky}, {Tarantino},
  {Veilleux}, {Walter}, {Wei{\ss}}, \& {Zwaan}}]{Levy21}
{Levy}, R.~C., {Bolatto}, A.~D., {Leroy}, A.~K., {et~al.} 2021, \apj, 912, 4,
  \dodoi{10.3847/1538-4357/abec84}

\bibitem[{{Levy} {et~al.}(2024){Levy}, {Bolatto}, {Mayya}, {Cuevas-Otahola},
  {Tarantino}, {Boyer}, {Boogaard}, {B{\"o}ker}, {Cronin}, {Dale}, {Donaghue},
  {Emig}, {Fisher}, {Glover}, {Herrera-Camus}, {Jim{\'e}nez-Donaire},
  {Klessen}, {Lenki{\'c}}, {Leroy}, {De Looze}, {Meier}, {Mills}, {Ott},
  {Rela{\~n}o}, {Veilleux}, {Villanueva}, {Walter}, \& {van der Werf}}]{Levy24}
{Levy}, R.~C., {Bolatto}, A.~D., {Mayya}, D., {et~al.} 2024, arXiv e-prints,
  arXiv:2408.04135, \dodoi{10.48550/arXiv.2408.04135}

\bibitem[{{Lopez} {et~al.}(2011){Lopez}, {Krumholz}, {Bolatto}, {Prochaska}, \&
  {Ramirez-Ruiz}}]{Lopez11}
{Lopez}, L.~A., {Krumholz}, M.~R., {Bolatto}, A.~D., {Prochaska}, J.~X., \&
  {Ramirez-Ruiz}, E. 2011, \apj, 731, 91, \dodoi{10.1088/0004-637X/731/2/91}

\bibitem[{{Lopez} {et~al.}(2014){Lopez}, {Krumholz}, {Bolatto}, {Prochaska},
  {Ramirez-Ruiz}, \& {Castro}}]{Lopez14}
{Lopez}, L.~A., {Krumholz}, M.~R., {Bolatto}, A.~D., {et~al.} 2014, \apj, 795,
  121, \dodoi{10.1088/0004-637X/795/2/121}

\bibitem[{{Mac Low} \& {McCray}(1988)}]{maclow88}
{Mac Low}, M.-M., \& {McCray}, R. 1988, \apj, 324, 776, \dodoi{10.1086/165936}

\bibitem[{{McKee} {et~al.}(1984){McKee}, {van Buren}, \& {Lazareff}}]{MVL84}
{McKee}, C.~F., {van Buren}, D., \& {Lazareff}, B. 1984, \apjl, 278, L115,
  \dodoi{10.1086/184237}

\bibitem[{{M{\'e}ndez-Delgado} {et~al.}(2024){M{\'e}ndez-Delgado}, {Skillman},
  {Aver}, {Morisset}, {Esteban}, {Garc{\'\i}a-Rojas}, {Kreckel}, {Rogers},
  {Rosales-Ortega}, {Arellano-C{\'o}rdova}, {Flury}, {Reyes-Rodr{\'\i}guez}, \&
  {Orte-Garc{\'\i}a}}]{MendezDelgado24}
{M{\'e}ndez-Delgado}, J.~E., {Skillman}, E.~D., {Aver}, E., {et~al.} 2024,
  arXiv e-prints, arXiv:2410.17381, \dodoi{10.48550/arXiv.2410.17381}

\bibitem[{{Menon} {et~al.}(2024{\natexlab{a}}){Menon}, {Burkhart},
  {Somerville}, {Thompson}, \& {Sternberg}}]{Menon24b}
{Menon}, S.~H., {Burkhart}, B., {Somerville}, R.~S., {Thompson}, T.~A., \&
  {Sternberg}, A. 2024{\natexlab{a}}, arXiv e-prints, arXiv:2408.14591,
  \dodoi{10.48550/arXiv.2408.14591}

\bibitem[{{Menon} {et~al.}(2022){Menon}, {Federrath}, \& {Krumholz}}]{Menon22}
{Menon}, S.~H., {Federrath}, C., \& {Krumholz}, M.~R. 2022, \mnras, 517, 1313,
  \dodoi{10.1093/mnras/stac2702}

\bibitem[{{Menon} {et~al.}(2023){Menon}, {Federrath}, \& {Krumholz}}]{Menon23}
---. 2023, \mnras, 521, 5160, \dodoi{10.1093/mnras/stad856}

\bibitem[{{Menon} {et~al.}(2024{\natexlab{b}}){Menon}, {Lancaster}, {Burkhart},
  {Somerville}, {Dekel}, \& {Krumholz}}]{Menon24a}
{Menon}, S.~H., {Lancaster}, L., {Burkhart}, B., {et~al.} 2024{\natexlab{b}},
  \apjl, 967, L28, \dodoi{10.3847/2041-8213/ad462d}

\bibitem[{{Murray} {et~al.}(2010){Murray}, {Quataert}, \&
  {Thompson}}]{Murray10}
{Murray}, N., {Quataert}, E., \& {Thompson}, T.~A. 2010, \apj, 709, 191,
  \dodoi{10.1088/0004-637X/709/1/191}

\bibitem[{{Nakamura} {et~al.}(2006){Nakamura}, {McKee}, {Klein}, \&
  {Fisher}}]{Nakamura06}
{Nakamura}, F., {McKee}, C.~F., {Klein}, R.~I., \& {Fisher}, R.~T. 2006, \apjs,
  164, 477, \dodoi{10.1086/501530}

\bibitem[{{Ngoumou} {et~al.}(2015){Ngoumou}, {Hubber}, {Dale}, \&
  {Burkert}}]{Ngoumou15}
{Ngoumou}, J., {Hubber}, D., {Dale}, J.~E., \& {Burkert}, A. 2015, \apj, 798,
  32, \dodoi{10.1088/0004-637X/798/1/32}

\bibitem[{{Ntormousi} {et~al.}(2011){Ntormousi}, {Burkert}, {Fierlinger}, \&
  {Heitsch}}]{Ntormousi11}
{Ntormousi}, E., {Burkert}, A., {Fierlinger}, K., \& {Heitsch}, F. 2011, \apj,
  731, 13, \dodoi{10.1088/0004-637X/731/1/13}

\bibitem[{{Ostriker} \& {Kim}(2022)}]{PRFM22}
{Ostriker}, E.~C., \& {Kim}, C.-G. 2022, \apj, 936, 137,
  \dodoi{10.3847/1538-4357/ac7de2}

\bibitem[{{Ostriker} {et~al.}(2010){Ostriker}, {McKee}, \& {Leroy}}]{OML10}
{Ostriker}, E.~C., {McKee}, C.~F., \& {Leroy}, A.~K. 2010, \apj, 721, 975,
  \dodoi{10.1088/0004-637X/721/2/975}

\bibitem[{{Ostriker} \& {Shetty}(2011)}]{OstrikerShetty2011}
{Ostriker}, E.~C., \& {Shetty}, R. 2011, \apj, 731, 41,
  \dodoi{10.1088/0004-637X/731/1/41}

\bibitem[{{Pandya} {et~al.}(2020){Pandya}, {Somerville},
  {Angl{\'e}s-Alc{\'a}zar}, {Hayward}, {Bryan}, {Fielding}, {Forbes},
  {Burkhart}, {Genel}, {Hernquist}, {Kim}, {Tonnesen}, \&
  {Starkenburg}}]{Pandya20}
{Pandya}, V., {Somerville}, R.~S., {Angl{\'e}s-Alc{\'a}zar}, D., {et~al.} 2020,
  \apj, 905, 4, \dodoi{10.3847/1538-4357/abc3c1}

\bibitem[{{Pandya} {et~al.}(2023){Pandya}, {Fielding}, {Bryan}, {Carr},
  {Somerville}, {Stern}, {Faucher-Gigu{\`e}re}, {Hafen},
  {Angl{\'e}s-Alc{\'a}zar}, \& {Forbes}}]{Pandya23}
{Pandya}, V., {Fielding}, D.~B., {Bryan}, G.~L., {et~al.} 2023, \apj, 956, 118,
  \dodoi{10.3847/1538-4357/acf3ea}

\bibitem[{{Pascale} {et~al.}(2023){Pascale}, {Dai}, {McKee}, \&
  {Tsang}}]{Pascale23}
{Pascale}, M., {Dai}, L., {McKee}, C.~F., \& {Tsang}, B. T.~H. 2023, \apj, 957,
  77, \dodoi{10.3847/1538-4357/acf75c}

\bibitem[{{Pellegrini} {et~al.}(2011){Pellegrini}, {Baldwin}, \&
  {Ferland}}]{pellegrini11}
{Pellegrini}, E.~W., {Baldwin}, J.~A., \& {Ferland}, G.~J. 2011, \apj, 738, 34,
  \dodoi{10.1088/0004-637X/738/1/34}

\bibitem[{{Pellegrini} {et~al.}(2007){Pellegrini}, {Baldwin}, {Brogan},
  {Hanson}, {Abel}, {Ferland}, {Nemala}, {Shaw}, \& {Troland}}]{Pellegrini07}
{Pellegrini}, E.~W., {Baldwin}, J.~A., {Brogan}, C.~L., {et~al.} 2007, \apj,
  658, 1119, \dodoi{10.1086/511258}

\bibitem[{{Pittard}(2013)}]{Pittard13}
{Pittard}, J.~M. 2013, \mnras, 435, 3600, \dodoi{10.1093/mnras/stt1552}

\bibitem[{{Pittard} {et~al.}(2022){Pittard}, {Kupilas}, \&
  {Wareing}}]{Pittard22Rad}
{Pittard}, J.~M., {Kupilas}, M.~M., \& {Wareing}, C.~J. 2022, \mnras, 510,
  2797, \dodoi{10.1093/mnras/stab3707}

\bibitem[{{Raga} {et~al.}(2012{\natexlab{a}}){Raga}, {Cant{\'o}}, \&
  {Rodr{\'\i}guez}}]{Raga12b}
{Raga}, A.~C., {Cant{\'o}}, J., \& {Rodr{\'\i}guez}, L.~F. 2012{\natexlab{a}},
  \rmxaa, 48, 149

\bibitem[{{Raga} {et~al.}(2012{\natexlab{b}}){Raga}, {Cant{\'o}}, \&
  {Rodr{\'\i}guez}}]{Raga12c}
---. 2012{\natexlab{b}}, \rmxaa, 48, 199

\bibitem[{{Rahner} {et~al.}(2017){Rahner}, {Pellegrini}, {Glover}, \&
  {Klessen}}]{Rahner17}
{Rahner}, D., {Pellegrini}, E.~W., {Glover}, S. C.~O., \& {Klessen}, R.~S.
  2017, \mnras, 470, 4453, \dodoi{10.1093/mnras/stx1532}

\bibitem[{{Rahner} {et~al.}(2019){Rahner}, {Pellegrini}, {Glover}, \&
  {Klessen}}]{Rahner19}
---. 2019, \mnras, 483, 2547, \dodoi{10.1093/mnras/sty3295}

\bibitem[{{Rosen} {et~al.}(2014){Rosen}, {Lopez}, {Krumholz}, \&
  {Ramirez-Ruiz}}]{Rosen14}
{Rosen}, A.~L., {Lopez}, L.~A., {Krumholz}, M.~R., \& {Ramirez-Ruiz}, E. 2014,
  \mnras, 442, 2701, \dodoi{10.1093/mnras/stu1037}

\bibitem[{{Rousseau-Nepton} {et~al.}(2019){Rousseau-Nepton}, {Martin},
  {Robert}, {Drissen}, {Amram}, {Prunet}, {Martin}, {Moumen}, {Adamo},
  {Alarie}, {Barmby}, {Boselli}, {Bresolin}, {Bureau}, {Chemin}, {Fernandes},
  {Combes}, {Crowder}, {Della Bruna}, {Duarte Puertas}, {Egusa}, {Epinat},
  {Ksoll}, {Girard}, {G{\'o}mez Llanos}, {Gouliermis}, {Grasha}, {Higgs},
  {Hlavacek-Larrondo}, {Ho}, {Iglesias-P{\'a}ramo}, {Joncas}, {Kam}, {Karera},
  {Kennicutt}, {Klessen}, {Lianou}, {Liu}, {Liu}, {de Amorim}, {Lyman},
  {Martel}, {Mazzilli-Ciraulo}, {McLeod}, {Melchior}, {Millan}, {Moll{\'a}},
  {Momose}, {Morisset}, {Pan}, {Pati}, {Pellerin}, {Pellegrini}, {P{\'e}rez},
  {Petric}, {Plana}, {Rahner}, {Ruiz Lara}, {S{\'a}nchez-Menguiano},
  {Spekkens}, {Stasi{\'n}ska}, {Takamiya}, {Vale Asari}, \&
  {V{\'\i}lchez}}]{SIGNALS}
{Rousseau-Nepton}, L., {Martin}, R.~P., {Robert}, C., {et~al.} 2019, \mnras,
  489, 5530, \dodoi{10.1093/mnras/stz2455}

\bibitem[{{Schneider} {et~al.}(2020){Schneider}, {Simon}, {Guevara},
  {Buchbender}, {Higgins}, {Okada}, {Stutzki}, {G{\"u}sten}, {Anderson},
  {Bally}, {Beuther}, {Bonne}, {Bontemps}, {Chambers}, {Csengeri}, {Graf},
  {Gusdorf}, {Jacobs}, {Justen}, {Kabanovic}, {Karim}, {Luisi}, {Menten},
  {Mertens}, {Mookerjea}, {Ossenkopf-Okada}, {Pabst}, {Pound}, {Richter},
  {Reyes}, {Ricken}, {R{\"o}llig}, {Russeil}, {S{\'a}nchez-Monge}, {Sandell},
  {Tiwari}, {Wiesemeyer}, {Wolfire}, {Wyrowski}, {Zavagno}, \&
  {Tielens}}]{SOFIA_FEEDBACK}
{Schneider}, N., {Simon}, R., {Guevara}, C., {et~al.} 2020, \pasp, 132, 104301,
  \dodoi{10.1088/1538-3873/aba840}

\bibitem[{{Silich} \& {Tenorio-Tagle}(2013)}]{SilichTT13}
{Silich}, S., \& {Tenorio-Tagle}, G. 2013, \apj, 765, 43,
  \dodoi{10.1088/0004-637X/765/1/43}

\bibitem[{{Skinner} \& {Ostriker}(2015)}]{Skinner15}
{Skinner}, M.~A., \& {Ostriker}, E.~C. 2015, \apj, 809, 187,
  \dodoi{10.1088/0004-637X/809/2/187}

\bibitem[{{Somerville} \& {Dav{\'e}}(2015)}]{SomervilleDave15}
{Somerville}, R.~S., \& {Dav{\'e}}, R. 2015, \araa, 53, 51,
  \dodoi{10.1146/annurev-astro-082812-140951}

\bibitem[{{Spitzer}(1978)}]{Spitzer78}
{Spitzer}, L. 1978, {Physical processes in the interstellar medium},
  \dodoi{10.1002/9783527617722}

\bibitem[{{Steigman} {et~al.}(1975){Steigman}, {Strittmatter}, \&
  {Williams}}]{Steigman75}
{Steigman}, G., {Strittmatter}, P.~A., \& {Williams}, R.~E. 1975, \apj, 198,
  575, \dodoi{10.1086/153636}

\bibitem[{{Str{\"o}mgren}(1939)}]{Stromgren39}
{Str{\"o}mgren}, B. 1939, \apj, 89, 526, \dodoi{10.1086/144074}

\bibitem[{{Sun} {et~al.}(2022){Sun}, {Leroy}, {Rosolowsky}, {Hughes},
  {Schinnerer}, {Schruba}, {Koch}, {Blanc}, {Chiang}, {Groves}, {Liu}, {Meidt},
  {Pan}, {Pety}, {Querejeta}, {Saito}, {Sandstrom}, {Sardone}, {Usero},
  {Utomo}, {Williams}, {Barnes}, {Benincasa}, {Bigiel}, {Bolatto}, {Boquien},
  {Chevance}, {Dale}, {Deger}, {Emsellem}, {Glover}, {Grasha}, {Henshaw},
  {Klessen}, {Kreckel}, {Kruijssen}, {Ostriker}, \& {Thilker}}]{Sun22}
{Sun}, J., {Leroy}, A.~K., {Rosolowsky}, E., {et~al.} 2022, \aj, 164, 43,
  \dodoi{10.3847/1538-3881/ac74bd}

\bibitem[{{Sun} {et~al.}(2024){Sun}, {He}, {Batschkun}, {Levy}, {Emig},
  {Rodr{\'\i}guez}, {Hassani}, {Leroy}, {Schinnerer}, {Ostriker}, {Wilson},
  {Bolatto}, {Mills}, {Rosolowsky}, {Lee}, {Dale}, {Larson}, {Thilker},
  {Ubeda}, {Whitmore}, {Williams}, {Barnes}, {Bigiel}, {Chevance}, {Glover},
  {Grasha}, {Groves}, {Henshaw}, {Indebetouw}, {Jim{\'e}nez-Donaire},
  {Klessen}, {Koch}, {Liu}, {Mathur}, {Meidt}, {Menon}, {Neumann}, {Pinna},
  {Querejeta}, {Sormani}, \& {Tress}}]{Sun24}
{Sun}, J., {He}, H., {Batschkun}, K., {et~al.} 2024, \apj, 967, 133,
  \dodoi{10.3847/1538-4357/ad3de6}

\bibitem[{{Tenorio-Tagle}(1979)}]{TT79}
{Tenorio-Tagle}, G. 1979, \aap, 71, 59

\bibitem[{{Vink} {et~al.}(2001){Vink}, {de Koter}, \& {Lamers}}]{Vink01}
{Vink}, J.~S., {de Koter}, A., \& {Lamers}, H.~J.~G.~L.~M. 2001, \aap, 369,
  574, \dodoi{10.1051/0004-6361:20010127}

\bibitem[{{Vink} \& {Sander}(2021)}]{VinkSander21}
{Vink}, J.~S., \& {Sander}, A. A.~C. 2021, \mnras, 504, 2051,
  \dodoi{10.1093/mnras/stab902}

\bibitem[{{Virtanen} {et~al.}(2020){Virtanen}, {Gommers}, {Oliphant},
  {Haberland}, {Reddy}, {Cournapeau}, {Burovski}, {Peterson}, {Weckesser},
  {Bright}, {van der Walt}, {Brett}, {Wilson}, {Millman}, {Mayorov}, {Nelson},
  {Jones}, {Kern}, {Larson}, {Carey}, {Polat}, {Feng}, {Moore}, {Vand erPlas},
  {Laxalde}, {Perktold}, {Cimrman}, {Henriksen}, {Quintero}, {Harris},
  {Archibald}, {Ribeiro}, {Pedregosa}, {van Mulbregt}, \& {SciPy 1. 0
  Contributors}}]{scipy}
{Virtanen}, P., {Gommers}, R., {Oliphant}, T.~E., {et~al.} 2020, Nature
  Methods, 17, 261, \dodoi{10.1038/s41592-019-0686-2}

\bibitem[{{Vishniac}(1983)}]{Vishniac83}
{Vishniac}, E.~T. 1983, \apj, 274, 152, \dodoi{10.1086/161433}

\bibitem[{{Vishniac}(1994)}]{Vishniac94}
---. 1994, \apj, 428, 186, \dodoi{10.1086/174231}

\bibitem[{{Vishniac} \& {Ryu}(1989)}]{VishniacRyu89}
{Vishniac}, E.~T., \& {Ryu}, D. 1989, \apj, 337, 917, \dodoi{10.1086/167161}

\bibitem[{{Voit} {et~al.}(2024{\natexlab{a}}){Voit}, {Carr}, {Fielding},
  {Pandya}, {Bryan}, {Donahue}, {Oppenheimer}, \& {Somerville}}]{Voit24b}
{Voit}, G.~M., {Carr}, C., {Fielding}, D.~B., {et~al.} 2024{\natexlab{a}},
  \apj, 976, 151, \dodoi{10.3847/1538-4357/ad81d5}

\bibitem[{{Voit} {et~al.}(2024{\natexlab{b}}){Voit}, {Pandya}, {Fielding},
  {Bryan}, {Carr}, {Donahue}, {Oppenheimer}, \& {Somerville}}]{Voit24a}
{Voit}, G.~M., {Pandya}, V., {Fielding}, D.~B., {et~al.} 2024{\natexlab{b}},
  \apj, 976, 150, \dodoi{10.3847/1538-4357/ad81d6}

\bibitem[{{Weaver} {et~al.}(1977){Weaver}, {McCray}, {Castor}, {Shapiro}, \&
  {Moore}}]{Weaver77}
{Weaver}, R., {McCray}, R., {Castor}, J., {Shapiro}, P., \& {Moore}, R. 1977,
  \apj, 218, 377, \dodoi{10.1086/155692}

\end{thebibliography}
\bibliographystyle{aasjournal}

\end{document}